 \documentclass[final,3p,times]{elsarticle}

\usepackage{dsfont,amsmath,amssymb}
\usepackage{epsfig}
\usepackage{color}

\usepackage[normalem]{ulem}

\def\b0{{\bf 0}}

\def\bfe{{\bf e}}
\def\bk{{\bf k}}

\def\bp{{\bf p}}
\def\br{{\bf r}}

\def\bx{{\bf x}}
\def\by{{\bf y}}

\def\bq{{\bf q}}
\def\bz{{\bf z}}

\def\cR{{\cal R}}

\def\grad{\hbox{$\boldmath{\nabla}$}}

\def\ul{\underline}
\def\wt{\widetilde}

\def\llangle{\left\langle}
\def\rrangle{\right\rangle}

\def\btimes{\hbox{\boldmath $\times$}}
\def\bdot{\hbox{\boldmath $\cdot$}}

\def\grad{\hbox{\boldmath $\nabla$}}
\def\bell{\hbox{\boldmath $\ell$}}
\def\bpi{\hbox{\boldmath $\pi$}}
\def\barphi{\hbox{\boldmath $\varphi$}}
\def\btheta{\hbox{\boldmath $\theta$}}
\def\bOmega{\hbox{\boldmath $\Omega$}}
\def\bzed{\hbox{\boldmath $0$}}

\def\be{\begin{equation}}
\def\ee{\end{equation}}
\def\lb{\label}

\def\dbar{\,\,{\mathchar'26\mkern-10mu d}}
\def\ddelta{\,\,{\mathchar'26\mkern-7mu \delta}}

\newtheorem{theorem}{Theorem}[section]

\newtheorem{prop}[theorem]{Proposition}

\newenvironment{pf}[1][Proof]{\begin{trivlist}
\item[\hskip \labelsep {\bfseries #1}]}{\end{trivlist}}

\journal{Physica D: Nonlinear Phenomenon}

\begin{document}

\begin{frontmatter}

\title{Resonance Van Hove Singularities in Wave Kinetics}

\author{ Yi-Kang Shi${\,\!}^{1}$ and Gregory L. Eyink${\,\!}^{1,2}$}

\address{(1) Department of Applied Mathematics \& Statistics and (2) Department of Physics \& Astronomy,\\
The Johns Hopkins University, Baltimore, MD, USA}
 
\begin{abstract}
Wave kinetic theory has been developed to describe the statistical dynamics of weakly nonlinear, dispersive waves. 
However, we show that systems which are generally dispersive can have resonant sets of wave modes with identical group velocities, leading to a local breakdown of dispersivity. This shows up as a geometric singularity of the resonant manifold and possibly as an infinite phase measure in the collision integral. Such singularities occur widely for classical wave systems, including acoustical waves, Rossby waves, helical waves in rotating fluids, light waves in nonlinear optics and also in quantum transport, e.g. kinetics of electron-hole excitations (matter waves) in graphene. These singularities are the exact analogue of the critical points found by Van Hove in 1953 for phonon dispersion relations in crystals. The importance of these singularities in wave kinetics depends on the dimension of phase space $D=(N-2)d$ ($d$ physical space dimension, $N$ the number of waves in resonance) and the degree of degeneracy $\delta$ of the critical points.
Following Van Hove, we show that non-degenerate singularities lead to finite phase measures for $D>2$ but produce divergences
when $D\leq 2$ and possible breakdown of wave kinetics if the collision integral itself becomes too large (or even infinite).
Similar divergences and possible breakdown can occur for degenerate singularities,  when $D-\delta\leq 2,$ as we find for several physical examples, 
including electron-hole kinetics in graphene. When the standard kinetic equation breaks down, then one must  develop a new singular wave kinetics.
We discuss approaches from pioneering 1971 work of Newell \& Aucoin on multi-scale perturbation theory for acoustic waves and field-theoretic 
methods based on exact Schwinger-Dyson integral equations for the wave dynamics. 
 %These divergences are not removed by nonlinear broadening of the resonances. Futhermore, we give a new discussion on the divergence of degenerate singularities based on the Hessian matrix and higher order derivatives near the critical points. We apply the results to show the local well-posedness of the wave kinetic equation and kinetic hierarchy following Lanford's idea on the Boltzmann equation.  
 \end{abstract}


\end{frontmatter}

%%%%%%%%%%%%%%%%%%%%%%%%%%%%%%%%%
%%%%%%%%%%%%%%%%%%%%%%%%%%%%%%%%%
\section{Introduction} 
%%%%%%%%%%%%%%%%%%%%%%%%%%%%%%%%%
%%%%%%%%%%%%%%%%%%%%%%%%%%%%%%%%%

Wave kinetic equations were first introduced by Peierls in 1929 to discuss thermal transport by phonons in crystals 
\cite{Peierls29} and have since been extended to a large number of quantum and classical wave systems \cite{Zakharovetal92,
Nazarenko11,NewellRumpf11,EyinkShi12}. However, many questions remain about the validity of these equations, 
in which wavenumber regimes they hold and under what precise assumptions on the underlying wave dynamics. 
Some authors based their derivation of the wave kinetic equation on the RPA (``random phases and amplitudes") assumption \cite{Nazarenko11},
but it has been cogently argued \cite{NewellRumpf11,Spohn06} that one can relax the RPA assumption to dispersivity of the waves. In this paper 
we argue that the dispersivity requirement is more stringent than what has commonly been understood. In particular, a wave system that 
is generally dispersive can experience a breakdown of dispersivity locally in the $N$-wave phase space which renders the wave kinetic equation ill-defined. 
Consider, for example, a general 3-wave equation for the evolution of the wave-action spectrum, $n(\bk,t):$
\begin{eqnarray}
 \partial_t n(\bk,t)
%+\grad_\bk\omega(\bk,\bx)\cdot\grad_\bx W(\bk,\bx,t)-\grad_\bx\omega(\bk,\bx)\cdot\grad_\bk W(\bk,\bx,t)\cr
&=&36\pi \sum_{\ul{s}=(-1,s_2,s_3)} \int d^dk_2 \int d^dk_3\ |H^{\ul{s}}_{\ul{\bk}}|^2
\delta(\ul{s}\cdot\omega(\ul{\bk}))\ddelta^d(\ul{s}\cdot\ul{\bk})\cr
&& \,\,\,\,\,\,\,\,\,\,\,\,\,\,\,\,\,\,\,\,\,\,\,\,\,\,\,
\times \Big\{n(\bk_2)n(\bk_3)-s_2 n(\bk)n(\bk_3)-s_3 n(\bk)n(\bk_2)\Big\}.  
\lb{wavkineq}\end{eqnarray}
Here we use the shorthand notations $\ul{\bk}=(\bk,\bk_2,\bk_3),\ul{s}=
(s,s_2,s_3)$ with $\ul{a}\cdot\ul{b}=ab+a_2b_2+a_3b_3.$ The integer $s$ labels the degree 
of degeneracy of the waves, or the number of frequencies $\omega_s$ corresponding to a given wavevector $\bk$. We have 
assumed above that $s=\pm 1$ and $\omega_s(\bk)=s \omega(\bk),$ appropriate for 
systems second-order in time where there are two waves traveling in opposite directions. 
Because of the Dirac delta functions, the collision integral is restricted 
to the {\it resonant manifold}:
$$ {\mathcal R}_{\bk}= \Big\{\bk_2: s\omega(\bk)+s_2\omega(\bk_2)+s_3 \omega(\bk_3)
\Big|_{\bk_3=-s_3(s\bk+s_2\bk_2)}=0\Big\}. $$
An immediate issue in making sense of the kinetic equation is the well-definedness
of the measure on ${\mathcal R}_{\bk}:$
\begin{eqnarray} 
 \int d^dk_2\int d^dk_3\ \delta(\ul{s}\cdot\omega(\ul{\bk})) \delta^d(\ul{s}\cdot\ul{\bk})
= \int_{{\mathcal R}_{\bk}} \frac{dS(\bk_2)}{|\grad \omega(\bk_2)-\grad \omega(\bk_3)|}.
\lb{meas} 
\end{eqnarray}
In the last equality we have used a standard formula to express the measure in terms of the surface area $S$ on the 
imbedded manifold in Euclidean space (e.g. see \cite{Hormander89}, theorem 6.1.5). The difficulty occurs at critical points where
$$ \grad \omega(\bk_2)=\grad \omega(\bk_3) $$
and the denominator vanishes. Since $\grad \omega(\bk)$ is the group velocity of wavepackets, such points 
correspond physically to wavevector triads at which the dispersivity of the wave system breaks down locally and for which two distinct
wavepackets from the triad propagate together for all times with the same velocity. The kinetic equation may be ill-defined for $\bk$ at which the denominator produces a non-integrable singularity. 

An exactly analogous problem was studied in 1953 by Van Hove \cite{VanHove53} for the phonon density of states in a crystalline solid,
which is the function defined by 
$$ g(\omega) = L^d \sum_s \int \dbar^dk \,\, \delta^d(\omega-\omega_s(\bk))
=\frac{L^d}{(2\pi)^d}\sum_s \int_{\omega_s(\bk)=\omega} \frac{dS(\bk)}{|\grad\omega_s(\bk)|}$$
where the the sum $s$ is over the branches of phonon frequencies and the $\bk$-integral is over a Brillouin zone 
of the crystal. Critical points of the phonon dispersion relations, where the group velocity $\grad\omega_s(\bk)$ vanishes, 
may lead to singularities in the density of states.
Indeed, by a beautiful application of the Morse inequalities \cite{Milnor63}, Van Hove showed that the homology groups 
of the Brillouin zone (topologically a $d$-torus) make such critical points inevitable. He showed further by a local analysis of the 
critical points using the Morse Lemma \cite{Milnor63,Lang12} that the resulting singularities are non-integrable for space dimension $d=2,$ 
giving rise to a logarithmic divergence in the density of states, and are integrable for $d=3,$ producing divergences 
only in the derivative (cusps). As we shall see in the following, the local analysis of Van Hove carries over to the 
corresponding point singularities in the measures (\ref{meas}) on the resonant manifolds, 
%allowing us under appropriate conditions to establish their finiteness. 
%An alternative construction of these measures 
%by Lukkarinen \& Spohn (\cite{LukkarinenSpohn07}, Appendix A) assumed that the dispersion relation is a Morse 
%function of the wavenumber, i.e. a $C^2$ function with only non-degenerate critical points. This assumption is too
%restrictive, however, as it rules out most physically relevant dispersion relations.
%%, which frequently are of power-law 
%%form $\propto k^\alpha$ at zero wavenumber and not $C^2$ when $\alpha<2.$ This is exactly the situation 
%%considered by Van Hove also, who had to deal separately with the acoustic branch of phonons whose dispersion 
%%law at $k=0$ has exponent $\alpha=1.$ 
%We therefore undertake a direct local analysis 
%of the critical points \footnote{This local analysis is really implicit in the approach of \cite{LukkarinenSpohn07}, which
%is based on decay of a ``free propagator'' $p_t(\bx)=\int d^dk \, e^{2\pi i\bx\bdot\bk} e^{-i t\omega(\bk)}\sim t^{-d/2}.$
%The proof of this decay for Morse functions (e.g. \cite{Hormander89}, Theorem 7.7.5) uses a partition of unity to isolate 
%the effects of the critical points.}\lb{propagator}, 
which we propose to call {\it resonance Van Hove singularities} because of their close 
physical and mathematical analogies with the singularities considered by Van Hove.  

It is important to emphasize that the global topological argument for existence of critical points given by Van Hove 
does not carry over to the wave kinetic problem, even when there is a spatial lattice and the reciprocal wavenumber 
space is a $d$-torus. The most essential difference is that the density of states function involves all level sets 
of the dispersion relation, whereas only the zero level sets are relevant for wave kinetics.
Nevertheless, topological criteria for existence of singularities can be sometimes exploited in wave kinetics.
For example, in the absence of singular points, the resonant manifold can deform continuously from $\cR_{\bk_1}$ to $\cR_{\bk_2}$ for any $\bk_1$ and $\bk_2$.
It is thus sufficient to establish the existence of singularity if one can find certain topological invariants of 
two manifolds $\cR_{\bk_1}$ and $\cR_{\bk_2}$ for $\bk_1\neq\bk_2$, e.g., the fundamental group or the homology group, which are different.
We shall not develop general criteria along this line but instead show by many concrete examples below that critical points often appear 
on the resonant manifold in practice.  Figure 1 plots a simple example \footnote{All of the 
resonant manifolds exhibited in this paper are plotted with {\tt MATLAB}, using {\tt contour} for $D=2$ and 
{\tt isosurface} for $D\geq 3$} in which physical space is a cubic lattice and wavenumber space is a 3-torus, 
which can have relevance for numerical simulations of wave turbulence on a rectangular grid (for more discussion, 
see section \ref{lattice}). This example illustrates the general geometric feature of resonant Van Hove singularities that  
the ``resonant manifold'' is no longer a true manifold, locally diffeomorphic  to Euclidean space. 
When zero is a regular value of the resonance condition, then 
the Regular Value Theorem/Submersion Theorem \cite{GuilleminPollack74}, guarantees that the ``resonant manifold" is indeed a smooth 
manifold, but this is not usually the case when zero is a critical value.
If the dispersion relation is not a Morse function (i.e. a smooth function with only non-degenerate critical points), 
then the singularities can occur also along critical lines and surfaces within the resonant manifold. 

% FIGURE HERE

\vspace{5pt} 

\begin{figure}[!h]
\begin{center}\lb{Fig1}
\includegraphics[width=3.5in,height=2.5in,keepaspectratio]{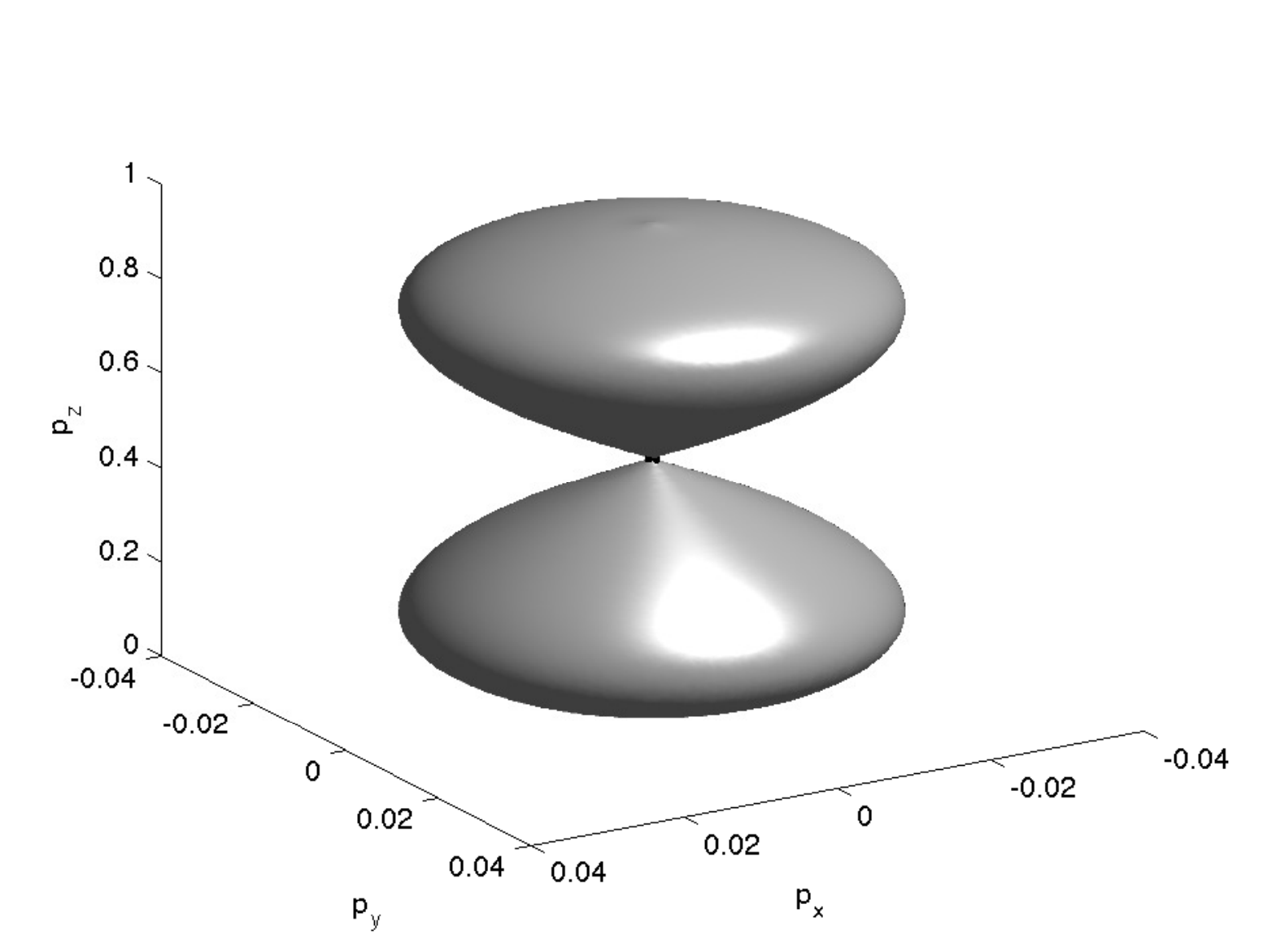}
\end{center}
\caption{Three-wave resonant manifold ${\mathcal R}_\bk=\{\bp:\,\omega(\bp)+\omega(\bk-\bp)-\omega(\bk)=0\}$ for dispersion relation
$\omega(\bk)=\left[4\sum_{i=1}^3 \sin^2\left(\frac{k_i}{2}\right)\right]^{\alpha/2},$ $\alpha=\frac{1}{1+\log_2\cos(1/4)},$
on the 3-torus $[-\pi,\pi]^3,$ and for the specific wavenumber $\bk=(0,0,1).$ There is a critical point (black dot) at $\bp=\bk/2.$
}
\end{figure}

The critical set on the singular manifold may have effects on the wave kinetic theory either little or drastic, depending on their character.
As in the analysis of Van Hove, isolated critical points have only very mild consequences for 3-wave resonances in space dimensions $d\geq 3,$ but lead to divergences for $d=1,2$, which may alter the naive predictions of the wave kinetic theory or vitiate the theory entirely.
For higher-order wave resonances and in higher space dimensions, critical lines and surfaces can have similar effects.
We shall give below examples of resonance Van Hove singularities for many concrete systems, including acoustic waves in compressible fluids, helical waves in rotating incompressible fluids, capillary-gravity waves on free fluid surfaces, one-dimensional optical waves, electron-hole matter waves in graphene, etc. A pioneering study of acoustic wave turbulence by Newell 
\& Aucoin \cite{NewellAucoin71} argued that the breakdown of dispersivity of sound waves on the resonant manifold leads to a different long-time asymptotics
and a modified kinetic equation. 
The breakdown for acoustic waves is extremely severe, in that all points on the resonant manifold are critical (for more discussion, see section \ref{iso-power}). 
As we shall discuss in this work, even a single critical point on the resonant manifold can lead to breakdown of standard wave kinetics, 
which then must be replaced with a new {\it singular kinetic equation} with a collision integral resulting only from resonant interactions in the critical set. 

The detailed contents of this work are as follows. In section \ref{example} we present many physical examples of wave systems 
with various degrees of divergence in the phase measure and collision integral due to local breakdown of dispersivity and 
we discuss some of the specific mechanisms which can produce such critical points. 
 Next we present in section \ref{finite} a few general results on the effects of these singularities on the local finiteness 
of the phase measure, including the argument of Van Hove for non-degenerate critical points and some analysis of the effects of degeneracy. 
Finally, in section \ref{singular}, we consider briefly the singular wave kinetics that becomes relevant when the standard kinetic equation breaks down, 
developing the multi-scale asymptotics suggested in \cite{NewellAucoin71} and also the field-theoretic approach to 
generalized kinetic equations suggested in \cite{Lvovetal97}.  The main example we consider in this section, of some physical interest 
by itself, is the kinetics of the electron-hole plasma in defectless graphene \cite{Kashuba08, Fritzetal08}. Full details 
of the derivation of the quantum kinetic equation, which are relatively standard in the wave turbulence community, are relegated 
to an Appendix.  

%[CHANGE] 

%The detailed contents of this paper are follows:

%%%%%%%%%%%%%%%%%%%%%%%%%%%%%%%%%%%%%%%%%%%%%%
%%%%%%%%%%%%%%%%%%%%%%%%%%%%%%%%%%%%%%%%%%%%%%
%%%%%%%%%%%%%%%%%%%%%%%%%%%%%%%%%%%%%%%%%%%%%%
\section{A Case Study of Resonance Van Hove Singularities}\lb{example}
%%%%%%%%%%%%%%%%%%%%%%%%%%%%%%%%%%%%%%%%%%%%%%
%%%%%%%%%%%%%%%%%%%%%%%%%%%%%%%%%%%%%%%%%%%%%%
%%%%%%%%%%%%%%%%%%%%%%%%%%%%%%%%%%%%%%%%%%%%%%

In this section, we want to consider singularities that occur for various dispersion relations commonly discussed in the literature.
For simplicity, we assume throughout that the dispersion relation depends only upon wavenumber, not position. 
We begin with the simplest case of 3-wave resonances.

%%%%%%%%%%%%%%%%%%%%%%%%%%%%%%%%%%%%%%%%%%%%%%
%%%%%%%%%%%%%%%%%%%%%%%%%%%%%%%%%%%%%%%%%%%%%%
%%%%%%%%%%%%%%%%%%%%%%%%%%%%%%%%%%%%%%%%%%%%%%
\subsection{Triplet Resonances} 
%%%%%%%%%%%%%%%%%%%%%%%%%%%%%%%%%%%%%%%%%%%%%%
%%%%%%%%%%%%%%%%%%%%%%%%%%%%%%%%%%%%%%%%%%%%%%
%%%%%%%%%%%%%%%%%%%%%%%%%%%%%%%%%%%%%%%%%%%%%%

The conditions of resonance for triplet interaction can be written, without loss of generality, as 
\be s_2\omega(\bp) + s_3 \omega(\bq) = \omega (\bk) \lb{res1} \ee
\be s_2 \bp + s_3 \bq = \bk \lb{res2} \ee
by taking $s'_i=-s_is\rightarrow s_i,$ and the condition for singularity as
\be \grad\omega(\bp)=\grad\omega(\bq). \lb{sing} \ee 

%%%%%%%%%%%%%%%%%%%%%%%%%%%%%%%%%%%%%%%%%%%%%%
%%%%%%%%%%%%%%%%%%%%%%%%%%%%%%%%%%%%%%%%%%%%%%
\subsubsection{Isotropic Power Law}\lb{iso-power}
%%%%%%%%%%%%%%%%%%%%%%%%%%%%%%%%%%%%%%%%%%%%%%
%%%%%%%%%%%%%%%%%%%%%%%%%%%%%%%%%%%%%%%%%%%%%%

In wave turbulence literature, the most commonly considered dispersion relation is an isotropic power-law: 
\be \omega(\bk)=Ck^\alpha. \lb{power} \ee
Resonance requires $\alpha\geqslant1$ \cite{Nazarenko11}. The singularity condition (\ref{sing}) becomes
\be p^{\alpha-2}\bp= q^{\alpha-2}\bq \lb{pow1} \ee
and thus 
\be p^{\alpha-1} = q^{\alpha-1}. \lb{pow2} \ee

We treat first the case $\alpha>1,$ where clearly (\ref{pow2}) implies $p=q$ and (\ref{pow1}) then implies $\bp=\bq.$
Substituting back into (\ref{res1}),(\ref{res2}) gives
$$ k^\alpha=(s_2+s_3)p^\alpha, \,\,\,\,\bk=(s_2+s_3)\bq. $$
If $p=q=0,$ then also $k=0,$ which is the trivial resonance with all members of the triad zero. If $p=q>0,$
then we can have either $s_2+s_3=2$ or $s_2+s_3=0.$ However, $s_2+s_3=2$
implies both $k^\alpha=2p^\alpha$ and $k=2p,$ which requires $\alpha=1,$ a contradiction. If instead 
$s_2+s_3=0,$ then $\bk=\bzed$ and any $\bp=\bq$ are allowed. The ``resonant manifold'' is 
here all of space. However, for any dynamics which preserves the space average of the wave field, the interaction 
leaves the zero-wavenumber Fourier amplitude invariant. The dynamics of the $\bk=0$ mode is then null and 
there is no interest in the kinetic equation for that wavenumber. (This conclusion is, however, a bit too glib,
as we shall discuss further in Section \ref{anisotropic} below.)  Thus, no singularities of a nontrivial type are allowed
for $\alpha>1.$ 

The case $\alpha=1$ is different, as has been long understood \cite{ZakharovSagdeev70,NewellAucoin71}. For this situation,
the condition (\ref{pow2}) gives no restriction on $p,q$ and (\ref{pow1}) is the condition of collinearity:
\be \hat{\bp}=\hat{\bq}. \lb{collin} \ee  
For $s_2=s_3=1,$ the conditions
$$ k=p+q, \,\,\,\, \bk=\bp+\bq $$
are satisfied by 
$$ \bp = \beta\bk, \,\,\,\, \bq = \bk-\bp = (1-\beta) \bk, \,\,\,\,\, \beta \in [0,1]$$
For $s_2=-s_3=1$, the conditions
$$ p=k+q, \,\,\,\, \bp=\bk+\bq $$
are satisfied by 
$$ \bp = \gamma \bk, \,\,\,\, \bq = \bp-\bk = (\gamma-1) \bk, \,\,\,\,\, \gamma \in [1,\infty)$$
Finally, for $-s_2=s_3=1$, we just reverse $\bp\leftrightarrow \bq$ in the last two equations. Because of our 
initial choice $s=-1,$ all of the above solutions have $\bp\bdot \bk>0.$ Taking $\bk\rightarrow -\bk,\,\,\,$ 
$|\bk|=k\rightarrow |-\bk|=k$ gives solutions with $\bp\bdot \bk<0.$ 
Note that the resonant ``manifolds" where 
\be E^{s_2s_3}(\bp;\bk)=s_2\omega(\bp)+s_3\omega(s_3(\bk-s_2\bp))-\omega(\bk)=0 \lb{resfun} \ee
are now straight line segments or rays, and not manifolds at all. The entire ``manifold'' is in the critical set 
where $\grad_{\bp}E^{s_2s_3}(\bp;\bk)=\bzed.$ It is a consequence of the Morse Lemma that non-degenerate 
critical points must be isolated, but the critical points here form a continuum. In fact, these critical points 
are all one-dimensionally degenerate, since the Hessian $\grad\otimes\grad E^{s_2s_3}(\bp;\bk)$ 
has $\hat{\bk}$ as an eigenvector with eigenvalue 0. 

The dispersion law (\ref{power}) with $\alpha=1$ occurs physically, for example for sound waves, and all of the above 
facts have been noted in previous discussions of acoustic wave turbulence \cite{ZakharovSagdeev70,NewellAucoin71}.
This was termed a ``semidispersive'' wave system, since sound waves are nondispersive on the resonant manifolds, along 
the direction of propagation, but there is lateral dispersion due to angular separation of wave packets. The usual wave kinetic equation 
is no longer applicable, as the standard phase measure on the resonant manifold becomes ill-defined. Various proposals 
have been made to derive related kinetic equations by generalizing the arguments for strictly dispersive waves \cite{NewellAucoin71}  
or by taking into account nonlinear broadening of the resonance \cite{Lvovetal97}.    
The breakdown of dispersivity is quite severe for sound waves, but, as we have seen, it is an isolated phenomenon 
for wave systems with power dispersion laws, occurring only for $\alpha=1.$ This may have led to an expectation that dispersivity breakdown and the 
associated singularity of the resonant manifold is otherwise absent. However, we shall now show by further examples  
that it occurs more widely, although in generally less severe forms than for sound waves. 

%%%%%%%%%%%%%%%%%%%%%%%%%%%%%%%%%%%%%%%%%%%%%%
%%%%%%%%%%%%%%%%%%%%%%%%%%%%%%%%%%%%%%%%%%%%%%
\subsubsection{Lattice Regularization of Power Laws}\lb{lattice}
%%%%%%%%%%%%%%%%%%%%%%%%%%%%%%%%%%%%%%%%%%%%%%
%%%%%%%%%%%%%%%%%%%%%%%%%%%%%%%%%%%%%%%%%%%%%%
 
The work of Van Hove suggests that the presence of a spatial lattice could facilitate the appearance 
of resonance singularities. We show that this expectation is correct, by considering a ``lattice 
regularization'' of the power-law dispersion of the previous section. If physical space is a cubic lattice $a{\mathbb Z}^d,$ 
then Fourier space is the $d$-torus $\frac{2\pi}{a}{\mathbb T}^d=\left[-\frac{\pi}{a},\frac{\pi}{a}\right]^d.$ The dispersion
law must be a periodic function on $\Lambda_*=\left[-\frac{\pi}{a},\frac{\pi}{a}\right]^d.$ A natural replacement for 
$k^2$ is therefore $\frac{4}{a^2}\sum_{i=1}^d \sin^2(\frac{k_i a}{2}),$ the discrete Fourier transform of the lattice Laplacian
$-\bigtriangleup_a$ defined by 2nd-order differences. We thus consider
\begin{eqnarray}
\omega(\bk)=C\left[\frac{4}{a^2}\sum_{i=1}^d \sin^2(\frac{k_i a}{2})\right]^{\alpha/2}. \lb{lpower} 
\end{eqnarray}
This dispersion law is a toy model that does not describe any physical system but that we use as the simplest example 
of the effect of a lattice. It could appear in a computational algorithm to solve a wave equation with a power-law dispersion relation 
on a regular space grid. A lattice formulation of the equations can also be useful for theoretical purposes as an 
``ultraviolet regularization'' to avoid high-wavenumber divergences, a mathematically cleaner alternative to the 
Fourier-Galerkin truncation of the wave dynamics  employed in \cite{Nazarenko11}. Of course, lattices exist physically 
in nature, e.g. in crystalline solids, and similar singularities must exist in certain cases on the resonant manifolds in the 
quantum Boltzmann equations which describe particle transport in crystals.
Since this a 4-wave resonance, we discuss it in more detail in subsection \ref{quartet} below. It would be interesting to investigate 
whether such resonance singularities exist more widely for quantum transport in crystals. 

We now turn to our simple example. Without loss of generality, we take lattice constant $a=1.$ We first show 
that, as for the continuous case, nontrivial resonance requires $\alpha>1:$
\begin{prop}
If $\alpha\leqslant 1,$ the resonant manifolds $\cR_\bk^{s_2s_3}=\{\bp: E^{s_2s_3}(\bp;\bk)=0\}$ contain 
only two points $\b0$ and $\bk$. 
\end{prop}
\begin{pf}  We introduce the notation 
$$ |\bk|_{\Lambda^*}=\left[4\sum_{i=1}^d \sin^2(k_i/2)\right]^{1/2}, \,\,\,\, \bk\in 2\pi{\mathbb T}^d$$
for the ``vector  norm'' on the torus.  We see that this function satisfies the triangle inequality, $|\bp+\bq|_{\Lambda^*}\leq |\bp|_{\Lambda^*}+|\bq|_{\Lambda^*},$ 
with equality only for $\bp=\b0$ or $\bq=\b0$ (mod $\Lambda^*$),   
because of an observation of \cite{Spohn06} that $|\bk|_{\Lambda^*}=|\bz(\bk)|$ for $\bz(\bk)\in {\mathbb C}^d$ defined by  
$z_j(\bk)=1-e^{ik_j}$ with $z_j(\bp+\bq)=z_j(\bp)+e^{ip_j}z_j(\bq).$ In particular note that $e^{-ip_j}z_j(\bp)=\lambda \cdot z_j(\bq)$ for some
real $\lambda\in [0,\infty]$ if and only if either $p_j=0$ or $q_j=0$ (mod $2\pi$). It follows from this fact that the standard proof of lack
of resonances for $\bk$ in Euclidean space \cite{Nazarenko11} carries over. For completeness we give the details. 

Consider first $s_2=s_3=1,$ so that $\bk=\bp+\bq.$ Then,
$$ |\bk|_{\Lambda^*}=|\bp+\bq|_{\Lambda^*}\leq |\bp|_{\Lambda^*}+|\bq|_{\Lambda^*} 
\leq \left[|\bp|^\alpha_{\Lambda^*}+|\bq|^\alpha_{\Lambda^*}\right]^{1/\alpha}, \,\,\,\, {\rm for} \,\,\,\, \alpha\leqslant 1.$$
so that $\omega(\bk)=|\bk|^\alpha_{\Lambda^*}\leq |\bp|^\alpha_{\Lambda^*}+|\bq|^\alpha_{\Lambda^*}=s_2\omega(\bp)+s_3\omega(\bq)$ 
with equality only for $\bp=\bzed$ or $\bp=\bk$.  

Now consider $s_2+s_3=0,$ for example $s_2=-s_3=1.$ In that case $\bp=\bk+\bq$
and the previous argument shows that $|\bk|^\alpha_{\Lambda^*}+|\bq|^\alpha_{\Lambda^*}\geq |\bp|^\alpha_{\Lambda^*},$ or
$$ \omega(\bk)=|\bk|^\alpha_{\Lambda^*}\geq  |\bp|^\alpha_{\Lambda^*}-|\bq|^\alpha_{\Lambda^*} = s_2\omega(\bp)+s_3\omega(\bq), $$
with inequality only for $\bp=\bk$ or $\bp=\bq.$ The second possibility requires $\bk=\bzed,$ however, and is not of interest. 
Repeating the argument for $s_3=-s_2=1$ leads to the same conclusion with $2,\bp\leftrightarrow 3,\bq$ so that equality requires
$\bq=\bk$ or $\bq=\bp.$ The first possibility is equivalent to $\bp=\bzed$ and the second to $\bk=\bzed,$ again not of interest. 
%\qed
\hfill $\Box$ 
\end{pf}

We now establish results on existence of critical points, separately for the cases $s_2\cdot s_3=1$ and $s_2\cdot s_3=-1:$
\begin{prop}
For any $\alpha>1$, there always exists $\bk$ for which the resonant manifold $\cR^{++}_\bk$ contains nontrivial critical points.
\end{prop}

\begin{pf}
The singularity condition 
\begin{eqnarray}
|\bp|^{\alpha-2}_{\Lambda^*}\sin(p_{i})=|\bq|^{\alpha-2}_{\Lambda^*}\sin(q_{i})\quad \text{for $i=1,\cdots,d$.}
\end{eqnarray}
is solved by either $\bp=\bq=\frac{1}{2}\bk$ or by $\bp=\bq=\frac{1}{2}\bk+ \bpi$, where $\bpi=(\pi,\pi,...,\pi).$ 
Observe that the notation ``$\bk/2$'' is actually ambiguous, because changing the components of $\bk$ by 
integer multiples of $2\pi$ yield different points depending on whether the integer is odd or even. We 
remove that ambiguity by requiring that $-\pi<k_j\leq \pi$, $j=1,\dots, d$ so that both $\frac{1}{2}\bk$ 
and $\frac{1}{2}\bk+ \bpi$ are well-defined (mod $\Lambda^*$). In either case, $\bp+\bq=\bk$ (mod $\Lambda^*$).  

The resonance condition $\omega(\bp)+\omega(\bq)=\omega(\bk)$ then yields for $\bq=\bp=\frac{1}{2}\bk$
\begin{eqnarray}
&&2 \left|\frac{\bk}{2}\right|_{\Lambda^*}^\alpha  = |\bk|_{\Lambda^*}^\alpha 
\cr
&\Longrightarrow& 2\left[\sum_{i=1}^d \sin^2\left(\frac{k_i}{4}\right)\right]^{\alpha/2}=\left[\sum_{i=1}^d \sin^2\left(\frac{k_i}{2}\right)\right]^{\alpha/2}=2^{\alpha} \left[\sum_{i=1}^d \sin^2\left(\frac{k_i}{4}\right)\cos^2\left(\frac{k_i}{4}\right)\right]^{\alpha/2}
\cr
&\Longrightarrow& 2^{2/\alpha-2}=\llangle\cos^2\left(\frac{\bk}{4}\right)\rrangle \lb{lattice-resonance}
\end{eqnarray}
where we define the average $\langle f(\bk) \rangle =\sum_{i=1}^d f(k_i)\sin^2(\frac{k_i}{4})/[\sum_{i=1}^d \sin^2(\frac{k_i}{4})]$. 
Note $\langle\cos^2(\frac{\bk}{4})\rangle=\frac{1}{2}$ when $\bk=\bpi$ and and approaches $1$ when $\bk\rightarrow\b0.$ Thus 
for any $1<\alpha\leqslant 2$, there always exists $\bk$ that solves the condition (\ref{lattice-resonance}) by the intermediate value theorem. 

On the other hand, for 
$\bp=\bq=\frac{1}{2}\bk+ \bpi$, the resonance condition $\omega(\bp)+\omega(\bq)=\omega(\bk)$ yields 
\begin{eqnarray}
&&2 \left|\frac{\bk}{2}+\bpi\right|_{\Lambda^*}^\alpha =|\bk|_{\Lambda^*}^\alpha 
\cr
&\Longrightarrow& 2\left[\sum_{i=1}^d \cos^2\left(\frac{k_i}{4}\right)\right]^{\alpha/2}=\left[\sum_{i=1}^d \sin^2\left(\frac{k_i}{2}\right)\right]^{\alpha/2}=2^{\alpha} \left[\sum_{i=1}^d \sin^2\left(\frac{k_i}{4}\right)\cos^2\left(\frac{k_i}{4}\right)\right]^{\alpha/2}
\cr
&\Longrightarrow& 2^{2/\alpha-2}=\llangle\sin^2\left(\frac{\bk}{4}\right)\rrangle \lb{lattice-resonance2}
\end{eqnarray}
where we now define $\langle f(\bk)\rangle = \sum_{i=1}^d f(k_i)\cos^2(k_i/4)/[\sum_{i=1}^d \cos^2(k_i/4)]$. 
Note $\langle\sin^2(\bk/4)\rangle=0$ for $\bk=\b0$ and equals $\frac{1}{2}$ for $\bk=\bpi.$ Thus for any $\alpha\geqslant 2$, 
there always exists $\bk$ that solves the condition (\ref{lattice-resonance2}).

Note that in the borderline case $\alpha=2,$ the singularity condition becomes simply 
\begin{eqnarray}
\sin(p_{i})=\sin(k_i-p_{i})\quad \text{for $i=1,\cdots,d$.}
\end{eqnarray}
and one can easily check that there are only resonant solutions for $\bk=\bpi.$ For this choice, every point of $\Lambda^*$ 
is in the resonant manifold, since 
$$\sin^2\left(\frac{p_i}{2}\right)+\sin^2\left(\frac{q_i}{2}\right)=\sin^2\left(\frac{p_i}{2}\right)+\cos^2\left(\frac{p_i}{2}\right)=
1=\sin^2\left(\frac{k_i}{2}\right). $$
Every point of the torus is a degenerate critical point in this case. \hfill $\Box$

\end{pf}

\noindent The condition (\ref{lattice-resonance}) provides the example in the Introduction.
For $\bk=(0,0,1)$, one gets $\langle \cos^2(\bk/4)\rangle=\cos^2(1/4)$, which reduces (\ref{lattice-resonance}) to $2^{2/\alpha-2}=\cos^2(1/4),$ or $\alpha=1/(1+\log_2\cos(1/4))\doteq 1.0477.$
One can likewise use condition (\ref{lattice-resonance2}) to construct examples for $\alpha>2$.
For instance, if one chooses $\bk=(3,3,3),$ then $\langle \sin^2(\bk/4)\rangle=\sin^2(3/4)$, and one obtains
from (\ref{lattice-resonance2}) that  $\alpha=1/(1+\log_2\sin(3/4))\doteq 2.2367$. This example is plotted in Fig.~2. 

% FIGURE HERE

\begin{figure}[!h]
\begin{center}\lb{Fig1}
\includegraphics[width=3.5in,height=2.5in]{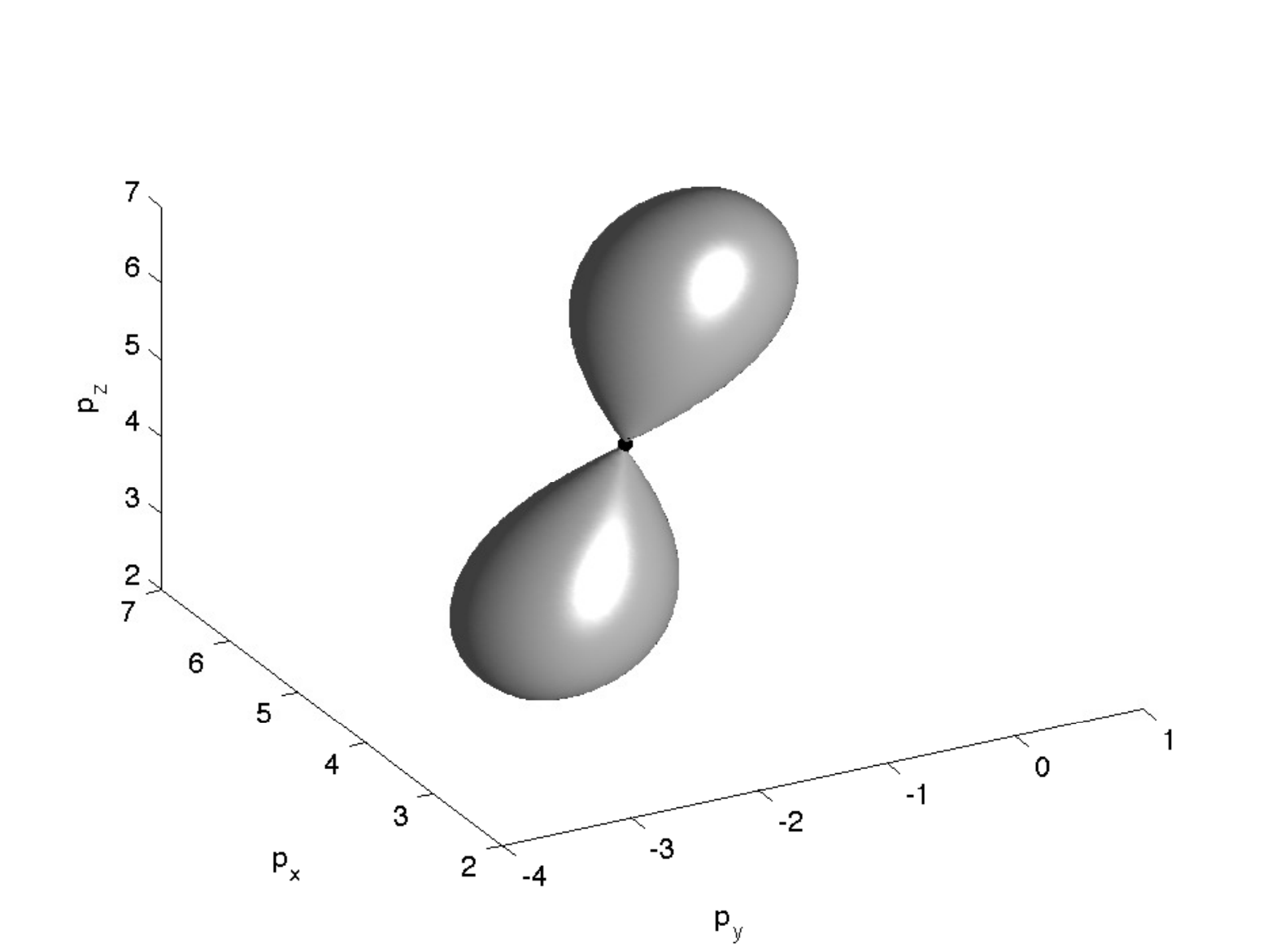}
\end{center}
\caption{Resonant manifold ${\mathcal R}^{++}_\bk=\{\bp:\,\omega(\bp)+\omega(\bk-\bp)-\omega(\bk)=0\}$ for dispersion relation
$\omega(\bk)=\left[4\sum_{i=1}^3 \sin^2\left(\frac{k_i}{2}\right)\right]^{\alpha/2},$ $\alpha=\frac{1}{1+\log_2\sin(3/4)},$
on the 3-torus $[\frac{1}{2}\pi,\frac{5}{2}\pi]\times [-\frac{3}{2}\pi,\frac{1}{2}\pi]\times [\frac{1}{2}\pi,\frac{5}{2}\pi],$ and for the specific wavenumber $\bk=(3,3,3)$. 
%indicated by a \textcolor{red}{red} arrow. 
There is a critical point (black dot) at $\bp=\bk/2-\bpi.$
}
\end{figure}

%\newpage

We finally discuss briefly the case $s_2\cdot s_3=-1$, or,  without loss of generality, $s_2=1, s_3=-1$. 
We present here no construction of critical points for all values of $\alpha>1,$ but we can establish their 
existence in specific instances.  A simple example is provided by exploiting a general criterion for the existence of critical points, namely, that there be 
a pair of distinct wavevectors $\bk_1$ and $\bk_2$ such that the topological invariants of $\cR_{\bk_1}$ and $\cR_{\bk_2}$ are different.
For example, in the case $\alpha=5/2$, critical points must be present because there is a change of topological invariants of the resonant 
manifold as $\bk$ is varied, see Fig.~3.
The manifold for $\bk_1=(.2,-.2)$ is connected and belongs to the trivial 1st homology class 
on the 2-torus, while the manifold for $\bk_2=(.18,-.2)$ has two connected components in the 
same non-trivial 1st homology class  (both winding around the same hole of the 2-torus).
Thus, as the wavevector is changed continuously from $\bk_1$ to $\bk_2,$ a critical point in $\cR_\bk$ 
must occur at some intermediate value of $\bk.$ Such a critical point is shown in the middle subplot of Fig.~3.

% FIGURE

\begin{figure}[!h]
\begin{center}\lb{resonance_lattice2topo}
\includegraphics[width=2.1in, height=2in,keepaspectratio]{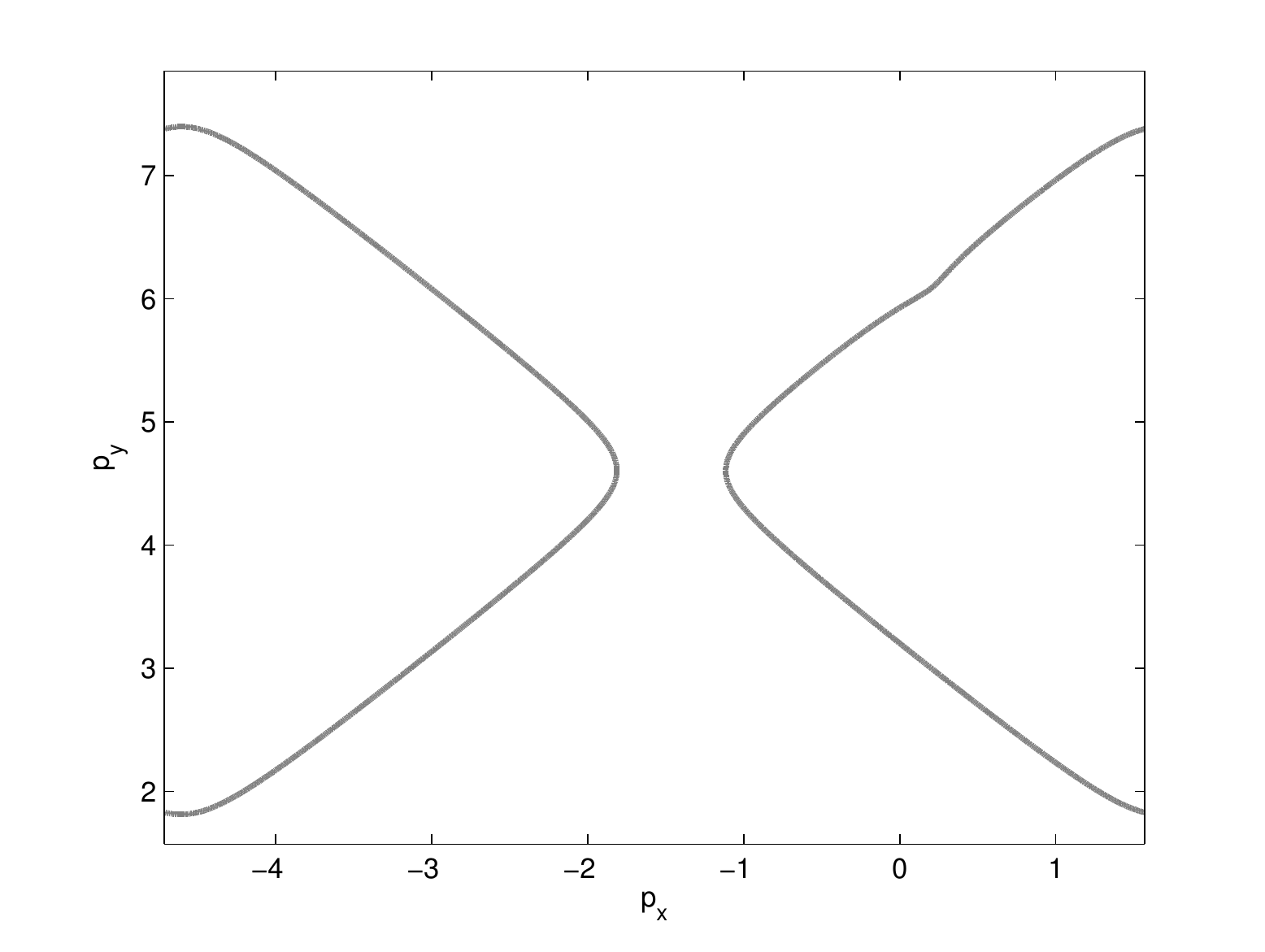}
\includegraphics[width=2.1in, height=2in,keepaspectratio]{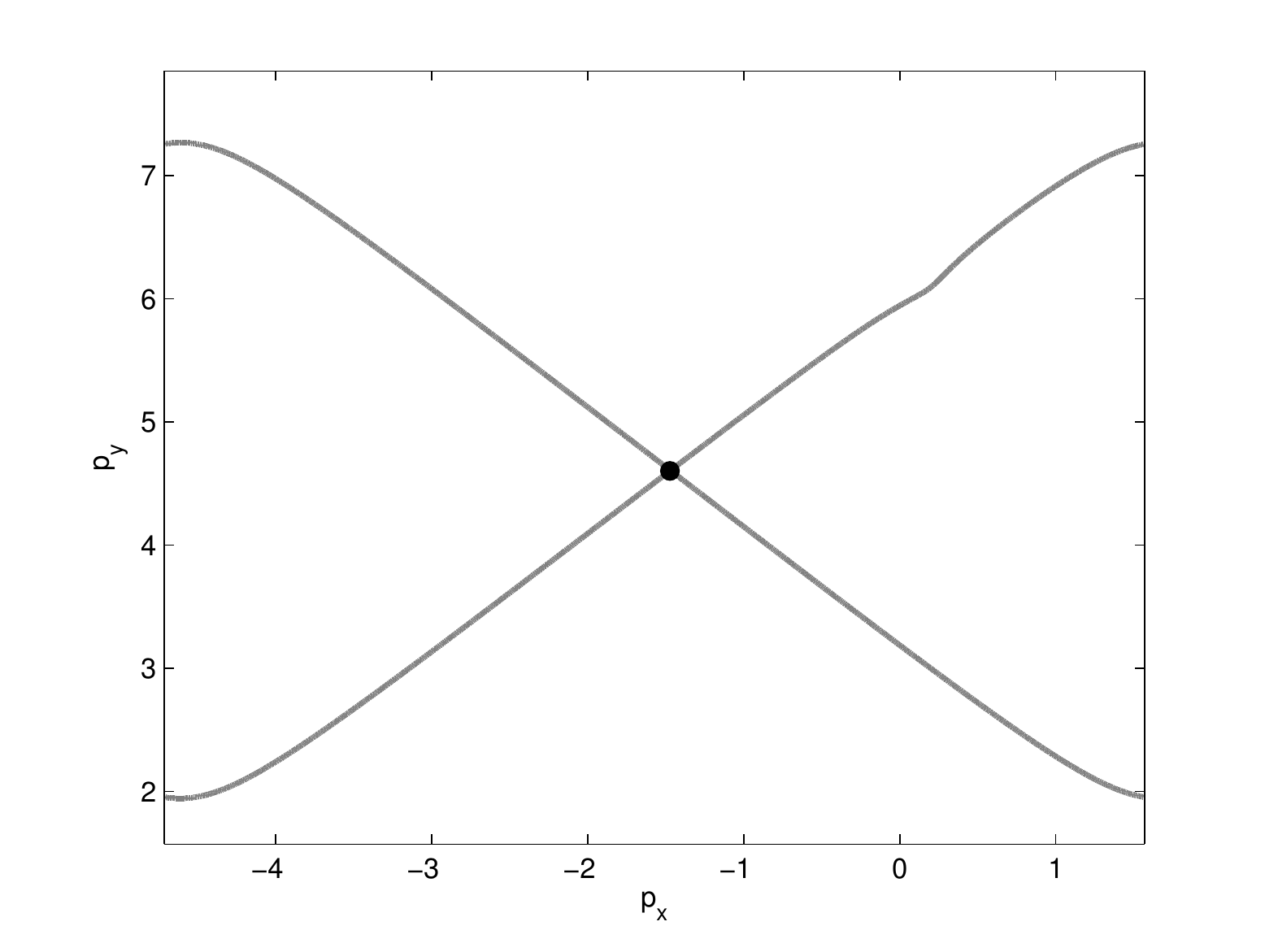}
\includegraphics[width=2.1in, height=2in,keepaspectratio]{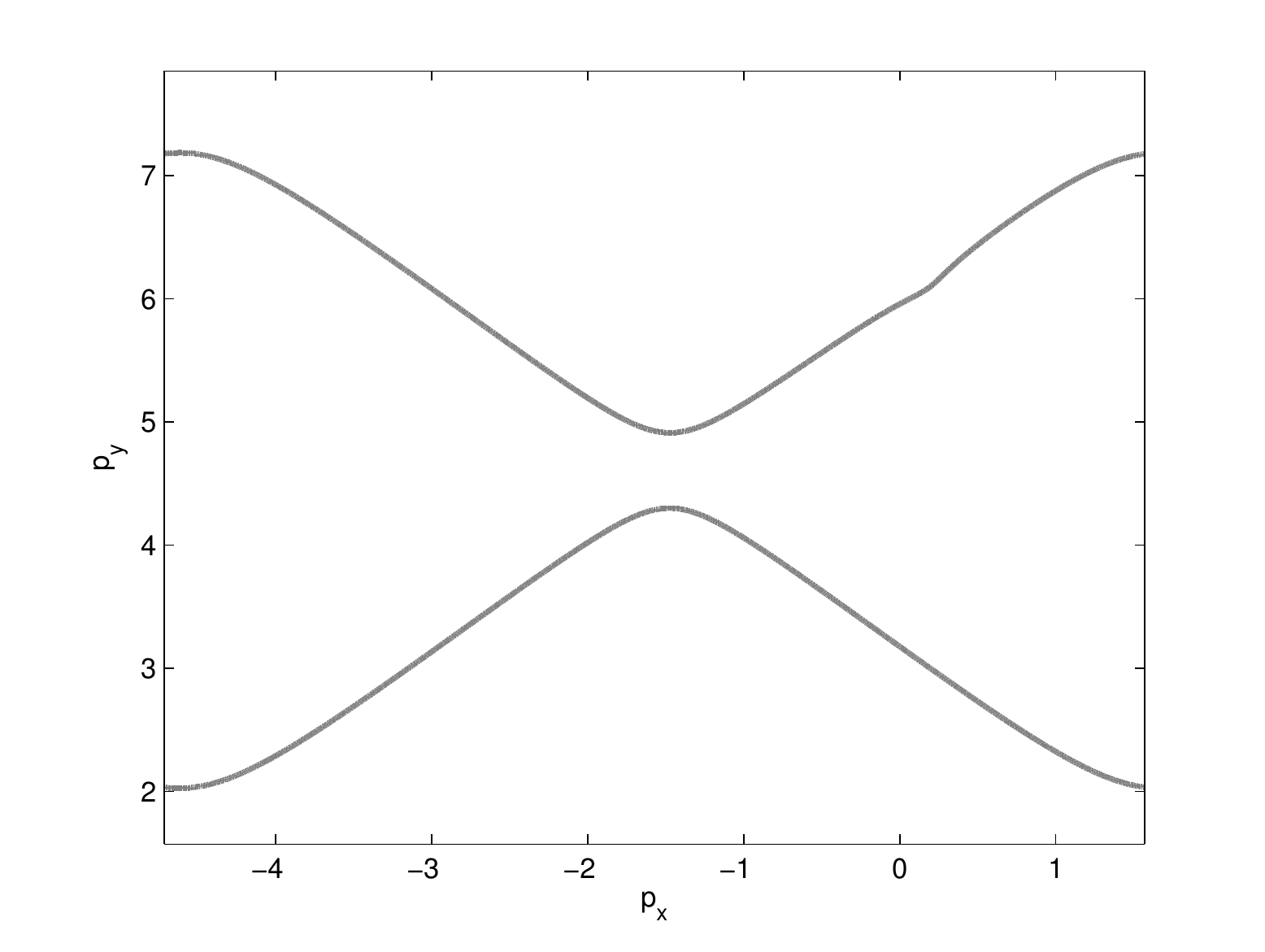}
\end{center}
\caption{Resonant manifold ${\mathcal R}_\bk^{+-}$ for dispersion relation
$\omega(\bk)=\left[4\sum_{i=1}^2 \sin^2\left(\frac{k_i}{2}\right)\right]^{\alpha/2}$ with $\alpha=2.5,$
on the 2-torus $[-\pi/2,3/2\pi]\times[\pi/2,5/2\pi]$), and for the specific wavenumber $\bk=(.2,-.2)$ on the left and $\bk=(.18,-.2)$ on the right. For the specific numerically approximated wavenumber $\bk=(.1887713,-.2)$ (middle figure),
there is a critical point (black dot) at 
$\bp=(-1.4716, 4.6016).$}
\end{figure}

The examples in this section are only toy problems meant to provide some analytical insight. 
In section \ref{finite} we shall show that non-degenerate singularities at a single point, like those 
in the examples of this section, can lead in appropriate circumstances to an infinite phase measure 
on the resonant manifolds. For 3-wave resonances, the phase measure is finite for dimensions $d>2$ 
but logarithmically divergent for $d=2.$ We see below that such non-degenerate singularities can arise 
from dispersion laws in physically relevant models. 
%In this example, 
%As we illustrate further with examples below, such divergences 
%can have important consequences for wave kinetics.
%We will also examine cases where the singularities are degenerate.

%%%%%%%%%%%%%%%%%%%%%%%%%%%%%%%%%%%%%%%%%%%%%%
%%%%%%%%%%%%%%%%%%%%%%%%%%%%%%%%%%%%%%%%%%%%%%
\subsubsection{Anisotropic Dispersion Relations}\lb{anisotropic} 
%%%%%%%%%%%%%%%%%%%%%%%%%%%%%%%%%%%%%%%%%%%%%%
%%%%%%%%%%%%%%%%%%%%%%%%%%%%%%%%%%%%%%%%%%%%%%

Many wave dispersion relations in physical systems are anisotropic, either power-laws or more general forms. Here we 
show by several examples that anisotropy can lead to critical points on the resonant manifolds. A simple class 
of examples are systems whose dispersion relation has the form 
$$ \omega(\bk) = k_1 \varphi(k) \lb{form1} $$
where $k=|\bk|$ is the Euclidean norm and $k_1$ is a component in a distinguished direction. 
%Some well-known examples are: 
%\begin{itemize}
%\item[] {\it (i) Rossby/drift waves}, where $d=2,$ the $1$-direction is zonal, $\varphi(k)=-\beta\rho^2/(1+\rho^2k^2)$
%with $\beta$ the beta parameter (meridional gradient of the Coriolis frequency) and $\rho$ the Rossby radius
%\cite{Balketal90}. 
%\item[] {\it (ii) inertial waves}, where $d=3,$ the $1$-direction is the rotation axis, $\varphi(k)=2\Omega/k$ 
%with $\Omega$ the rotation rate \cite{Galtier03}. 
%\end{itemize}
In cases such as this, for any ``slow mode'' $\bk$ with $k_1=0,$
$$ {\mathcal R}_\bk^{++}=\{ \bp: \, E^{++}(\bp;\bk)=p_1(\varphi(p)-\varphi(q))=0, \,\, q=|\bk-\bp|\}. $$
Since 
$$ \grad_\bp E^{++}(\bp;\bk) = ( \varphi(p)-\varphi(q) ) \hat{\bfe}_1 + p_1 \bigg(\varphi'(p)\hat{\bp} + \varphi'(q)\hat{\bq}\bigg) , $$
it follows that the subset 
$$ {\mathcal R}_\bk^{++*}=\{ \bp: \, p_1=0\,\,\,\, \& \,\,\,\, \varphi(p)=\varphi(q), \,\, q=|\bk-\bp|\}\subset {\mathcal R}_\bk^{++} $$
consists of critical points satisfying $\grad\omega(\bp)=\grad\omega(\bq)\propto \hat{\bfe}_1$.  
Note also that when $p_1=0$ the Hessian matrix becomes
$$
\grad\otimes\grad E^{++}(\bp;\bk) = \varphi'(p)(\hat{\bfe}_1\otimes\hat{\bp}+\hat{\bp}\otimes\hat{\bfe}_1) +(\bp\leftrightarrow\bq) 
% \grad\otimes\grad E^{++}(\bp;\bk) = (\varphi'(p)+\varphi'(q))(\hat{\bfe}_1\otimes\hat{\bp}+\hat{\bp}\otimes\hat{\bfe}_1)
%+ (p_1\varphi''(p)+q_1\varphi''(q))\hat{\bp}\otimes\hat{\bp}, 
$$
so that the critical points $\bp\in {\mathcal R}_\bk^{++*}$ with $\varphi'(p),\varphi'(q)\neq 0$ are non-degenerate for $d=2,$ 
while for $d>2$ the Hessian matrix is degenerate. A null eigenvector of the Hessian for any $d>2$ when $\hat{\bp}, \hat{\bq}$ are non-parallel
is given by the vector $\varphi'(p)\hat{\bq}^\perp-\varphi'(q)\hat{\bp}^\perp,$ where $\hat{\bp}^\perp$ is the component of $\hat{\bp}$ perpendicular
to $\hat{\bq}$ and vice versa for $\hat{\bq}^\perp.$ When $\hat{\bp}\| \hat{\bq},$ then any vector orthogonal to both $\hat{\bfe}_1$
and $\bp$ (or $\bq$) is a null eigenvector.  For $d>3,$ furthermore, any vector orthogonal to all three vectors $\hat{\bfe}_1$,
$\bp$, and $\bq$ (or $\bk$) is a null eigenvector. 

\paragraph{Rossby/drift waves}\lb{Rossby}

A well-known example of an isotropic dispersion relation occurs for Rossby/drift waves, where $d=2,$ the $1$-direction is zonal, 
$\varphi(k)=-\beta\rho^2/(1+\rho^2k^2)$
with $\beta$ the beta parameter (meridional gradient of the Coriolis frequency) and $\rho$ the Rossby radius \cite{Balketal90}. 
In that case, the condition $\varphi(p)
=\varphi(q)$ can be easily solved to give $\bk\bdot(\bp-\bk/2)=0.$ For a general ``slow'' wavevector 
$\bk=(0,k_y),$ the resonant manifold ${\mathcal R}_\bk^{++}$ is the union of the two lines $p_x=0$ and $p_y=k_y/2$, see Fig.~4, 
middle panel.
The point of intersection $(0,k_y/2)$ is a nondegenerate resonance Van Hove singularity. Note that for generic $\bk,$ 
the manifold ${\mathcal R}_\bk^{++}$ for the system of Rossby waves is instead diffeomorphic to a circle,  
as shown in Fig.~4, left and right panels. 
%(e.g. see Fig.~1 in \cite{Balketal90}).

%FIGURE HERE

\begin{figure}[!h]
\begin{center}\lb{Fig2}
\includegraphics[width=2.1in,height=2.5in,keepaspectratio]{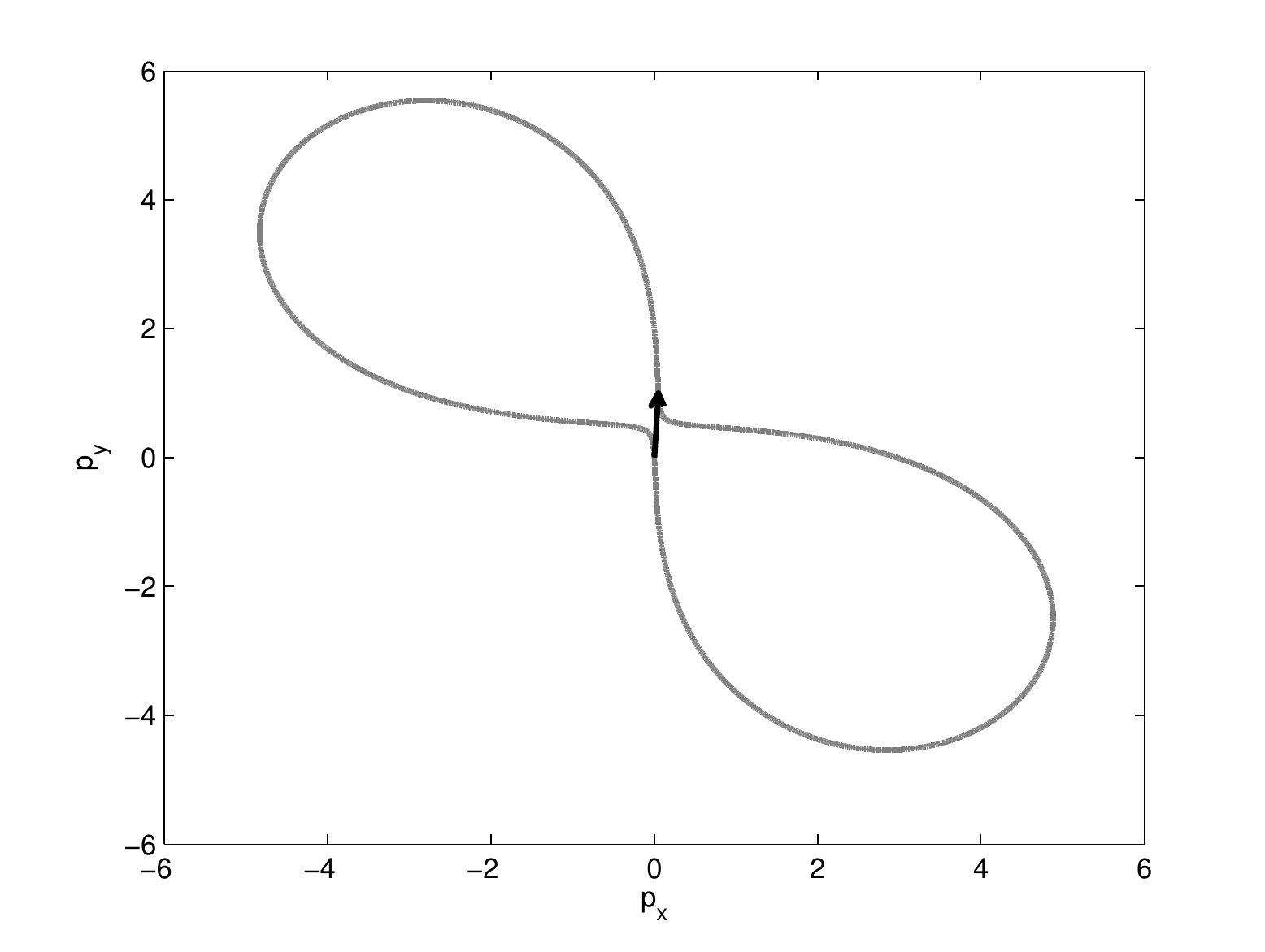}
\includegraphics[width=2.1in,height=2.5in,keepaspectratio]{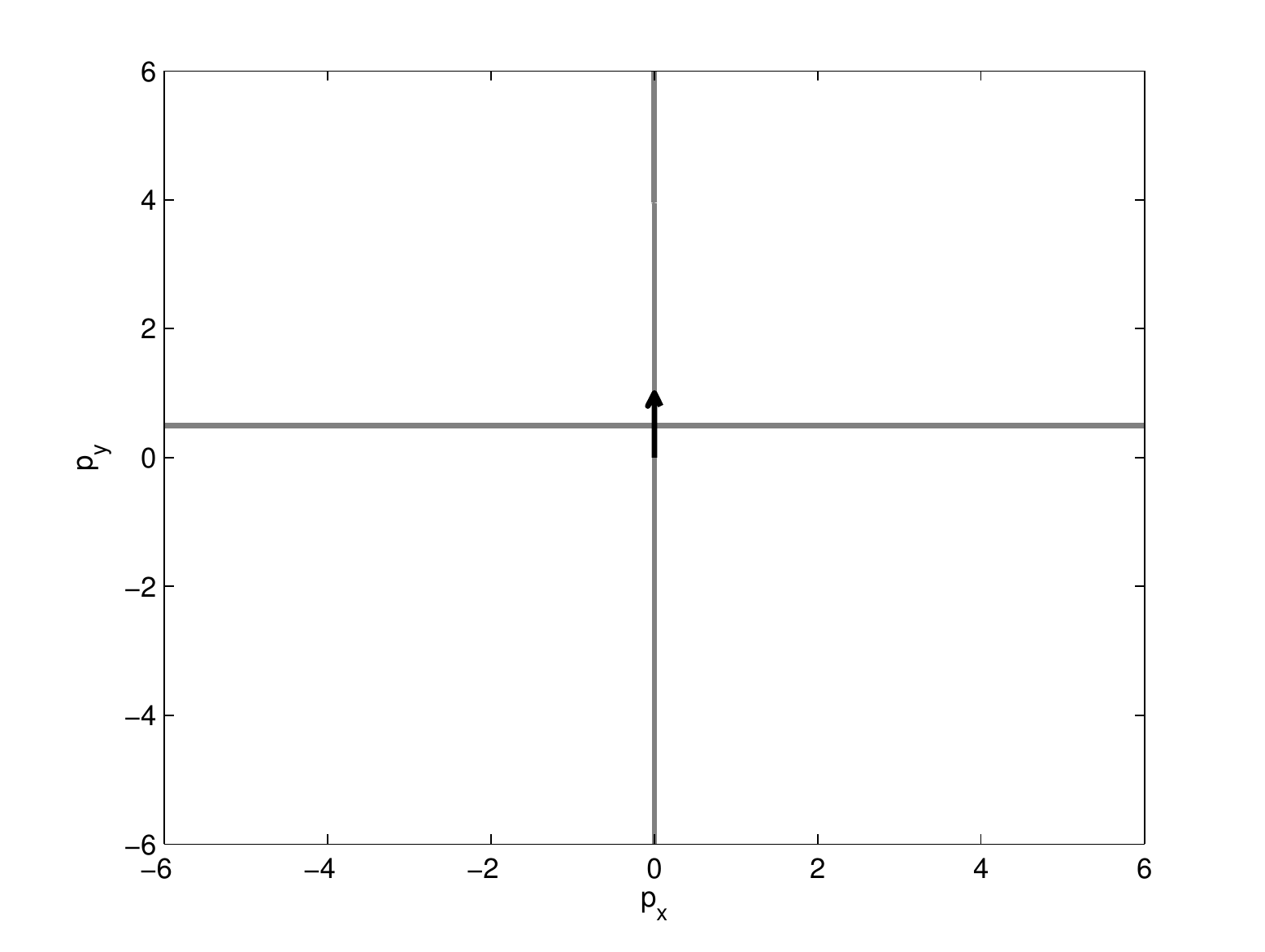}
\includegraphics[width=2.1in,height=2.5in,keepaspectratio]{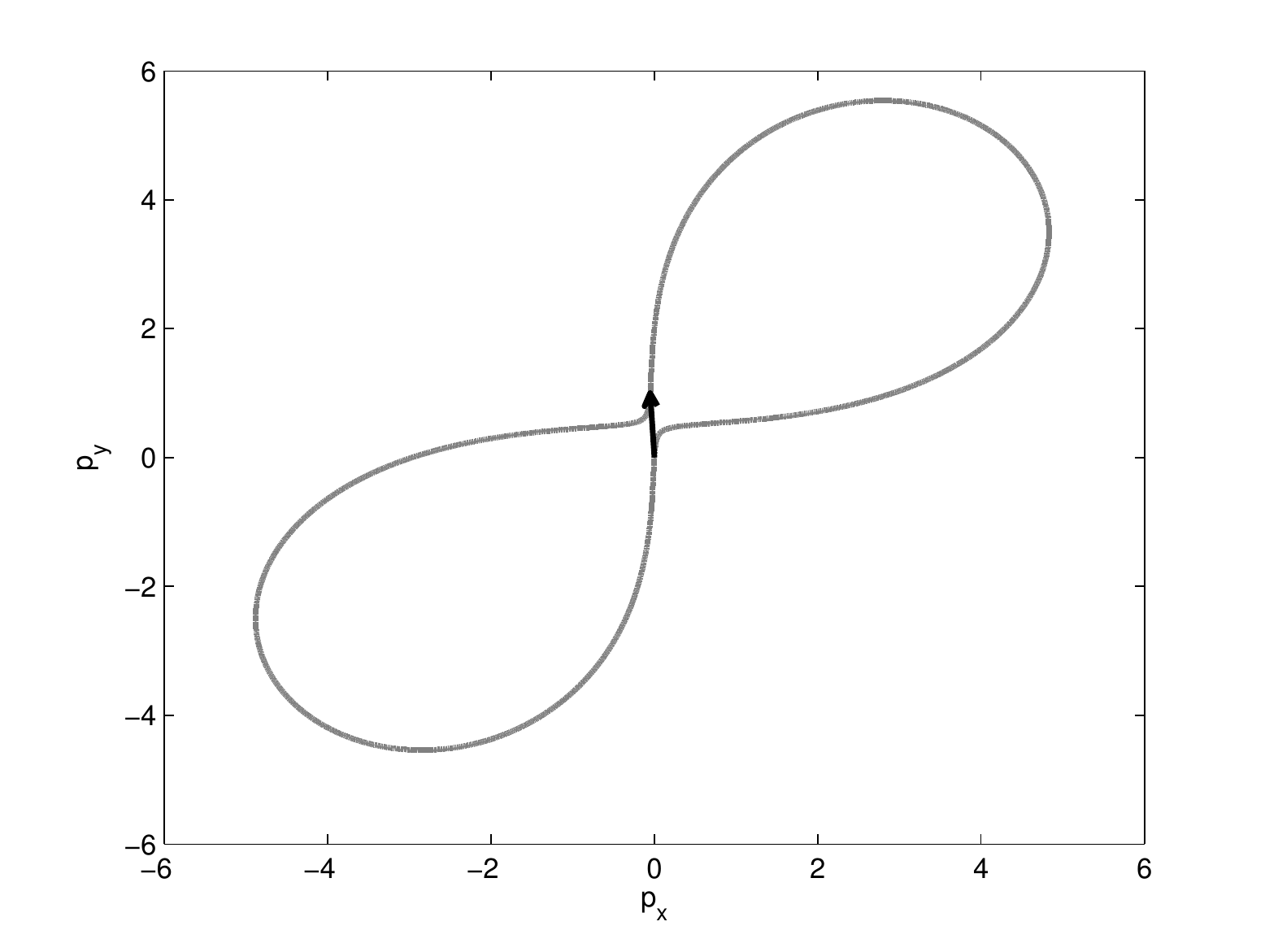}
\end{center}
\caption{Resonant manifold ${\mathcal R}_\bk^{++}, \,\,\,\, \bk=(\cos\theta,\sin\theta) $ for Rossby/drift waves, plotted
as gray lines. The black arrow is the vector $\bk.$ Here $\theta=\frac{\pi}{2}-0.05$ for the left panel, $\theta=\frac{\pi}{2}$ for the middle panel, and $\theta=\frac{\pi}{2}+0.05$ for the right panel.}
\end{figure}

It is easy in this example to exhibit explicitly the logarithmic divergence in the phase measure 
$dS/|\grad E|$ on the resonant manifold, which is expected for $d=2.$ A simple calculation gives 
$$ \grad_\bp E^{++}(\bp;\bk) =  \frac{2\beta \rho^4k_y}{(1+\rho^2p^2)(1+\rho^2q^2)}\left[ \left(p_y-\frac{k_y}{2}\right)\hat{\bx} + p_x\hat{\by}\right] $$
on the resonant manifold for $\bk=(0,k_y)$, vanishing approaching the singular point.
Since $E^{++}(\bp;\bk)\propto p_x(p_y-k_y/2),$ the divergence has the general form of the integral 
$$\iint dx \,dy\, \delta(xy) = \iint dx\, dy\, \left[\frac{1}{|x|}\delta(y)+\frac{1}{|y|}\delta(x)\right].$$
One might dismiss this singularity as dynamically irrelevant, since the zonal flows with $k_x=0$ have vanishing
nonlinearity. This may be easily verified for the Charney-Hasegawa-Mima equation \cite{Balketal90}, 
$$ \partial_t(\rho^2\bigtriangleup\psi-\psi)-\beta \partial_x\psi + A J(\psi, \rho^2\bigtriangleup\psi) =0, $$
with $J(f,g)=f_xg_y-f_yg_x$ the Jacobian, which vanishes whenever one of the functions is independent of $x$
(or of $y$). This argument is correct, but must be made carefully. 

The delicate point is that the critical 
point for the resonant manifold ${\mathcal R}_\bk^{++}$ with $k_x=0$ implies not only divergent phase measure 
on that manifold but also extremely large phase measures on adjacent manifolds with $k_x$ very small. Note for such $k_x$
that the resonant manifold locally for $\bp$ near $(0,k_y/2)$ is a hyperbola whose equation is, to leading order,
\begin{eqnarray*}
(p_x-\frac{k_x}{2})(p_y-\frac{k_y}{2}) = \frac{3}{8} k_x k_y \frac{1+\frac{1}{4}\rho^2 k_y^2}{1+\rho^2 k_y^2}.
\end{eqnarray*}
Near $\bp=(0,k_y/2)$, to leading order,
\begin{eqnarray*}
\grad E^{++}(\bp;\bk)=\frac{\beta \rho^4 }{(1+\frac{\rho^2}{4}k_y^2)^2} \Big[ \left(3k_x(2p_x -k_x) +k_y (2p_y-k_y)\right) \hat{\bx} + \left( k_x(2p_y-k_y) +k_y(2p_x-k_x)  \right)\hat{\by}  \Big]
\end{eqnarray*}
Therefore on the resonant manifold
\begin{eqnarray*}
|\grad E^{++}(\bp;\bk)| \geqslant \frac{ \beta\rho^4}{ (1+\frac{\rho}{4}k_y^2)^2 }  \sqrt{16|k_x k_y (k_y-2p_y) (k_x-2p_x)|}
= \frac{ \beta\rho^4}{ (1+\frac{\rho}{4}k_y^2)^2 }  \sqrt{6 k_x^2 k_y^2 \frac{1+\frac{1}{4}\rho^2 k_y^2}{1+\rho^2 k_y^2}},
\end{eqnarray*}
which implies a phase measure which is $O(|k_x|^{-1}).$  Fortuitously, however, all of the standard dynamical models 
of Rossby/drift waves \cite{Balketal90} have interaction coefficients $H^{s,s_1,s_2}_{\bk,\bp,\bq}$ vanishing proportional 
to $|k_x p_x q_x|^{1/2}$ for any of the three wavevectors in the ``slow'' set.  In these models, the singularity 
in the phase measure for small $k_x$ is 
%We shall discuss this in detail in the next section.
cancelled by the interaction coefficient $|H_{\bk,\bp,\bq}|^2\sim |k_x p_x q_x|$, rendering the 
collision integral finite even as $k_x$ tends to zero. If the interaction coefficient had vanished more slowly 
then $|k_x|^{1/2}$ in the limit, then the collision integral for near-zonal flows could become large, 
threatening the validity of the kinetic description \cite{Newelletal01,Bivenetal01,Nazarenko11}. These considerations apply more generally,
e.g. to the isotropic power-law dispersion relations discussed in section \ref{iso-power} \footnote{In the case 
of an isotropic power-law dispersion relation for $|\bk|\ll|\bp|$
$$
\grad_\bp E^{+-}(\bp;\bk) = \alpha |\bp|^{\alpha-2}\bp - \alpha |\bp-\bk|^{\alpha-2}(\bp-\bk)
\approx \alpha |\bp|^{\alpha-2} \left[\bk^\perp+(\alpha-1) k_{\|}\right],
$$
where $\bk^\perp,k_{\|}$ are components of $\bk$ perpendicular and parallel to $\bp,$ resp. A lower bound follows 
that $|\grad_\bp E^{+-}(\bp;\bk)|\geq const.|\bp|^{\alpha-2}|\bk|$ and the limit $\bk\to\b0$ gives a finite collision integral 
if the interaction coefficient vanishes no slower than $|\bk|^{1/2}$}. 
% Thus, in these models the logarithmic singularity in the phase measure is not dynamically relevant. 

\paragraph{Inertial waves}\lb{inertial}

Another example is  inertial waves, where $d=3,$ the $1$-direction is the rotation axis, and $\varphi(k)=2\Omega/k$ 
with $\Omega$ the rotation rate \cite{Galtier03}. 
This case is geometrically quite similar, but extended to $d=3$. For a generic wavevector $\bk$
the resonant manifold ${\mathcal R}_\bk^{++}$ is diffeomorphic to a sphere. For a ``slow'' mode with $\bk\bdot\bOmega=0,$ 
however, the resonant manifold ${\mathcal R}_\bk^{++}$ is a union of two planes, one orthogonal to $\bOmega$ and one 
orthogonal to $\bk.$ For example, if $\bOmega=\Omega\hat{\bz}$
and $\bk=(k_x,0,0),$ these are the planes $p_z=0$ and $p_x=k_x/2.$ See Fig.~5. Consistent with our general discussion
for $d=3,$ the critical subset of the resonant manifold is 1-dimensionally degenerate, given here by the intersection of the two planes.
The phase measure on the resonant manifold is logarithmically divergent in the vicinity of the singular line.
This result shows by example that, while non-degenerate critical points produce integrable singularities for $d>2,$ 
line singularities can lead to divergences in three dimensions.
Although geometrically quite similar to the case of drift waves, the situation is dynamically very different. 
While for drift waves the nonlinearity vanishes for the ``slow'' modes, in the case of inertial waves the 
``slow'' modes with $\bk\bdot\bOmega=0$ correspond to a strongly interacting system described by 2D 
Navier-Stokes dynamics. It has been argued convincingly that the kinetic theory for inertial waves must break down 
in the vicinity of this 2D plane of ``slow'' modes, as there is there no separation of time scales between fast linear 
and slow nonlinear dynamics \cite{Galtier03}. Here we see that there is also a breakdown in the fundamental assumption
of dispersivity of waves. There is an infinite set of wavevector pairs of ``slow'' modes with identical group velocities 
along the rotation axis and triads formed from these pairs produce a diverging contribution to the phase measure. 
Thus, the kinetic equation for inertial waves is not even well-defined for wave action non-zero in the vicinity of the ``slow'' 2D modes.

%FIGURE HERE

\begin{figure}[!h]
\begin{center}\lb{Fig3}
\includegraphics[width=3.4in,height=2.5in,keepaspectratio]{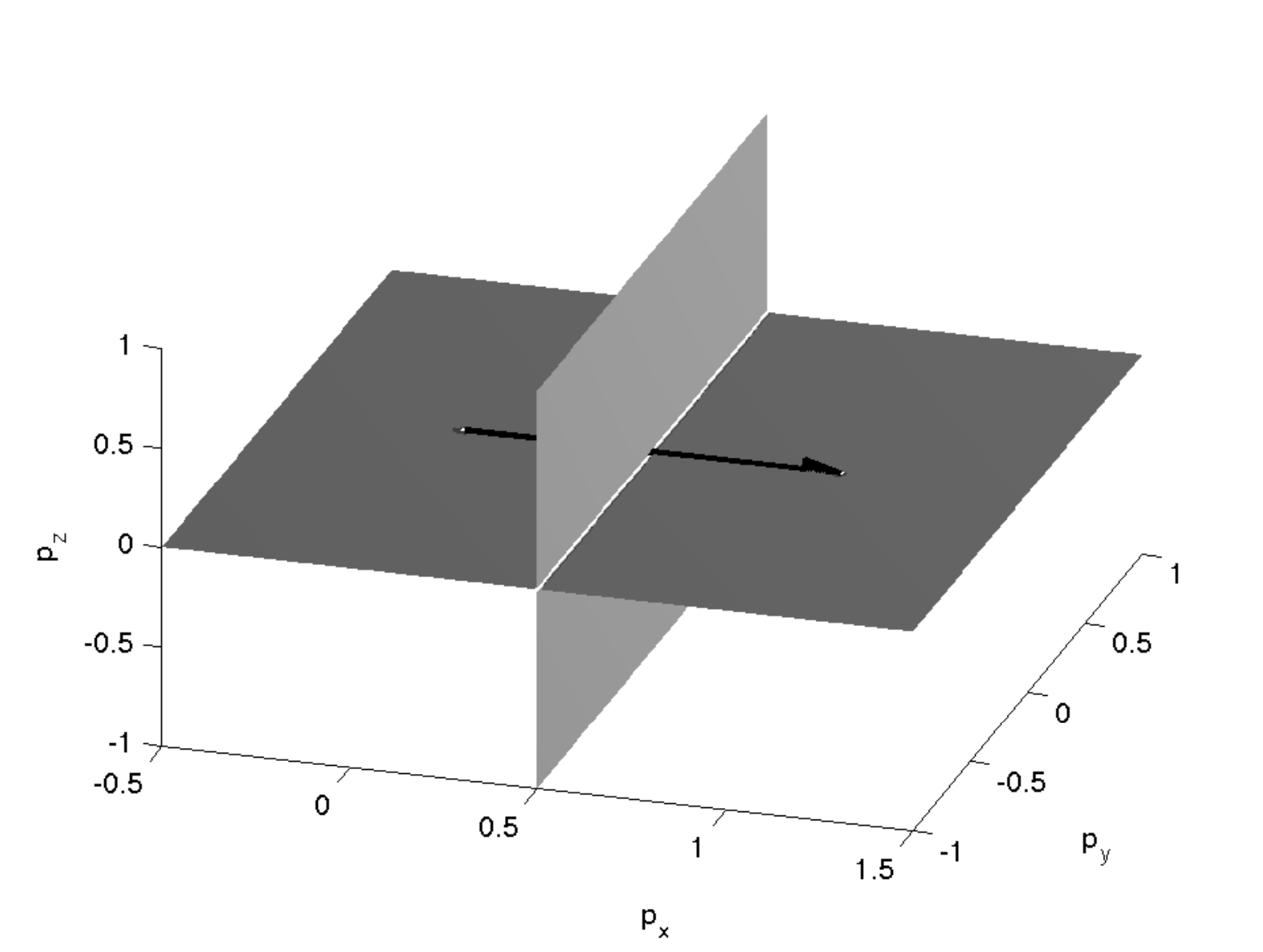}
\end{center}
\caption{Resonant manifold ${\mathcal R}^{++}_\bk, \,\,\,\, \bk=(1,0,0) $ for inertial waves with rotation about the 
z-axis. The black arrow is the vector $\bk.$}
\end{figure}

\paragraph{Internal gravity waves}

 A third example of an anisotropic dispersion law in $d=3$ of a slightly different sort is internal gravity waves, where 
the $1$-direction is vertical (the direction of gravity), and $\omega(\bk)=N k_H/k,$ with $k_H=\sqrt{k^2-k_1^2}$ the 
magnitude of the horizontal component and $N$ the Brunt-V\"ais\"al\"a frequency \cite{CaillolZeitlin00}. In this case 
$$ \grad\omega(\bk) = \frac{Nk_1}{k_Hk}\, \hat{\bk}\btimes (\hat{\bk}\btimes \hat{\bfe}_1). $$
% where $\hat{\bE}_1^\perp=\hat{\bE}_1-(\hat{\bk}\bdot\hat{\bE}_1)\hat{\bk}$ is the component of $\hat{\bE}_1$ orthogonal to $\bk.$ 
For a ``slow mode'' with only vertical variation ($k_H=0$), it is straightforward to see that
$ {\mathcal R}_\bk^{++} = \{ \bp: \,\, p_H=0 \}, $
i.e. the resonant manifold is the $1$-axis or the set of slow modes. Since in that case $\bp \parallel \hat{\bfe}_1,$ $\grad\omega(\bp)=\b0$ 
and the entire resonant set ${\mathcal R}_\bk^{++}$ consists of (degenerate) critical points. As in the case of Rossby waves,
however, the nonlinear interaction coefficient of the Euler-Boussinesq system vanishes rapidly near the set of slow modes
(see \cite{CaillolZeitlin00}, eqs.(62) and (63)) and this singular manifold is not dynamically relevant in wave kinetics. 

Another interesting phenomenon is seen in the resonant manifold of internal gravity waves
when $\bk$ is a 2D mode, with $k_1=0$. It is easy to show in that case that the only possible critical points in 
${\mathcal R}_\bk^{++}$ are also 2D modes and, because of the restriction $p_H/p+q_H/q=1,$ the only 
allowed values are $\bp=\b0$ and $\bp=\bk.$ A plot of the resonant manifold for $\bk=\bfe_1$ in Fig.6 below
shows that geometric singularities indeed occur at $\bp=\b0,\bk.$  However, these do not correspond to ordinary 
critical points where $\grad_\bp E^{++}(\bp;\bk)=\b0$ but to points $\bp$ instead where $|\grad_\bp E^{++}(\bp;\bk)|=\infty.$  
In this example, the singularities are cube-root cusps, since the equation for the resonant manifold is given near $\bp=\b0$ 
in cylindrical coordinates by $p_1=\pm (2k^2p_H)^{1/3}$, to leading order. Although we have considered here a 2D mode $\bk$, the 
resonant manifold ${\mathcal R}_\bk^{++}$ of internal gravity waves exhibits similar cusps at $\bp=\b0$ and $\bp=\bk$ for generic $\bk,$
because of the divergence of $\grad\omega$ at the origin. 

\vspace{-10pt}
%FIGURE HERE

\begin{figure}[!h]
\begin{center}\lb{Fig4}
\includegraphics[width=3.4in,height=2.5in,keepaspectratio]{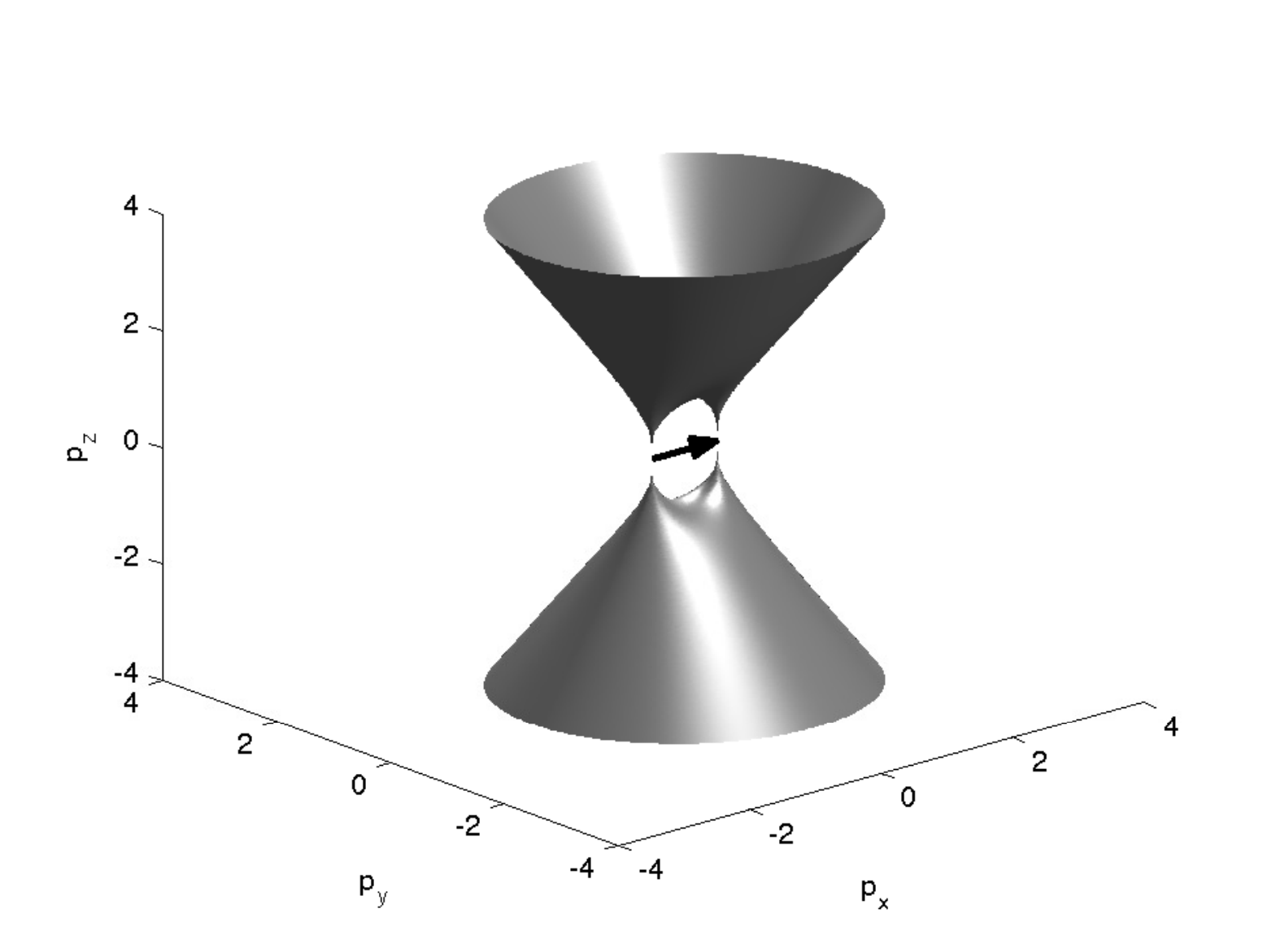}
\end{center}
\vspace{-5pt}
\caption{Resonant manifold ${\mathcal R}^{++}_\bk, \,\,\,\, \bk=(1,0,0) $ for internal gravity waves with vertical direction along the 
z-axis. The black arrow is the vector $\bk.$}
\end{figure}

In general, we shall use the term ``pseudo-critical point" for any 
point $\bp$ on a resonant manifold ${\mathcal R}_\bk^{s_1s_2}$ where $E^{s_1s_2}(\bp;\bk)$ is non-smooth in $\bp.$ 
Although such points may give rise to geometric singularities, they do not usually produce an infinite phase measure.
In fact, the density of the phase measure with respect to surface area (Hausdorff measure) vanishes at 
points where $|\grad_\bp E^{++}(\bp;\bk)|=\infty,$ and thus the phase measure is locally finite whenever the 
Hausdorff measure is locally finite. The latter condition may easily be checked for the pseudo-critical points 
in Fig.~6 by using the standard formula for element of surface area in cylindrical coordinates to obtain 
near $\bp=\b0$ that $dA=2\pi (2k^2)^{1/3} p_H^{1/3} dp_H,$ which has locally a finite integral. 
Note furthermore for the Euler-Boussinesq system  
that the pseudo-critical points are not dynamically relevant in wave kinetics,  since the nonlinear interaction coefficient 
vanishes when all modes in the triad have zero vertical wavenumber \cite{CaillolZeitlin00}. 

\paragraph{Summary}

As these examples show, anisotropy---whether power-law or other type---can readily lead to critical points.
In most of the common cases that we have examined, the singularities in the resonant manifold are protected
by vanishing nonlinearity from having any dynamical effects. Such protection is by no means guaranteed. The case 
of inertial waves presents an opposite case, where the singularity is associated to strong nonlinearity and 
a breakdown of the wave kinetic theory. More generally, the singularities can have intermediate effects between 
none at all and complete breakdown of wave kinetics. We shall present examples of this in the next section.

%%%%%%%%%%%%%%%%%%%%%%%%%%%%%%%%%%%%%%%%%%%%%%
%%%%%%%%%%%%%%%%%%%%%%%%%%%%%%%%%%%%%%%%%%%%%%
\subsection{Quartet Resonances}\lb{quartet}
%%%%%%%%%%%%%%%%%%%%%%%%%%%%%%%%%%%%%%%%%%%%%%
%%%%%%%%%%%%%%%%%%%%%%%%%%%%%%%%%%%%%%%%%%%%%%

Resonance Van Hove singularities also occur in 4-wave systems, for which the collision integral has the standard form:
\begin{eqnarray}
C_\bk[n]&=&\int d^dp\,d^dq \,d^d\ell\,\,\delta^d(\bp+\bq-\bk-\bell)\delta(\omega(\bp)+\omega(\bq)-\omega(\bk)-\omega(\bell))\,\,
\left|H^{\bk,\text{\scriptsize{$\bell$}}}_{\bp,\bq}\right|^2
\cr
&& \,\,\,\,\,\,\,\,\,\,\,\,\,\,\,\,\,\,\,\,\,\,\,\,\,\,\,\,\,\,
\times n(\bp)n(\bq)n(\bell)n(\bk)\left[\frac{1}{n(\bell)}+\frac{1}{n(\bk)} - \frac{1}{n(\bp)}-\frac{1}{n(\bq)}\right]
%\cr
%&=& \int d^dp_3 \int_{\{\omega(\bp_2)+\omega(\bk+\bp_3-\bp_2)=\omega(\bk)+\omega(\bp_3)\}} dS(\bp_2) 
\lb{coll4} \end{eqnarray} 
When the wave dynamics is Hamiltonian and 3-wave resonances are absent, the collision integral can be brought 
to the above form by a canonical transformation of the Hamiltonian system \cite{Zakharovetal92}. 
The resonant manifold is now ${\mathcal R}_\bk^{(4)}= \{(\bp,\bq)\,:\, E^{(4)}(\bp,\bq;\bk)=0 \}$, with 
\be E^{(4)}(\bp,\bq;\bk)=\omega(\bp)+\omega(\bq)-\omega(\bk)-\omega(\bp+\bq-\bk). \lb{E4} \ee  
Thus $\cR^{(4)}$ can be expected to be a $(2d-1)$-dimensional surface embedded in a Euclidean space of dimension 
$D=2d.$ As a matter of fact, this is only true if one disregards the ``trivial" part $\cR^{(4)}_{{\rm triv}}=\{(\bp,\bq):\,\, \bp=\bk  
\,\,\,\,{\rm or}\,\,\,\, \bq=\bk\},$  which is the union of two $d$-dimensional hyperplanes. This ``trivial'' part gives a vanishing direct 
contribution to the collision integral (\ref{coll4}) because either $n(\bp)=n(\bk),n(\bq)=n(\bell)$ or 
$n(\bp)=n(\bell),n(\bq)=n(\bk),$ and it is thus generally ignored. However, we shall see below 
that it may be of indirect importance because any intersection of the ``non-trivial'' part with $\cR^{(4)}_{{\rm triv}}$ 
leads to sets of critical points, generically of dimension $d-1$, and possible divergences.  

The condition for a critical point on the 4-wave resonant manifold is 
\be  \grad\omega(\bp)=\grad\omega(\bq)= \grad\omega(\bell) \ee
with the group velocity of all three wavevectors  $\bp,\bq,\bell$ the same. 
Degeneracy depends upon the rank of the $D\times D$ Hessian matrix
\be \grad\otimes\grad E^{(4)} = \left(\begin{array}{cc}
                                                  \grad\otimes\grad\omega(\bp)-\grad\otimes\grad\omega(\bell) 
                                                  & -\grad\otimes\grad\omega(\bell) \cr
                                                  -\grad\otimes\grad\omega(\bell) &
                                                  \grad\otimes\grad\omega(\bq)-\grad\otimes\grad\omega(\bell)
                                                  \end{array}\right), \lb{Hess4} \ee
the critical point being non-degenerate if this matrix has full rank and otherwise degenerate.                                                   

It is worth discussing the case of a general isotropic dispersion law $\omega(k),$ as a preliminary to some specific 
examples below. The condition for a critical point 
$$ \omega'(p)\hat{\bp}=\omega'(q)\hat{\bq}=\omega'(\ell)\hat{\bell} $$
can be met in one of two ways: 
\begin{itemize}
\item[] {\it (i)} \,\, $|\omega'(p)|=|\omega'(q)|=|\omega'(\ell)|\neq 0$ and the vectors 
$\hat{\bp},\hat{\bq},\hat{\bell}$ are all collinear with $\hat{\bk}$ (parallel or anti-parallel depending 
on the sign of $\omega'$), 
\item[] {\it (ii)} \,\, $\omega'(p)=\omega'(q)=\omega'(\ell)=0$ and no restriction on the vectors 
$\hat{\bp},\hat{\bq},\hat{\bell}$.
\end{itemize} 
We shall refer to the first case as a {\it non-null critical point} with non-vanishing group velocity of the waves and to the second 
as a {\it null critical point} with zero group velocities. We include for completeness the third case
\begin{itemize}
\item[] {\it (iii)} At least one of $\omega'(p),$ $\omega'(q)$, $\omega'(\ell)$ is infinite.  
\end{itemize} 
This is what we earlier termed a {\it pseudo-critical point}. These present generally no difficulty since the density of the phase measure 
with respect to surface area on the resonant manifold vanishes at pseudo-critical points, with an infinite gradient. Note also for the isotropic dispersion that 
$$ \grad\otimes\grad \omega(\bk) = \omega''(k) \hat{\bk}\otimes\hat{\bk} + \frac{\omega'(k)}{k}({\bf I}-\hat{\bk}\otimes\hat{\bk}), $$
from which it is easy to determine the rank of the Hessian matrix (\ref{Hess4}). 

We now consider several concrete examples: 

\subsubsection{Surface gravity-capillary waves}\lb{grav-cap}

An illustrative example is surface gravity-capillary waves with dispersion relation $\omega(k) = \sqrt{gk+\sigma k^3},$ where $g$ is the  
acceleration due to gravity and $\sigma=S/\rho$, where $S$ is surface tension. It turns out that most of the relevant features 
appear already for the idealized one-dimensional case $d=1.$

Consider first pure surface gravity waves with dispersion law $\omega=\sqrt{gk}$. 
This case has no null critical points. Because any non-null critical points must have wavevectors all parallel to $\hat{\bk},$ we may 
check for their existence in the simplest case $d=1$. 
%The 4-wave interaction coefficient is identically zero on the resonant manifold in $d=1$ \cite{DyachenkoZakharov94}, but
As a matter of fact, the resonant manifold is analytically known and explicitly parameterized for $d=1,$ and 
consists of points for which one of the wavenumbers out of $p,q,k,\ell$ has an opposite sign from the others \cite{DyachenkoZakharov94}.   
There are thus also no non-null critical points in $d=1$ and therefore none for $d>1$. There are, however, pseudo-critical points 
in $d=1$ where the nontrivial portion of the resonant manifold intersects the trivial parts, as seen in Fig.~7 below. These occur at the points where 
either $p=0$ or $q=0$ and $\omega'$ diverges. The non-trivial part of the resonant manifold is a union of three smooth pieces, joined 
at the pseudo-critical points, but the phase measure on it is locally finite 
%\footnote{
%We only consider $(p,q)=(k,0)$ with $k>0$ and write $(p,q)=(k+\delta p, \delta q)$. Introducing new coordinates
%$$
%\begin{cases}
%x = \mbox{sign}(\delta p+\delta q) \left(\sqrt{k+\delta p } - \sqrt{k} + \frac{\delta q}{2\sqrt{k}} - \sqrt{|\delta p +\delta q|} \right)^2,
%\cr
%y = \mbox{sign}(\delta q) (\sqrt{|\delta q|} - \frac{\delta q}{2\sqrt{k}})^2,
%\end{cases}
%$$
%with Jacobian of transformation $\left|\frac{\partial(x,y)}{\partial(\delta p,\delta q)}\right|=1$ at $(k,0)$, one can write
%$$
%\tilde{E} = \sqrt{|y|} - \sqrt{|x|}.
%$$
%and $|\grad\tilde{E}| = 1/\sqrt{2x}$ on the resonant manifold.
%Without loss of generality, we consider the neighborhood of the pseudo-critical point $V=[-\eta,\eta]^2$ so that the integral is given by
%$$ 
%J = \int_{|x|\le \eta, |y|\le\eta} dxdy\ \delta(\tilde{E}(x,y)) = \int_{-\eta}^\eta dy \int_{y=x} \cH^0(dx) \frac{1}{\sqrt{2x}}
%= (const.)  \int_{-\eta}^\eta y^{-1/2} dy.
%$$
%Unlike the pseudo-critical point for the internal gravity waves, the density of phase measure is not zero at the critical point, but the phase measure is still finite.
%}
 (and the 4-wave interaction coefficient zero \cite{DyachenkoZakharov94}).   

%FIGURE HERE

\vspace{15pt}
\begin{figure}[!h]
\begin{center}\lb{FigGrav2}
\includegraphics[width=2.5in,height=2in,keepaspectratio]{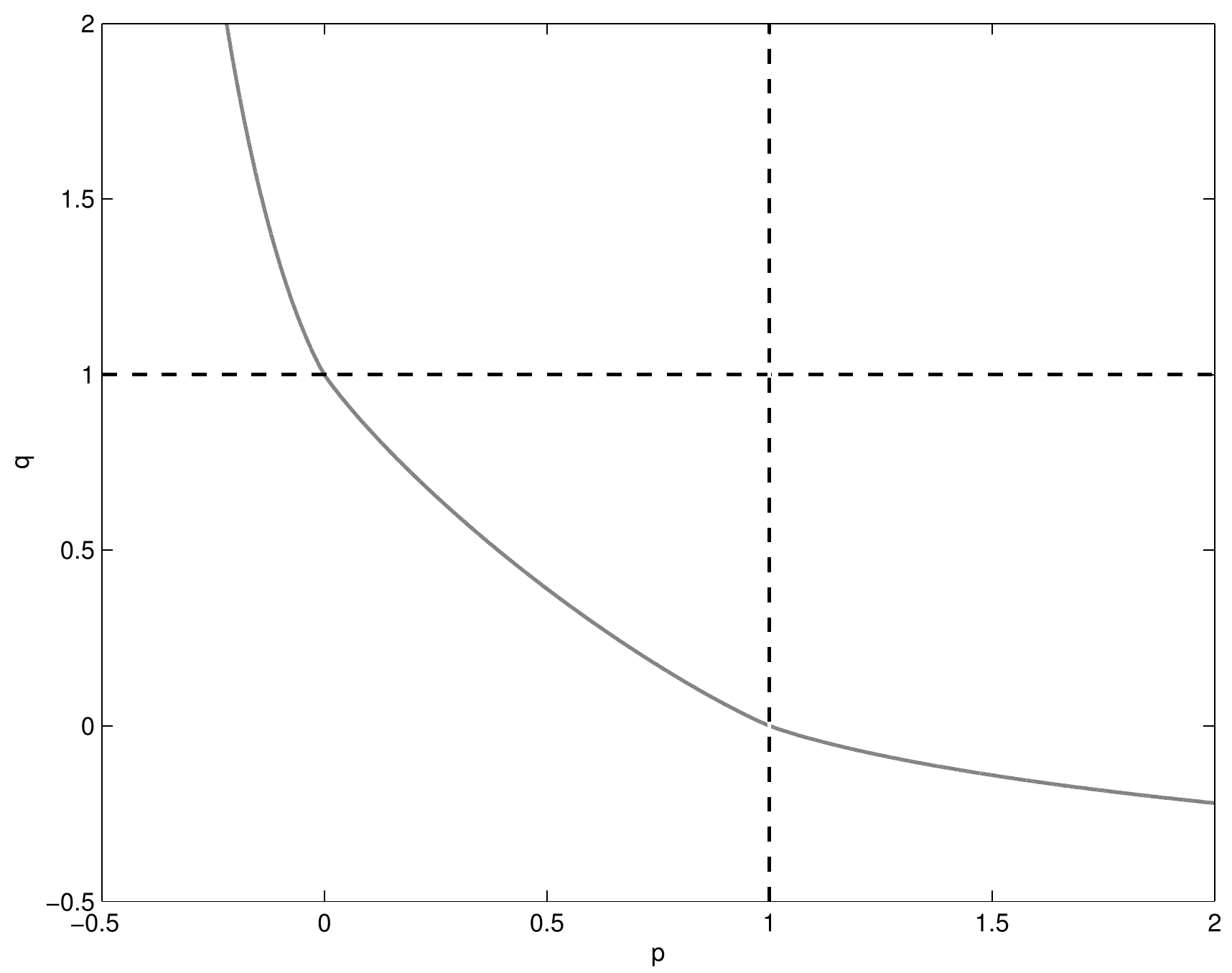}
\end{center}
\caption{Resonant manifold ${\mathcal R}^{(4)}_k, \,\,\,\, k=1$ for surface gravity waves in $d=1$, with the non-trivial
part plotted in gray and the trivial part in dashed black.}
\end{figure}

However, if the surface tension effect is included, there will be critical points as well as pseudo-critical points, as seen in Fig.~8, 
due to the inflection point of $\omega(k)$ at $k=k_*:=\sqrt{\frac{2-\sqrt{3}}{\sqrt{3}}\frac{g}{\sigma}}$. Again, null critical points are not present. 
% By condition (i) and the property of $\omega'(k)$, 
However, for $k\neq k_*$ there is a distinct wavenumber $\ell\neq k$ satisfying $\omega'(k)=\omega'(\ell)$, 
implying that there are two non-null critical points for $\bp=\bk,\bq=\bell$ and $\bq=\bk,\bp=\bell$ 
with $\bell=\ell\hat{\bk}.$ 
To determine their degeneracy, we again examine the Hessian (\ref{Hess4}). 
At least one of the diagonal term vanishes and the off-diagonal term $-\grad\otimes\grad\omega(\bell) $
has eigenvalue $-\omega''(\ell)$ with multiplicity $1$ and $-\omega'(\ell)/\ell$ with multiplicity $d-1$. 
If $k\neq k_*$ such that $\omega''(k)\neq 0$, then $\omega''(\ell) \neq 0$ and there exist two non-degenerate critical points; if $k=k_*$ such that $\omega''(k)=0$, then $\omega''(\ell)=0$ and there exists one degenerate critical point. In Fig.~8, we show for $d=1$ the typical resonant manifold $\cR_k^{(4)}$ for $k<k_*$ (left panel), 
$k=k_*$ (middle panel), and $k>k_*$ (right panel). In general, the degeneracy $\delta=0$ for $k\neq k_*$ and $\delta=1$ for $k=k_*$. 
As we shall discuss below, this implies that the phase measure remains finite for the physically relevant case $d=2$.

%FIGURE HERE

\vspace{15pt} 
\begin{figure}[!h]
\begin{center}\lb{Fig6}
\includegraphics[width=2.1in,height=3in,keepaspectratio]{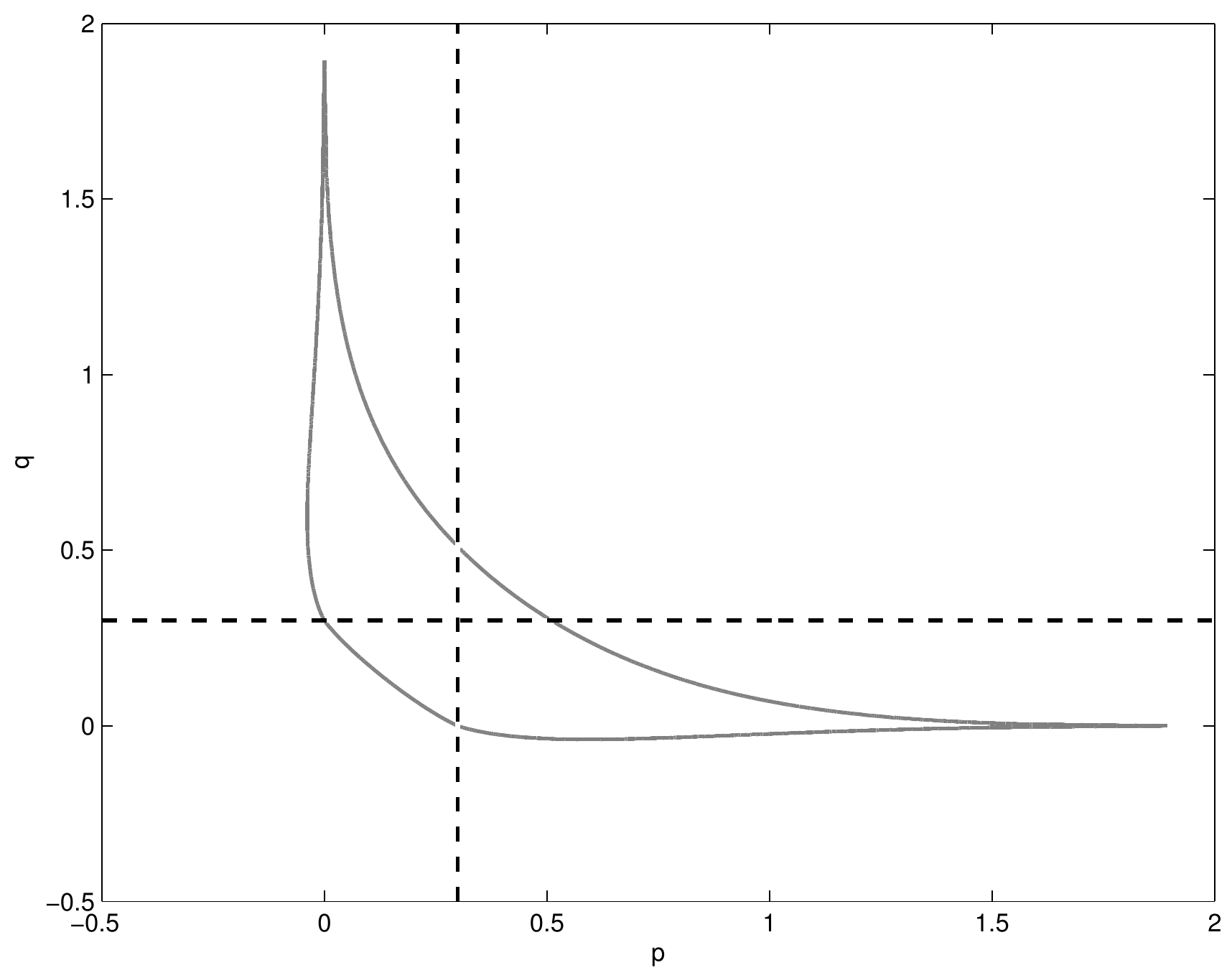}
\includegraphics[width=2.1in,height=3in,keepaspectratio]{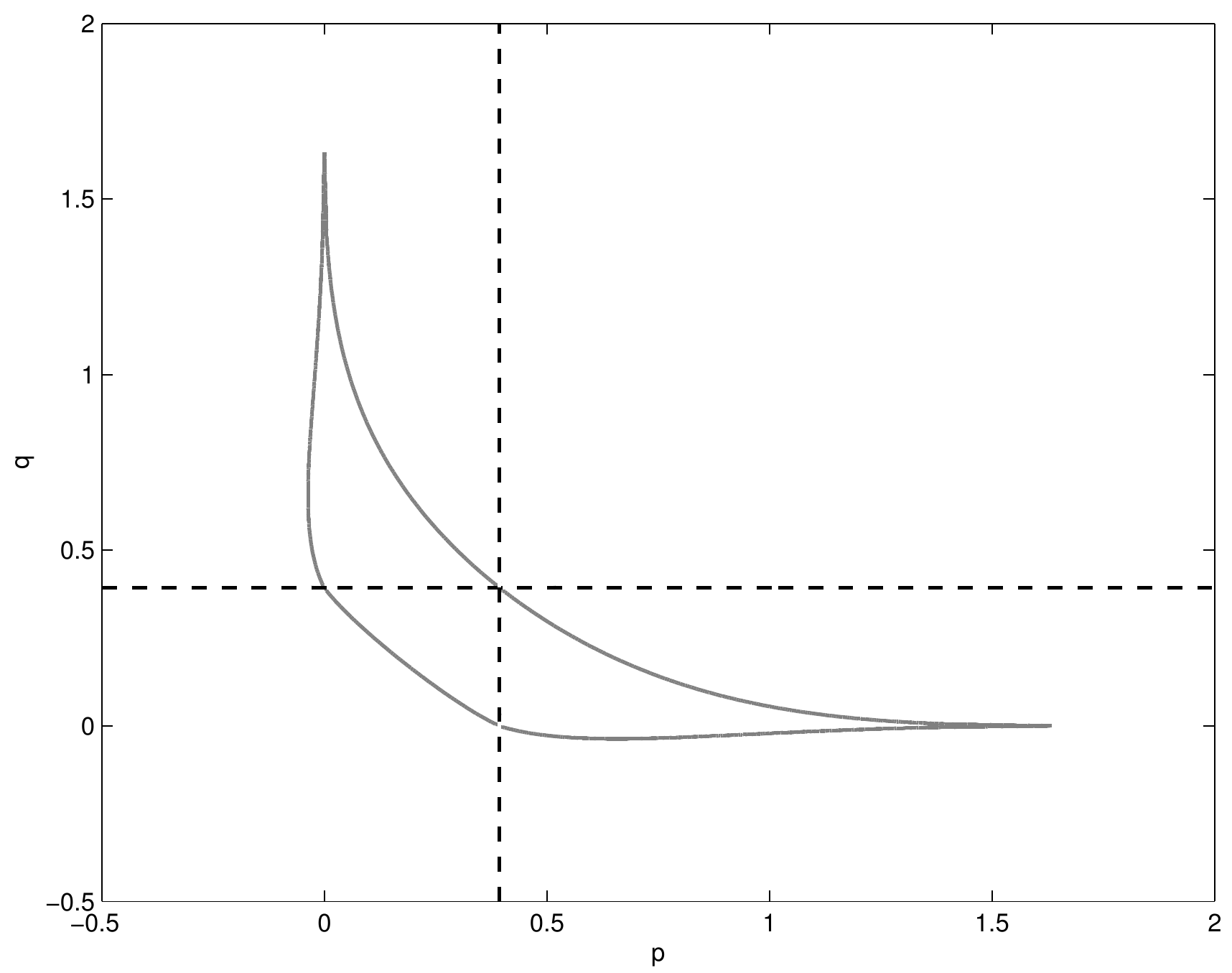}
\includegraphics[width=2.1in,height=3in,keepaspectratio]{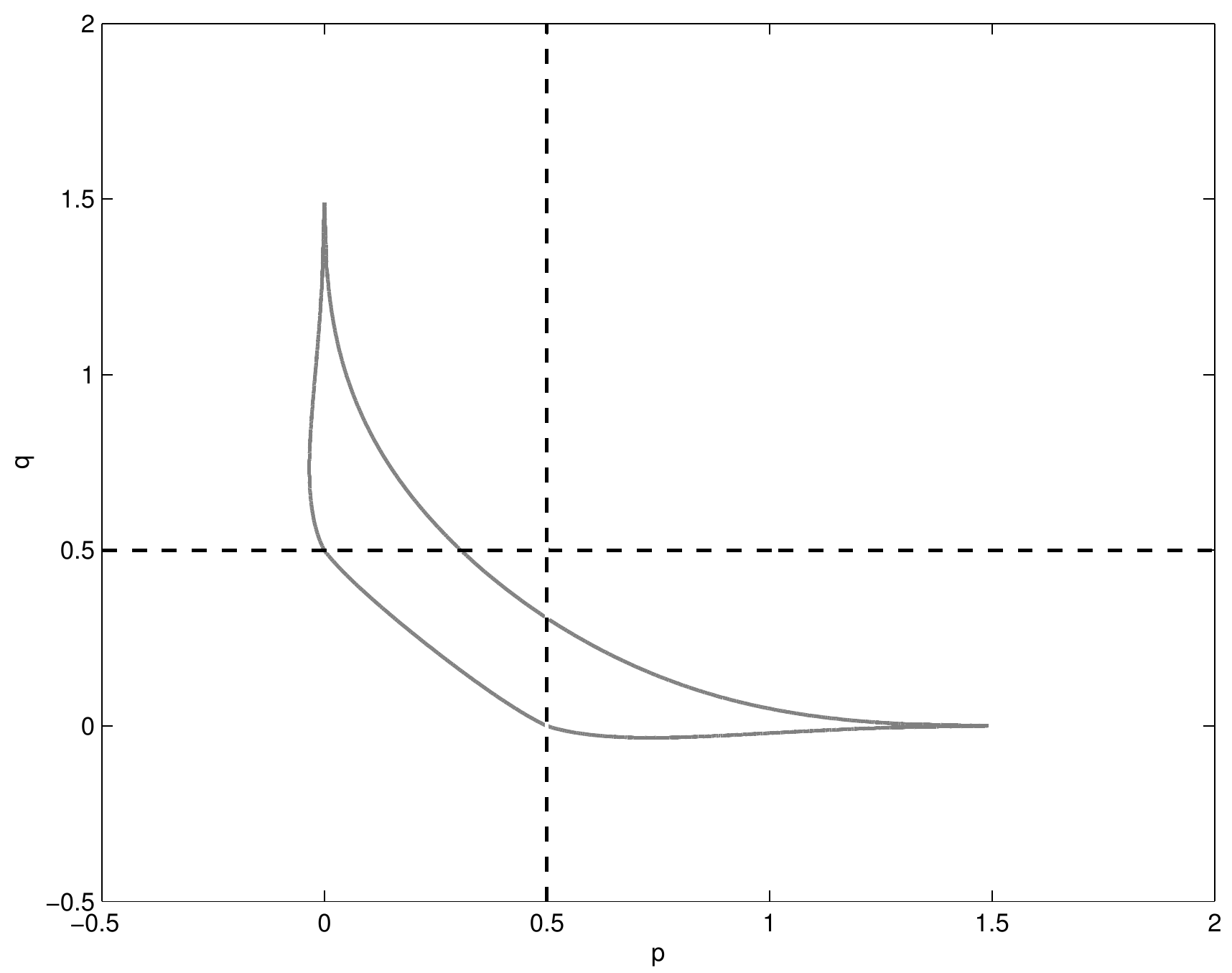}
\end{center}
\caption{The figures show the resonant manifold ${\mathcal R}^{(4)}_k$ for surface gravity waves in $d=1$ for $g=\sigma$, with the non-trivial part plotted in gray and the trivial part in dashed black. Here $k=0.3$ for the left panel; $k=\sqrt{(2-\sqrt{3})/\sqrt{3}}$ for the middle panel; $k=0.5$ for the right panel.}
\end{figure}

It should be noted that the wavenumber $k_*\approx 0.4 (g/\sigma)^{1/2}$ is expected to lie in the transition range between two energy cascades, 
one at low wavenumbers driven by quartet resonances of gravity waves and another at high wavenumbers driven by triplet resonances of capillary waves 
\cite{Bivenetal01,Newelletal01,NewellZakharov08}. The quartet at the critical points in Fig.~8 has two wavevectors with each magnitude $k,\ell$, and 
one of the magnitudes $k,\ell$ is always $<k_*$ and the other $>k_*.$ Thus, any possible observable effect of the critical points would presumably 
be found in the transition region where $k\simeq k_*,$ perhaps in laboratory experiments where there is no large scale-separation between 
the gravity and capillary wave regimes. However, finiteness of the phase measure for $d=2$ makes it unlikely that there are any appreciable effects.

\subsubsection{Wave propagation along an optical fiber}\lb{optics} 
%%%%%%%%%%%%%%%%%%%%%%%%%%%%%%%%%%%%%%%%%%%%%%

A very similar situation to the previous one, but for $d=1,$ occurs for optical wave propagation along a fiber, modeled by a 
$1d$ nonlinear Schr\"odinger (NLS) equation with third-order dispersion \cite{Micheletal10,Suretetal10}.
The dispersion law is\footnote{As usual in application of the NLS equation to optics, the space variable $z$ and the time variable $t$
have their roles exchanged, and thus also the roles of wavenumber $k$ and frequency $\omega.$ However, here 
we revert to the notations used elsewhere in our paper.} 
\begin{eqnarray}
\omega(k)=s k^2+\alpha k^3,
\end{eqnarray}
for $s=\pm 1,$ with an inflection point at $k_*= -s/3\alpha.$
% which has an inflection point at $k_*=-s/(3\alpha)$.
The collision integral is 
\begin{eqnarray}
C_k[n] &=&\frac{1}{\pi} \int dp \, dq \, d\ell \,\,  \delta(p+q-k-\ell)\delta\big(\omega(p)+\omega(q)-\omega(k)-\omega(\ell)\big)
\cr
&&\,\,\,\,\,\,\,\,\,\,\,\times n(k)n(\ell)n(p)n(q)\left[\frac{1}{n(k)}+\frac{1}{n(\ell)}-\frac{1}{n(p)}-\frac{1}{n(q)}\right].
\end{eqnarray}
It can be easily shown here that 
$$ E^{(4)}(p,q;k)\equiv \omega(p)+\omega(q)-\omega(k)-\omega(p+q-k)=3\alpha (p+q-2k_*)(p-k)(q-k). $$ 
The non-trivial part of the resonant manifold $\cR^{(4)}_k$ is the straight line $p+q=2k_*$ in the $pq$-plane, independent
of $k,$  and there are non-null critical points at the intersections with the trivial part, at $(p,q)=(k,2k_*-k)$ and $(2k_*-k,k).$
The collision integral on the non-trivial part becomes
\begin{eqnarray}
C_k[n]&=&\frac{1}{3\pi|\alpha|}\int dp \ \frac{n(k)n(2k_*-k)n(p)n(2k_*-p)}{|p-k|\,|p+k-2k_*|}
\cr
&&\,\,\,\,\,\,\,\,\,\,\,\,\,\,\,\,\,\,\,\,\,\,\times\left[\frac{1}{n(k)}+\frac{1}{n(2k_*-k)}-\frac{1}{n(p)}-\frac{1}{n(2k_*-p)}\right].
\end{eqnarray}
As expected for $D=2d=2,$ these critical points produce logarithmic divergences in the phase measure at the points $p=k,2k_*-k.$ For the 
special choice $k=k_*,$ there is a double pole at $p=k_*$ when $\grad\otimes\grad E^{(4)}$ becomes identically zero at the critical point 
$(p,q)=(k_*,k_*)$. This degenerate critical point corresponds to a triple intersection between all three smooth pieces of the resonant manifold;
see Fig.~9 below.  

%FIGURE HERE

\vspace{20pt}
 
\begin{figure}[!h]
\begin{center}\lb{Fig6}
\includegraphics[width=2.5in,height=2in,keepaspectratio]{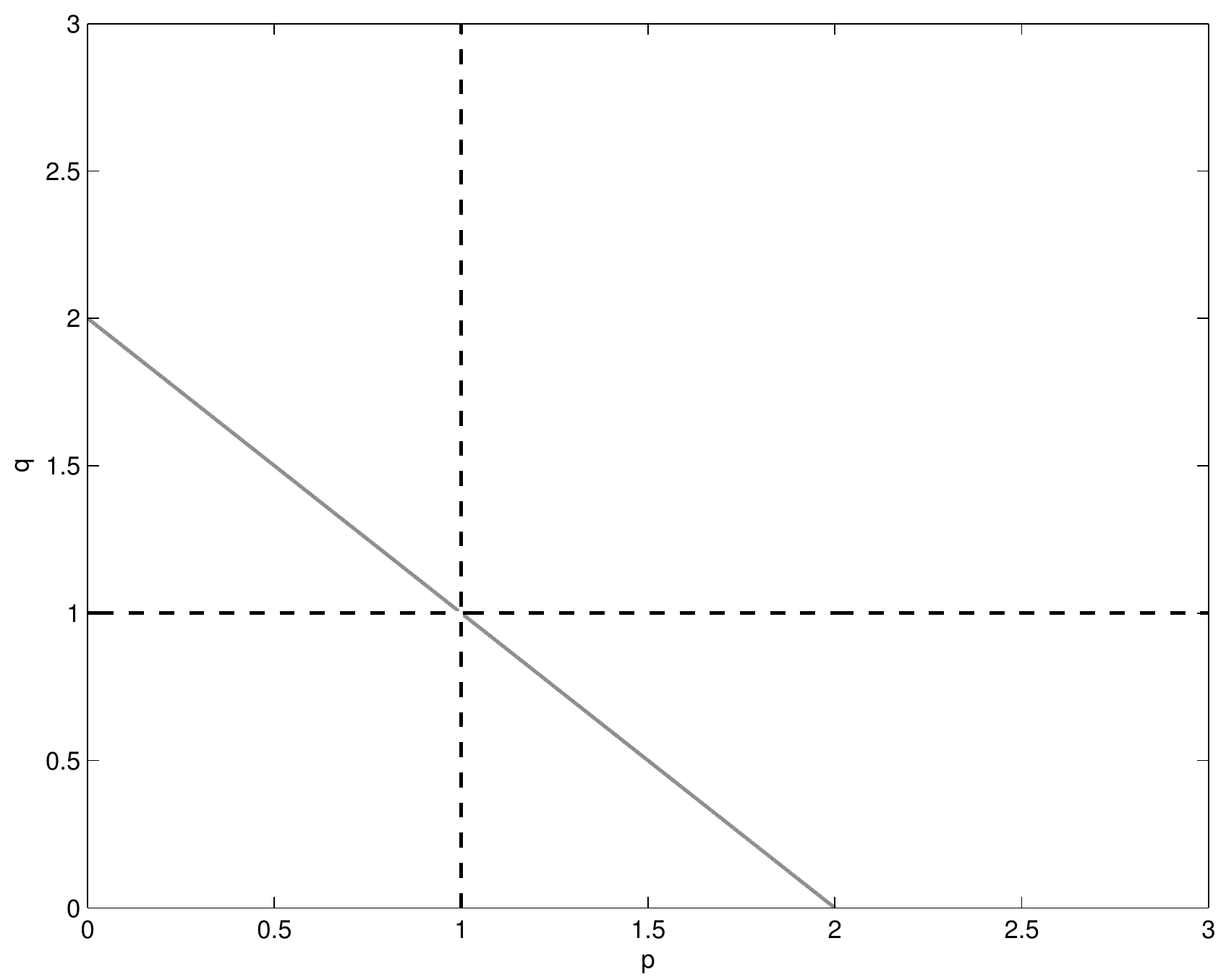}
\end{center}
\caption{Resonant manifold ${\mathcal R}^{(4)}_k, \,\,\,\, k=1$ of one-dimensional 3rd-order dispersive optical waves for $s=-1$, $\alpha=3,$ 
so that $k_*=1,$ with the non-trivial part of the manifold plotted in gray and the trivial part in dashed black. 
The triple intersection is a degenerate critical point.}
\end{figure}

Because the critical points correspond to intersections with the trivial part of the resonant manifold, the integrand of 
the collision integral vanishes at those points and thus the integrals may be finite. For example, assume that $n(p)$ is twice-differentiable
in the vicinity of $k$ and let $\delta p$ be the scale of variation of $n(p)$ near that point. Then one easily finds by Taylor expansion that 
the contribution to $C_k[n]$ from integrating over $|p-k|<\delta p$ is 
\be C_k^{(p\approx k)}[n] =\frac{1}{6\pi|\alpha|}\left[ \left(\frac{n''(k)n^2(2k_*-k)}{2}-\frac{(n'(k))^2n^2(2k_*-k)}{n(k)}\right)
+ (k\leftrightarrow 2k_*-k) \right] \frac{(\delta p)^2}{|k-k_*|} \lb{1pole} \ee
with a similar contribution coming from $p\approx 2k_*-k.$ When $|k-k_*|<\delta p$ this is replaced by a contribution from the double pole:
\be C_{k\approx k_*}^{(p\approx k_*)}(n) =\frac{1}{3\pi|\alpha|}\left[ n''(k_*)n^2(k_*)-2(n'(k_*))^2n^2(k_*)\right](\delta p). \lb{2pole} \ee
These are finite as long as $n(p)$ is twice continuously differentiable near the poles. More generally, the contribution to the 
collision integral is finite if $n(p)$ has cusp-like singularities near $p=k_*,$ $n(p)-n(k_*)\sim A |p-k_*|^c$ with $c>1.$ 

Although the collision integral remains finite under the assumptions stated above, it may nevertheless become large, in the sense 
that the nonlinear frequency $\Gamma_k(n)=C_k(n)/n(k)$ could be of the same order as (or greater than) the linear frequency $\omega(k).$
If so, this violates the condition $\Gamma_k(n)\ll \omega(k)$ required for validity of the kinetic equation \cite{Newelletal01,Bivenetal01,Nazarenko11}. 
Indeed, it was shown in the numerical study \cite{Micheletal10} that, for certain values of the cubic coefficient $\alpha,$
the ratio $\Gamma_k(n)/\omega(k)$ exceeded 1 near $k=k_*$ and, in that case, there was no longer quantitative agreement between 
the predictions of the kinetic equation and ensemble-averaged solutions of the NLS equation (see Figs.2 and 5 in \cite{Micheletal10}). 
This is a physically interesting example which shows that resonance Van Hove singularities can lead to a breakdown in validity of kinetic 
theory, even when the collision integral remains finite. 

%Furthermore,  if if $n(p)$ has cusp singularities near $p=k_*$ with $c\leq 1$ then the collision integral diverges due to the poles. 
%Such non-smooth solutions can be prepared by choosing non-smooth initial values of the quantity $J(k)=n(k)-n(2k_*-k)$ 
%which is an exact invariant of the kinetic equation for this problem \cite{Micheletal10,Suretetal10}. In that case, 
%the standard kinetic equation is no longer even well-defined and it certainly cannot describe the statistics of the 
%NLS system. We discuss this situation further in section \ref{optical}.  

%\newpage

\subsubsection{Electrons and holes in graphene}\lb{graph}

We have so far considered classical 4-wave systems, but resonance Van Hove singularities can also occur in quantum wave kinetics.
The collision integral of the quantum kinetic equation has the typical form 
\begin{eqnarray}
&& C_{\bk s}[n]=\sum_{s',s'',s'''}\int d^dp\,d^dq \,d^d\ell\,\,\delta^d(\bk+\bell-\bp-\bq)\delta(\omega_s(\bk)+\omega_{s'}(\bell)-\omega_{s''}(\bp)-\omega_{s'''}(\bq))\,\,
\cr 
&& \hspace{30pt}  \times R^{ss's''s'''}_{\bk\mbox{\scriptsize{$\bell$}}\bp\bq} 
\Big[ (1\pm n_s(\bk))(1\pm n_{s'}(\bell))n_{s''}(\bp)n_{s'''}(\bq) \cr
&& \hspace{220pt} - n_s(\bk)n_{s'}(\bell)(1\pm n_{s''}(\bp))(1\pm n_{s'''}(\bq))\Big] 
%&=& \int d^dp_3 \int_{\{\omega(\bp_2)+\omega(\bk+\bp_3-\bp_2)=\omega(\bk)+\omega(\bp_3)\}} dS(\bp_2) 
%&& \, \hspace{50pt} \cr
\lb{coll4} \end{eqnarray} 
with $\pm$ for Bose/Fermi particles, respectively. (E.g. see section 2.1.6 of \cite{Zakharovetal92}). The presence or not 
of resonance Van Hove singularities in the quantum case is thus governed by the same considerations as for classical wave kinetics. 

A concrete example of physical interest is the dispersion law $\omega_s(k)=csk,$ $s=\pm 1$ which for $d=2$ 
describes the band energies of electron-hole excitations in graphene near the Dirac points. The quantum kinetic equation has been used 
to predict electron transport properties of pure, undoped samples of graphene \cite{Kashuba08,Fritzetal08,Muelleretal09}.
The resonant manifold is a three-dimensional surface embedded in the Euclidean space of dimension $D=2d=4.$
Any critical points are clearly non-null with all wavevectors collinear. 
Because $\omega''(k)\equiv 0,$ the Hessian matrix (\ref{Hess4}) has null eigenvectors $(\hat{\bk},\b0)$, $(\b0,\hat{\bk})$
and is co-rank at least 2. Thus the critical points are all doubly degenerate. In fact, the critical points lie on a 2-dimensional surface. For example,
if $\bk=(k,0)$ with $k>0,$ then the critical set for $s,s',s'',s'''=1$ consists of $\bp=(p_1,0),$ $\bq=(q_1,0)$ satisfying $p_{1}>0,$ 
$q_{1}>0,$ and $p_{1}+q_{1}>k.$

%FIGURE HERE

\begin{figure}[!h]
\begin{center}\lb{FigGraphene}
\includegraphics[width=2.1in,height=3in,keepaspectratio]{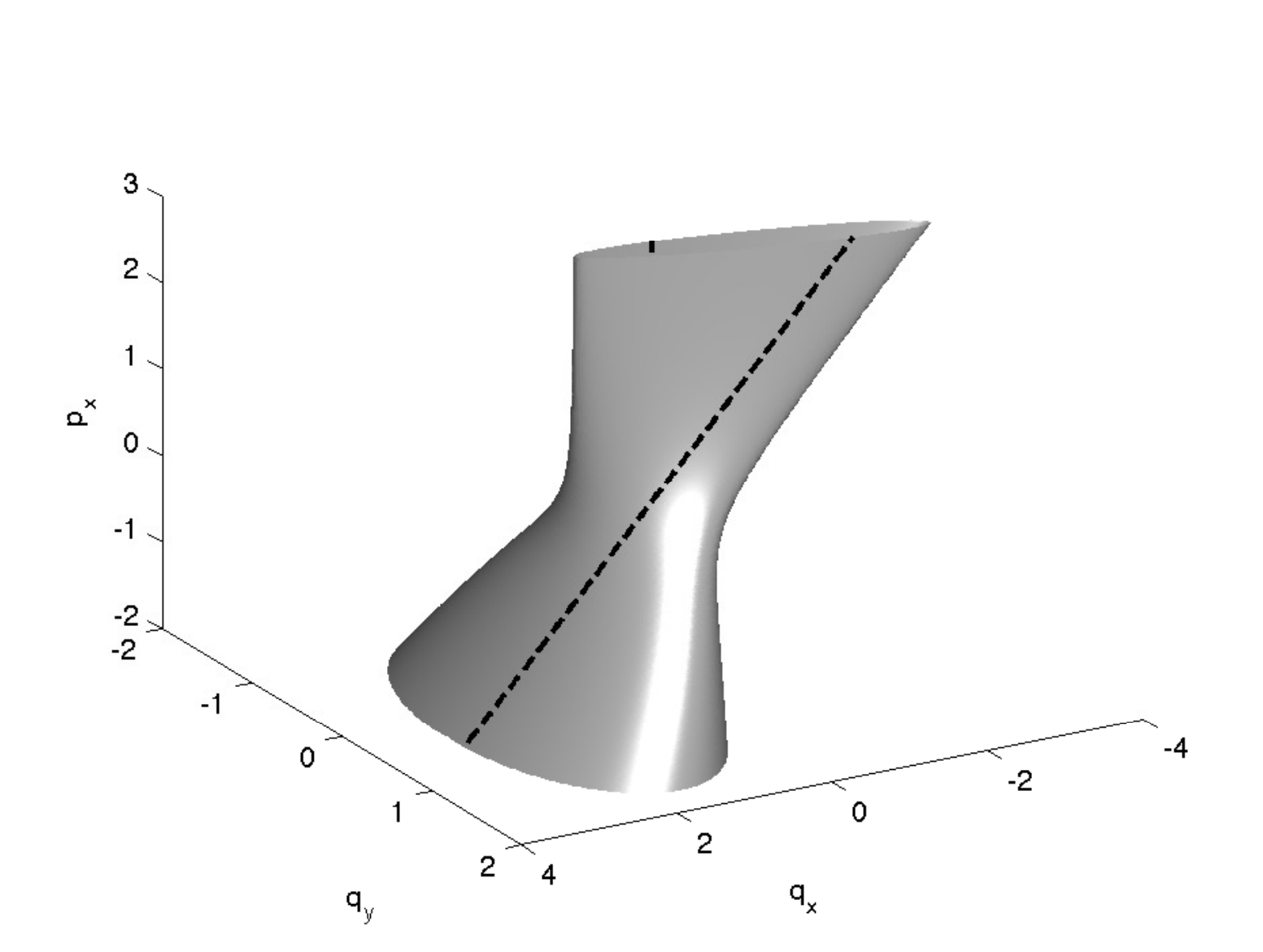}
\includegraphics[width=2.1in,height=3in,keepaspectratio]{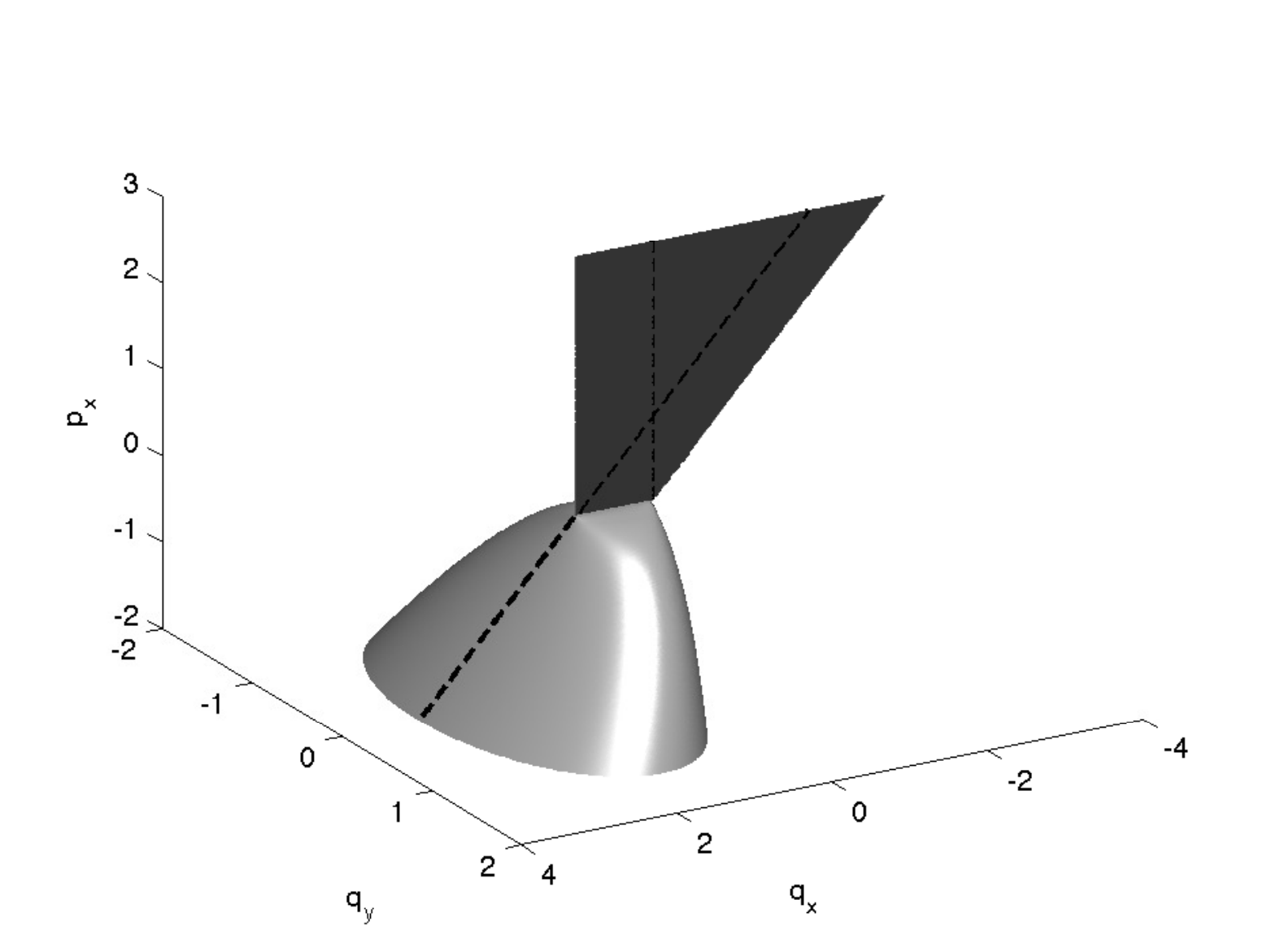}
\includegraphics[width=2.1in,height=3in,keepaspectratio]{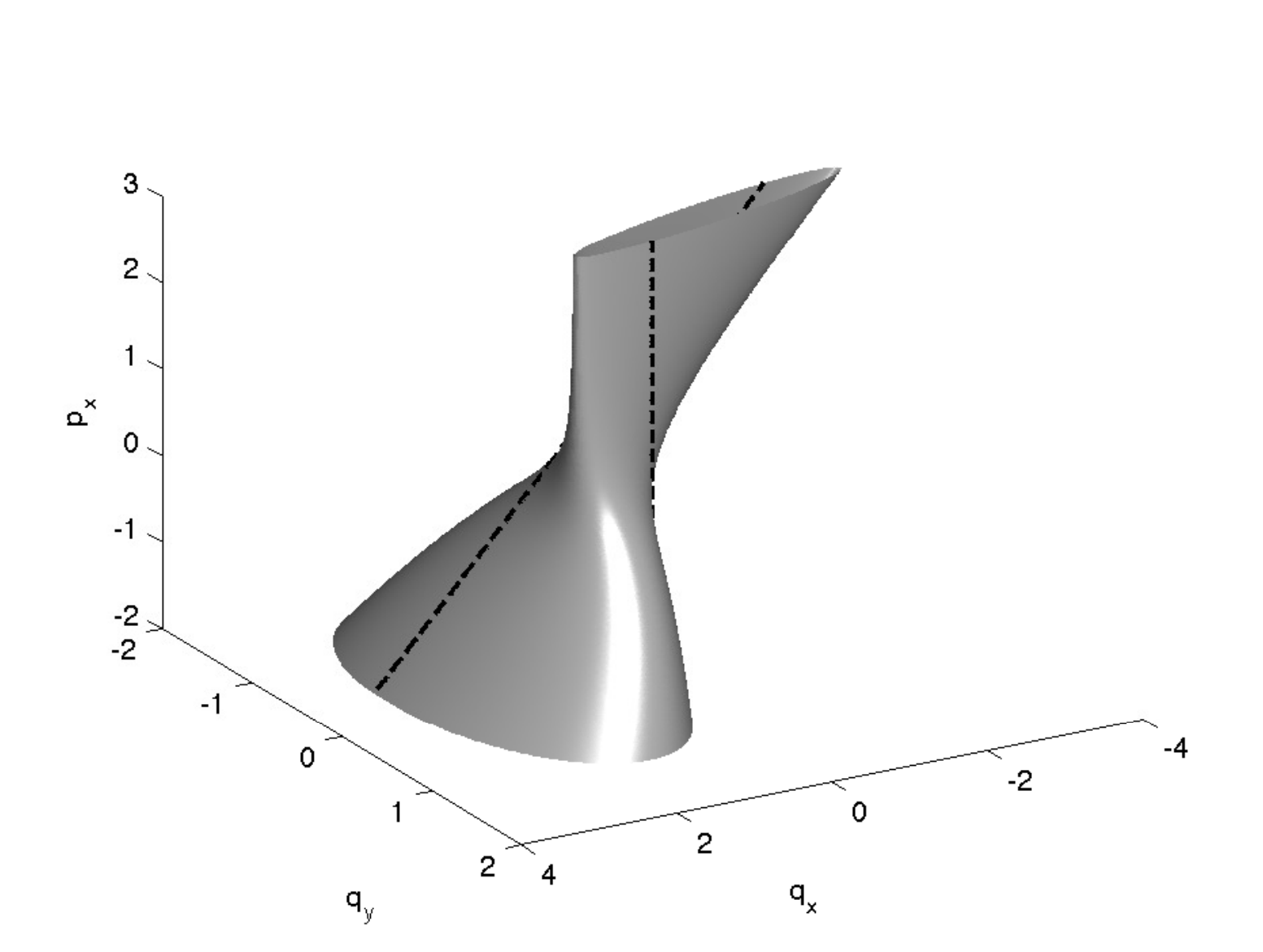}
\end{center}
\caption{The section of the resonant manifold ${\mathcal R}^{(++++)}_\bk, \,\,\,\, \bk=(1,0)$ for fixed $p_y$ 
($p_y<0$ for the left panel; $p_y=0$ for the middle panel; $p_y>0$ for the right panel), with the non-trivial
part plotted in gray and the trivial part in dashed black. The 2D critical set in the middle panel is plotted in dark gray.
For $p_y\neq 0$ the horizontal sections of the resonant manifold at fixed $p_x$-values are ellipses, but for 
$p_y=0$ and $p_x\geq 0$ these sections are line-segments.}
\end{figure}

In Fig.~10, we plot the resonant manifold $\cR_\bk^{(++++)}$, $\bk=\bfe_1$ for 4-wave interactions in graphene.
For visualization, we employ the notation in \cite{Sachdev98,Kashuba08, Fritzetal08} that the four resonant wave vectors $\bk,\bell,\bp,\bq$ are instead denoted by $\bk,\bp, \bk+\bq,\bp-\bq$. 
The resonance condition in these variables becomes $|\bq+\bk|+|\bq-\bp|=k+p,$ which, as noted in \cite{Sachdev98}, corresponds at fixed $\bp$ values
to ellipses in the $\bq$-planes, with foci at $-\bk$ and $+\bp.$  When $\bk\|\bp,$ these ellipses degenerate to line-segments $[-\bk,+\bp]$, 
whose set union comprises the critical set. In Fig.~10 we plot 2D sections of the resonant manifold in the 3D space $(q_x,q_y,p_x)$ at three 
different values of $p_y.$ For the chosen value of $\bk=\bfe_1,$ the 2D critical set is located on the $p_y=0$ section, as shown in the middle panel.

For graphene with $D=4,$ this 2-dimensional critical surface produces a logarithmic singularity in the phase measure, 
as has been previously noted \cite{Sachdev98,Kashuba08,Fritzetal08}. For example, for $\bk=k \bfe_1,$ $k>0,$
using the notation $\bk,\bp, \bk+\bq,\bp-\bq$ for the wavenumber quartet, and writing $\bp=(p_1,p_\perp),$ 
$\bq=(q_1,q_\perp),$ it was pointed out in \cite{Fritzetal08} that to quadratic order in the transverse variables 
near the critical set 
\begin{eqnarray}
k+p-|\bq+\bk|-|\bq-\bp| &\simeq & \frac{p_\perp^2}{2p}-\frac{q_\perp^2}{2(k+p)}-\frac{(p_\perp-q_\perp)^2}{2(p-q)} \cr
&=& -\frac{k+p}{2(k+q)(p-q)}(q_\perp-\zeta_1p_\perp)(q_\perp-\zeta_2p_\perp)
\end{eqnarray} 
where $\zeta_{1,2}$ are the roots of the quadratic polynomial in $q_\perp/p_\perp$ defined by the first line. At each fixed 
value of $p_1,q_1$ that corresponds to points in the critical set, the integral over the transverse variables $p_\perp,q_\perp$
is logarithmically divergent. This can be seen most easily by changing variables to 
$$r_\perp = q_\perp-\zeta_1p_\perp, \ \ \ s_\perp=q_\perp-\zeta_2 p_\perp$$ 
with the Jacobian of transformation 
$$ \left|\frac{\partial(r_\perp,s_\perp)}{\partial(q_\perp,p_\perp)}\right|=|\zeta_1-\zeta_2| = \frac{2}{k+p}\sqrt{\frac{k(k+q)(p-q)}{p}}.$$
Thus,
$$ \int dp_\perp\int dq_\perp \ \delta(k+p-|\bq+\bk|-|\bq-\bp| ) = \sqrt{\frac{p(k+q)(p-q)}{k}}
\int dr_\perp \int ds_\perp \ \delta(r_\perp s_\perp) $$
and the latter integral exhibits, for each fixed value of $p_1,q_1$ corresponding to points in the critical set, the same type 
of logarithmic divergence that was observed for Rossby/drift waves in $d=2.$ However, electron-hole interactions do not vanish 
near the critical set, unlike the case for Rossby/drift waves, so that the divergence is unregulated.   

The local logarithmic divergence can be removed by resonance-broadening of the delta function 
(e.g. \cite{Nazarenko11}, section 6.5.2). Assuming for simplicity a constant resonance width $\gamma$, 
the delta function $\delta(r_\perp s_\perp)$ is replaced by the Lorentzian (Cauchy distribution) 
$$\delta_\gamma(r_\perp s_\perp)=\frac{\gamma}{\pi} \frac{1}{r_\perp^2s_\perp^2+\gamma^2}. $$ 
Since $\delta_\gamma(r_\perp s_\perp)\sim 1/\pi\gamma$ for $r_\perp s_\perp\ll \gamma,$ an integral over a small 
neighborhood of $r_\perp=s_\perp=0$ is now finite. The quantitative behavior can be seen from the following integral 
$$ I=\int_{-a}^a dr_\perp \int_{-b}^b ds_\perp\  \delta_\gamma(r_\perp s_\perp) = 2 \int_{0}^a \frac{dr_\perp}{r_\perp} 
\left[ \frac{\gamma/r_\perp}{\pi}\int_{-b}^{b} \frac{ds_\perp}{s_\perp^2+(\gamma/r_\perp)^2}\right], 
$$
with cutoffs $a,b$. 
The region of integration $r_\perp\ll \gamma/b$ gives a $\gamma$-independent contribution, whereas for values $r_\perp\gg \gamma/b$
the inner integral is $\simeq 1$ and thus 
$$I\sim 2\ln(ab/\gamma), \,\,\,\, \gamma\ll ab.$$ 
This asymptotic evalation can easily be made rigorous by noting that $I=\frac{4}{\pi}{\rm Ti}_2(ab/\gamma)$ 
where ${\rm Ti}_2(x)=\int_0^x du\ \frac{\arctan u}{u}$ is the inverse tangent integral (\cite{Lewin81}, Ch.VII, \S 1.2). 
The cutoffs $a,b$ may come from limits to the magnitude of all wavevectors, e.g. the size of the Brillouin zone in graphene. 
The cutoffs $a,b$ may also be associated to the maximum size of the region where $E(\bp,\bq;\bk)\propto r_\perp s_\perp.$  
We shall discuss in further detail in section \ref{singular} how the logarithmic singularity is understood to affect the electron-hole 
kinetics in graphene. 

\subsection{Summary}

This section has shown by various concrete examples that critical points in the resonance condition 
(resonance Van Hove singularities) occur in many common wave kinetic equations. They lead to geometric 
singularities in the ``resonant manifold", which is thus no longer a true manifold. The singularities may furthermore 
lead to local non-finiteness of the phase measure appearing in the collision integral, which is associated physically 
to the infinite scattering time for locally non-dispersive waves. Such a diverging phase measure may nevertheless 
produce a finite collision integral, e.g. due to a vanishing interaction coefficient (Rossby/drift waves, $d=2$)
or due to cancellations between terms in the collision integrand (waves in an optical fiber, $d=1$). When the collision integral 
itself diverges or even if it is finite but large, standard kinetic theory may break down (optical waves, $d=1$; 
electron-holes in graphene, $d=2$). Before we discuss this latter situation in section \ref{singular}, we first discuss 
in the following section more generally the conditions under which a critical point leads to a locally infinite phase measure 
at the singularity.

%%%%%%%%%%%%%%%%%%%%%%%%%%%%%%%%%%%%%%%%%%%%%%
%%%%%%%%%%%%%%%%%%%%%%%%%%%%%%%%%%%%%%%%%%%%%%
%%%%%%%%%%%%%%%%%%%%%%%%%%%%%%%%%%%%%%%%%%%%%%
\section{Phase Measures and Their Finiteness}\lb{finite}
%%%%%%%%%%%%%%%%%%%%%%%%%%%%%%%%%%%%%%%%%%%%%%
%%%%%%%%%%%%%%%%%%%%%%%%%%%%%%%%%%%%%%%%%%%%%%
%%%%%%%%%%%%%%%%%%%%%%%%%%%%%%%%%%%%%%%%%%%%%%

While the previous section considered a rather disparate set of examples, the present section develops some 
quite general results about the effect of critical points on the local finiteness of phase measures.  
For the case of non-degenerate critical points we can carry over essentially unchanged the considerations of 
Van Hove in his classic study \cite{VanHove53}. Thereafter we discuss briefly the case of degenerate critical points. 

We begin with a mathematical issue that we have neglected until now: the resonance function $E(\ul{\bp};\bk)$ is generally 
not smooth enough in the phase-space variable $\ul{\bp}=(\bp_1,...,\bp_{N-2})$ in order to make the delta-function 
$\delta(E(\ul{\bp};\bk))$ meaningful in any naive sense. Thus, the phase measure that we have defined formally 
by $d\mu=d^D\ul{p} \ \delta(E(\ul{\bp};\bk))$ has no actual mathematical meaning. To give it a proper definition,
one must return to the derivation of the wave kinetic equation. A standard multiscale  perturbation argument 
\cite{BenneyNewell69,Nazarenko11} shows that what appears in the kinetic equation is actually an approximate 
delta function of the form 
\be \delta_T(E(\ul{\bp};\bk))  =\frac{T}{2\pi}\mbox{sinc}^2\left(\frac{T}{2} E(\ul{\bp};\bk)  \right), \lb{sinc-delta} \ee
for a time $T$ chosen so that $\omega(\bk)\gg 1/T\gg \Gamma_\bk(n),$ where ``sinc'' denotes 
the cardinal sine function ${\rm sinc}(x)=\sin x/x.$ A physically motivated definition of the phase measure 
is thus as a suitable limit 
\be d\mu= \lim_{T\rightarrow\infty} d^D\ul{p} \ \delta_T(E(\ul{\bp};\bk)). \ee 
We can show using a Daniell integral method that this limit yields a well-defined measure which is 
unique among those absolutely continuous with respect to $D$-dimensional Hausdorff measure ${\mathcal H}^{D-1}$
on the resonant manifold ${\mathcal R}_\bk$ and that this measure satisfies 
\be d\mu= \frac{d{\mathcal H}^{D-1}(\ul{\bp})}{|\grad E(\ul{\bp};\bk)|}. \lb{basic} \ee 
Because the details are rather technical, we provide them in another paper \cite{ShiEyink15a}. 
An alternate approach is suggested by field-theoretic derivations of wave kinetics \cite{Lvovetal97}, which 
instead yield an approximate delta function of Lorentzian (Cauchy) form:
\be \delta_\gamma (E(\ul{\bp};\bk))  =\frac{\gamma}{\pi} \frac{1}{E^2(\ul{\bp};\bk)+\gamma^2}, \lb{Lorentzian} \ee 
and which suggests to take a similar limit $\gamma\rightarrow 0.$ This was the starting point of Lukkarinen \& Spohn 
\cite{LukkarinenSpohn07} to define the phase measure, in a slightly different context.  We 
also consider this alternate approach in \cite{ShiEyink15a} and compare with the method 
of \cite{LukkarinenSpohn07}. 

Now using the relation (\ref{basic}) we study the local finiteness of the phase measure $\mu$ 
in the neighborhood of a non-degenerate critical point, following the basic idea of Van Hove \cite{VanHove53},
who analyzed this question for the energy density of states using the Morse Lemma. Since $\bk$ appears 
simply as a parameter in the argument, we omit it and write simply $E(\ul{\bp}).$ We recall here 
the statement of the Morse Lemma \cite{Milnor63,Lang12}: if $E(\ul{\bp})$ has a non-degenerate critical point $\ul{\bp}_*$ such that 
$E$ is $C^{k+2}$, $k\geq 1$  with respect to $\ul{\bp}$ in a neighborhood of $\ul{\bp}_*,$ then there is 
a $C^k$ diffeomorphism $\barphi$ of a neighborhood $U$ of $\ul{\bp}_*$ with a neighborhood $V$ 
of $\ul{\b0}=\barphi(\ul{\bp}_*)$ such that $\tilde{E}(\ul{\bq})=E(\barphi^{-1}(\ul{\bq}))$ has the canonical form 
$$ \tilde{E}(\ul{\bq}) = E(\ul{\bp}_*) - |\ul{\bq}^-|^2 + |\ul{\bq}^+|^2 $$
where $(\ul{\bq}^-,\ul{\bq}^+)=\ul{\bq}=\barphi(\ul{\bp})$ with $\ul{\bq}^\pm\in {\mathbb R}^{D_\pm}$ and $D_-+D_+=D.$
It is possible that $D_-=0,$  in which case $ \tilde{E}(\ul{\bq}) = E(\ul{\bp}_*) + |\ul{\bq}|^2$
or that $D_+=0,$  in which case $ \tilde{E}(\ul{\bq}) = E(\ul{\bp}_*) -|\ul{\bq}|^2.$ Note that $D_\pm$ 
are just the number of positive/negative eigenvalues of the Hessian of $E$ at $\ul{\bp}_*.$ Van Hove 
assumed further that $\barphi$ can be chosen to preserve $D$-dimensonal volume (Lebesgue measure), 
in which case
$$ I=\int_U d^D\ul{p}\ \delta(E(\ul{\bp})) =\int_V d^D\ul{q}\ \delta(\tilde{E}(\ul{\bq})), $$
and the righthand side can then be shown to be finite/infinite by a direct calculation. It has indeed 
been proved subsequently that such a volume-preserving (``isochoric") choice of $\barphi$ is possible, if $E$ is a 
$C^\infty$ function \cite{ColinVey79}. However, even if $E$ is only $C^{k+2},$ the neighborhoods $U,V$ can 
always be chosen so that the Jacobian determinant of $\barphi$ satisfies
$$ 0<c\leq |{\rm det}(D\barphi)(\ul{\bp}))|\leq C<\infty, \,\,\,\,\ul{\bp}\in U $$
and since 
$$ \int_V d^D\ul{q}\ \delta(\tilde{E}(\ul{\bq})) = \int_U d^D\ul{p}\  |{\rm det}(D\barphi)(\ul{\bp}))|\ \delta(E(\ul{\bp})) $$
the two integrals $J=\int_V d^D\ul{q}\ \delta(\tilde{E}(\ul{\bq}))$, $I=\int_U d^D\ul{p}\ \delta(E(\ul{\bp}))$ are either both 
finite or both infinite. Thus, the question again reduces to an elementary calculation of the integral $J$. 

Taking therefore a critical point $\ul{\bp}_*$ on the resonant manifold, satisfying $E(\ul{\bp}_*)=0,$ the 
condition for the resonant manifold ${\mathcal R}$ in the $\ul{\bq}$-coordinates in $V$ becomes simply
$$ |\ul{\bq}^-|=|\ul{\bq}^+|.$$
When either $D_-=0$ or $D_+=0,$ the resonant manifold reduces to the isolated point $\bq=\b0$, and one can 
easily show that $J=0.$ Thus we assume $D_\pm\neq 0.$ We use   
$$ J=\int_V d^D\ul{q}\ \delta(\tilde{E}(\ul{\bq})) = \int_{V\cap {\mathcal R}} \frac{d{\mathcal H}^{D-1}(\ul{\bq})}{|\grad\tilde{E}(\ul{\bq})|} $$
and $\grad\tilde{E}(\ul{\bq})=2(-\ul{\bq}^-,\ul{\bq}^+),$ so that $|\grad\tilde{E}(\ul{\bq})|=2\sqrt{2}|\ul{\bq}^-|$ on ${\mathcal R}.$ 
To simplify the calculation, without loss of generality, we take the neighborhood $V$ to be a Cartesian product of two balls
of radius $\eta$, $V=B(\ul{\b0}^-,\eta)\times B(\ul{\b0}^+,\eta).$ Using hyperspherical coordinates for both $\ul{\bq}^-$ and
$\ul{\bq}^+$, the calculation reduces to 
\footnote{
The Fubini-like theorem that we use here follows from the co-area
formula of geometric measure theory. E.g. see Theorem 3.2.22, \cite{Federer69}: For any $W\subset {\mathbb R}^M$ 
which is $M$-rectifiable and  ${\mathcal H}^{M}$-measurable,
$Z\subset {\mathbb R}^{N},$ $N<M$ which is $N$-rectifiable and  ${\mathcal H}^{N}$-measurable, a Lipschitz map $f:W\rightarrow Z,$
and any non-negative, ${\mathcal H}^{M}$-measurable function $g:W\rightarrow {\mathbb R},$ 
$\int_W g(w) (Jf)(w) d{\mathcal H}^{M}(w)=\int_Z d{\mathcal H}^{N}(z)\int_{f^{-1}(z)} g(w') d{\mathcal H}^{M-N}(w'). $
 We apply that theorem with $M=D-1,$ $N=D_-,$ 
$W=V\cap {\mathcal R},$ 
$V=Z\times Y$, 
$Z=B(\uline{\b0}^-, \eta)$, 
$Y=B(\uline{\b0}^+,\eta),$ $f:W\rightarrow Z$ is the restriction 
to $W$ of $\pi:V\rightarrow Z,$ the projection onto the first factor of $V,$ and $g=1/|\grad \tilde{E}|$. 
Note that for each $\uline{\bq}^-\in Z,$ $f^{-1}(\uline{\bq}^-)=S(\uline{\b0}^+,q_-),$ the $\uline{\bq}^+$-sphere of radius $q_-$
centered at $\uline{\b0}^+,$ and that the Jacobian $Jf\equiv 1.$ Finally, $W$ is obviously $(D-1)$-rectifiable and $Z$ is $D_{-}$-rectifiable.
Thus, $\int_{V\cap {\mathcal R}} d{\mathcal H}^{D-1}(\uline{\bq})/|{\grad} \tilde{E}(\uline{\bq})| =\int_{{B({\uline{\b0}}^-,\eta)}}
d{\mathcal H}^{D_{-}}(\uline{\bq}^-) \int_{{S({\uline{\b0}}^+,q_-)}}  d{\mathcal H}^{D_+-1}(\uline{\bq}^+)/|{\grad} \tilde{E}(\uline{\bq})| $.
}
$$ J=\int_{B(\ul{\b0}^-,\eta)} d{\mathcal H}^{D_-}(\ul{\bq}^-) \ \frac{1}{2\sqrt{2}q_-}\int_{q_+=q_-} d{\mathcal H}^{D_+-1}(\ul{\bq}^+)
= (const.) \int_0^\eta q_-^{D-3}dq_-, $$
where a factor $\propto q_-^{D_+-1}$ arises from the $(D_+-1)$-dimensional Hausdorff measure of the $\ul{\bq}^+$-sphere
of radius $q_-.$ Clearly, $J<\infty$ if $D>2$ and $J=\infty$ if $D\leq 2,$ exactly as had been concluded by 
Van Hove for the energy density of states. 

In particular, we find a logarithmic divergence for $D=2$, when one can write simply $\ul{\bq}=(q_-,q_+).$
Introducing the new coordinates 
$$ r= q_+-q_-, \ \ s=q_+-q_- $$
with Jacobian of transformation $\left|\frac{\partial(r,s)}{\partial(q_+,q_-)}\right|=2$
one can write 
$$ \tilde{E}=q_+^2-q_-^2=rs$$
and thus
$$J= \int_{|\ul{\bq}|<\eta} dq_+dq_- \ \delta(q_+^2-q_-^2)=\frac{1}{2}\int_{r^2+s^2<2\eta^2}dr\ ds \ \delta(rs). $$
We see from this formula that our previous considerations on the effect of resonance broadening for graphene
carry over to the general case, with the local divergence removed and replaced by a logarithmically large value 
$\propto \ln(ab/\gamma)$ for resonance width $\gamma.$

We have thus obtained quite general results on the local finiteness of the phase measure 
in the vicinity of a non-degenerate critical point. These general considerations explain the 
specific results we found in earlier examples, such as the logarithmically divergent phase measure 
for 3-wave resonance of Rossby/drift waves in $d=2$ ($D=2$) and the finite phase measure for 4-wave 
resonance of capillary-gravity waves in $d=2$ ($D=4$). 

Let us now consider briefly the effect of degeneracy. The simplest situation is when the critical 
points lie on a $\delta$-dimensional submanifold where $\grad E(\ul{\bp})=\b0,$ which immediately
implies a degeneracy degree of at least $\delta$ at each such critical point.  This corresponds to the situation where
in a neighborhood $U$ of each critical point $\ul{\bp}_*,$ with $E(\ul{\bp}_*)=0$, there is a 
diffeomorphism $\barphi$ with a neighborhood $V$ of $\ul{\b0}=\barphi(\ul{\bp}_*)$ such that 
$$ \tilde{E}(\ul{\bq}) = - |\ul{\bq}^-|^2 + |\ul{\bq}^+|^2 $$
where $(\ul{\bq}^0,\ul{\bq}^-,\ul{\bq}^+)=\ul{\bq}=\barphi(\ul{\bp})$ with $\ul{\bq}^\pm\in {\mathbb R}^{D_\pm}$ 
and $\ul{\bq}^0\in {\mathbb R}^{\delta}.$ In this case, $D_++D_-=D-\delta\equiv D'.$ 
One can take the neighborhood $V$, without loss of generality, to be of the form $V=V_0\times V',$ where 
$V_0$ is a neighborhood of $\ul{\b0}^0\in {\mathbb R}^\delta$ and $V'$ is a neighborhood of 
$(\ul{\b0}^-,\ul{\b0}^+)\in {\mathbb R}^{D'}.$ In this case, it is seen that 
$$ J={\mathcal L}^\delta(V_0)J', \ \ \ \ J'= \int_{V'} d^{D'}\ul{q}' \ \delta(\tilde{E}(\ul{\bq}')) $$
with $\ul{\bq}'=(\ul{\bq}^-,\ul{\bq}^+)$ and ${\mathcal L}^\delta$ the $\delta$-dimensional Lebesgue measure (volume).
Here $J'$ has the same form as did $J$ for the non-degenerate case, but with $D'$ replacing $D.$ Thus, the measure 
is locally finite at each critical point for $D'>2$ but locally infinite for $D'\leq 2.$ 

This analysis explains the results we obtained in several concrete examples, such as the logarithmically
divergent phase measures for 3-wave resonance of inertial waves in $d=3$ ($D=3,\delta=1$) and for 
4-wave resonance of electron-hole excitations of graphene in $d=2$ $(D=4,\delta=2)$, both with 
$D'=2.$ 

Another case of interest is an isolated critical point with degeneracy degree $\delta$ 
(co-rank of the Hessian matrix). The classification of isolated critical points for differentiable 
functions $E$ belongs to the field of singularity theory; see \cite{Arnoldetal12}. We shall 
not discuss such a classification in detail here, but we briefly mention in this light the doubly 
degenerate critical point obtained for  wave propagation along an optical fiber, from 
section \ref{optics}. For the distinguished value $k=k_*$ (known as the ``zero-dispersion frequency"
in the nonlinear optics community) one can express $E(p,q;k_*)$ locally at the degenerate
critical point $(k_*,k_*)$ in terms of the deviation variables $\delta p=p-k_*,$ $\delta q=q-k_*$ as 
$$ E(\delta p,\delta q;k_*) = 3\alpha (\delta p +\delta q)\ \delta p \ \delta q. $$
After a linear transformation $\delta p=(x-y)/(6\alpha)^{1/3},$ $\delta q=-(x+y)/(6\alpha)^{1/3}$
this becomes 
$$ E(x,y;k_*) = x^2y-y^3. $$
The latter is the normal form for the family $D_4^-$ in the classification of ``simple" singularities for differentiable 
real functions in \cite{Arnoldetal12}. The general possibilities are quite rich and complex, and still 
the subject of mathematical investigation. However, we note from the optics example that the effect 
of degeneracy is again to worsen the divergence of the phase measure, whose density now exhibits a 
double pole rather than the simple poles (leading to logarithmic divergences) found for the 
non-degenerate critical points when $k\neq k_*.$ 

The cases of degenerate points that we have discussed here are by no means exhaustive. For example,
one could have a set of critical points with degeneracy degree $\delta$ comprising a $D_0$-dimensional 
submanifold with $1<D_0<\delta.$ There is also the possibility of pseudo-critical points, but, as discussed 
earlier, these will produce no divergence of phase measure unless the geometric singularity is so severe 
that the resonant manifold develops a locally infinite Hausdorff measure. 

The general morals to be drawn from our discussion are as follows. Pseudo-critical points should usually yield  
a locally finite phase measure and are ``harmless'' for kinetic theory. True critical points are potentially ``dangerous"
and can lead to locally infinite phase measure, especially in situations of low dimensions $d,$ low-order $N$ 
of resonance, and/or high degeneracy degree $\delta.$ 
The divergence of phase measure due to such singularities can be rendered harmless by vanishing 
interaction coefficients or by cancellations in the collision integral. In the next section we explore the opposite 
situation when the standard collision integral with exact resonances remains divergent.  

%%%%%%%%%%%%%%%%%%%%%%%%%%%%%%%%%%%%%%%%%%%%%%
%%%%%%%%%%%%%%%%%%%%%%%%%%%%%%%%%%%%%%%%%%%%%%
%%%%%%%%%%%%%%%%%%%%%%%%%%%%%%%%%%%%%%%%%%%%%%
\section{Singular Wave Kinetics}\lb{singular}
%%%%%%%%%%%%%%%%%%%%%%%%%%%%%%%%%%%%%%%%%%%%%%
%%%%%%%%%%%%%%%%%%%%%%%%%%%%%%%%%%%%%%%%%%%%%%
%%%%%%%%%%%%%%%%%%%%%%%%%%%%%%%%%%%%%%%%%%%%%%

We have seen several examples (inertial waves in $d=3$ fluids, optical wave propagation along a 
$d=1$ fiber, and Dirac electron-hole excitations in $d=2$ graphene) where an ``unprotected'' 
resonance Van Hove singularity leads to a breakdown of standard wave kinetics. This is analogous 
to the situation in the theory of low-amplitude acoustic wave turbulence, except that now the breakdown 
of dispersivity of the waves occurs only locally on the resonant manifold rather than for all resonances. 
In the case of acoustic waves, generalized equations were derived to describe the ``singular wave kinetics'' 
of semi-dispersive waves, both by multiple time-scale perturbation theory \cite{NewellAucoin71} and by 
field-theoretic methods \cite{Lvovetal97}. One can expect that such singular kinetic theories will apply 
more generally, even when the critical set is only a subset of the entire resonant manifold. We shall briefly
illustrate this situation with the example of electron-hole kinetics in graphene.   

We first discuss graphene from the point of view of multiple time-scale perturbation theory, as 
originally developed in \cite{NewellAucoin71}. It should be stressed at the outset that the applicability of 
perturbation theory is itself a nontrivial result, because electron-hole excitations in graphene 
on most substrates are not weakly coupled (nor infinitely-strongly coupled). It is instead believed that 
the coupling becomes weak for sufficiently low wavenumbers because of a globally attractive, asymptotically-free 
renormalization-group (RG) fixed point. See \cite{Gonzalezetal99,Son07,Hofmannetal14} for discussions 
of this issue. Thus, the perturbation theory argument must be applied to a low-wavenumber renormalized 
theory. A further very significant complication is that the Coloumb interaction is long-range and many-electron effects 
such as dynamical screening are expected in graphene \cite{Kotovetal12}, and these effects do not appear at any 
finite order in naive weak-coupling perturbation theory. Although such a naive perturbation theory analysis is therefore 
very incomplete, we find that it provides useful insight. 

The perturbative derivation of the quantum Boltzmann equation is much the same as the derivation of the 
kinetic equation for weakly-coupled classical waves, e.g. see \cite{Spohn06,Zakharovetal92}. Start with 
a general quantum Hamiltonian $H=H_0+H_1$ with free part
\be  H_0 = \sum_{s,a} \int\dbar^dk\ \omega_s(k) \gamma^\dagger_{s a}(\bk) \gamma_{s a}(\bk) \lb{H0} \ee
and interaction part
%\newpage
\begin{eqnarray}  
H_1 &=& \frac{\alpha}{2} \sum_{s_1,s_2,s_3,s_4}\sum_{a,b}
\int \dbar^dk_1 \dbar^dk_2 \dbar^dk_3 \dbar^dk_4 \ddelta^d(\bk_1+ \bk_2-\bk_3-\bk_4)
\cr
&&\,\,\,\,\,\,\,\,\,\,\,\,\,\,\,\,\,\,\,\,\,\,\,\,\,\,\,\,\,\,\,\, \times\ T^{s_1s_2s_3s_4}_{\bk_1\bk_2\bk_3\bk_4} \gamma^\dagger_{s_4 b}(\bk_4)\gamma^\dagger_{s_3 a}(\bk_3)
\gamma_{s_2 a}(\bk_2)\gamma_{s_1 b}(\bk_1), 
\lb{H1} \end{eqnarray}
where $\gamma^\dagger_{s a}(\bk), \gamma_{s a}(\bk)$ are standard creation/annihilation operators.  We assume here that 
these operators obey canonical anti-commutation relations, as appropriate for fermionic electron \& hole excitations.   
The indices $a,b$ are summed over integers $1,...,N,$ to incorporate a possible $N$-fold degeneracy of the fermions.  
For an infinite-volume system, define the mean occupation number $n_{s}(\bk,t)$ by $\langle \gamma^\dagger_{s a}(\bk,t)\gamma_{s' a'}(\bk',t)\rangle
=n_{s}(\bk,t)\delta_{ss'}\delta_{aa'}\ddelta^d(\bk-\bk').$ Perturbation theory in the small parameter $\alpha$ yields the result that 
\begin{eqnarray}
&& n_{s_1}(\bk_1, t) - n_{s_1}(\bk_1, 0) \cr 
&& = - \alpha^2 \sum_{s_2,s_3,s_4}\int \dbar^dk_2\dbar^dk_3 \dbar^dk_4\  
\ddelta^d(\bk_1+\bk_2-\bk_3-\bk_4)
\frac{2\bigg(1-\cos(E^{s_1s_2s_3s_4}_{\bk_1\bk_2\bk_3\bk_4} t)\bigg)}{(E^{s_1s_2s_3s_4}_{\bk_1\bk_2\bk_3\bk_4})^2} 
\cr
&&\,\,\,\,\,\,\,\,\,\,\,\,\,\,\,\,  \times 
R^{s_1s_2s_3s_4}_{\bk_1\bk_2\bk_3\bk_4} \left[ n_{s_1}(\bk_1)  n_{s_2}(\bk_2)\wt{n}_{s_3}(\bk_3)\wt{n}_{s_4}(\bk_4) 
- \wt{n}_{s_1}(\bk_1)\wt{n}_{s_2}(\bk_2) n_{s_3}(\bk_3) n_{s_4}(\bk_4)\right]
\lb{q-pert} \end{eqnarray}
where
\begin{eqnarray}
E^{s_1s_2s_3s_4}_{\bk_1\bk_2\bk_3\bk_4} &=& \omega_{s_1}(\bk_1)+\omega_{s_2}(\bk_2)-\omega_{s_3}(\bk_3)-\omega_{s_4}(\bk_4),
\\
R^{s_1s_2s_3s_4}_{\bk_1\bk_2\bk_3\bk_4} &=& \frac{1}{2}
|T^{s_1s_2s_3s_4}_{\bk_1\bk_2\bk_3\bk_4} - T^{s_1s_2s_4s_3}_{\bk_1\bk_2\bk_4\bk_3} |^2 
+ \frac{N-1}{2}|T^{s_1s_2s_3s_4}_{\bk_1\bk_2\bk_3\bk_4} |^2
+ \frac{N-1}{2}|T^{s_1s_2s_4s_3}_{\bk_1\bk_2\bk_4\bk_3} |^2. 
\end{eqnarray}
and 
\be
\wt{n}_s(\bk,t) = 1-n_s(\bk,t).
\ee
See \ref{perturbation} for the details of the derivation. An elementary calculation gives 
$$ \frac{1-\cos(Et)}{E^2}= \pi t\ \delta_t(E) $$
where $\delta_t(E)$ is the approximate delta function in eq.(\ref{sinc-delta}). When the condition $E^{s_1s_2s_3s_4}_{\bk_1\bk_2\bk_3\bk_4}=0$ 
defines a non-degenerate resonance, the righthand side of (\ref{q-pert}) exhibits a secular growth $\propto \alpha^2 t.$ This secular behavior 
is removed by choosing the occupation number to satisfy the quantum kinetic equation on the slow time scale $\tau=\alpha^2 t,$
with collision integral (\ref{coll4}).  

This standard derivation fails for electron-hole kinetics in graphene\footnote{The physical dimensions require a bit 
of discussion. The 1-particle energies in the free part of the Hamiltonian are $E_s(\bk)=\hbar \omega_s(\bk)$,
so our units correspond to $\hbar=1$. The Coulomb potential is $V(\br)=e^2/\kappa r,$ where $\kappa$ is 
the background dielectric constant. Hence, the parameter $\alpha$ in the interaction Hamiltonian for graphene corresponds 
to the dimensionless ``fine-structure constant'' $\alpha=e^2/\kappa \hbar v_F$, in units where $v_F=1.$ 
We have kept the factor $v_F$ in the dispersion relation, although it is strictly just 1 in our units.}, when $d=2$ and  
$$ \omega_s(k)= s v_F k, \,\,\,\, s=\pm 1.$$
The index $N=4,$ for the two electron spins and two Dirac points (valleys) in the Brillouin zone.
For the expression for $T^{s_1s_2s_3s_4}_{\bk_1\bk_2\bk_3\bk_4}$ arising from Coloumb interaction, see eq.(3.9) 
of \cite{Fritzetal08}. The degeneracy of the condition $E^{s_1s_2s_3s_4}_{\bk_1\bk_2\bk_3\bk_4}=0$ changes 
the asympotics of the integral in  eq.(\ref{q-pert}). Note that momentum and energy conservation allow non-trivial
resonances only for electron-electron/hole-hole collisions (all $s$'s of the same sign) or electron-hole collisions 
(one $s=+1$ and one $s=-1$ in both incoming and outgoing states). For simplicity, we discuss explicitly here 
only the first case. To obtain the long-time asymptotics, it is useful to take two derivatives with respect to time, 
to obtain the contribution with all $s_i=s$:
\be \frac{d^2}{dt^2}[n_{s}(\bk, t) - n_{s}(\bk, 0)] = - 2\alpha^2 \int \dbar^2k_3\dbar^2k_4\  
\cos(E^{++++}_{\bk \bk_2\bk_3\bk_4} t) M^{++++}(\bk_3,\bk_4), 
\lb{q-pert2} \ee
where $\bk_2=\bk_3+\bk_4-\bk$ and 
\begin{eqnarray}
&& M^{++++}(\bk_3,\bk_4) = R^{++++}_{\bk\bk_2\bk_3\bk_4} \cr
&& \times \left[ n_{s}(\bk)  n_{s}(\bk_2)\wt{n}_{s}(\bk_3)\wt{n}_{s}(\bk_4) 
- \wt{n}_{s}(\bk)\wt{n}_{s}(\bk_2) n_{s}(\bk_3) n_{s}(\bk_4)\right], 
\end{eqnarray} 
\be
R^{++++}_{\bk\bk_2\bk_3\bk_4} =\frac{1}{2}
|T^{++++}_{\bk\bk_2\bk_3\bk_4} - T^{++++}_{\bk\bk_2\bk_4\bk_3} |^2 
+ (N-1)|T^{++++}_{\bk\bk_2\bk_3\bk_4} |^2. 
\ee
Using polar coordinates $\bk=k(\cos\theta,\sin\theta)$ with fixed direction angle $\theta$ of wavevector $\bk$ and $\bk_i=k_i(\cos(\theta+\theta_i),\sin(\theta+\theta_i)),$
$i=3,4,$ this can be rewritten as 
\begin{eqnarray} 
&& \frac{d^2}{dt^2}[n_{s}(\bk, t) - n_{s}(\bk, 0)] \cr
&& = - \frac{\alpha^2}{2\pi^2} {\rm Re} \int_0^\infty k_3\dbar k_3 \int_0^\infty k_4\dbar k_4 \int_{-\pi}^{\pi} d\theta_3\int_{-\pi}^{\pi} d\theta_4\
e^{i tE^{++++}_{\bk, \bk_3+\bk_4-\bk,\bk_3,\bk_4}} M^{++++}(k_3,k_4,\theta_3,\theta_4),
\lb{q-pert3} \end{eqnarray} 
where a simple calculation gives 
\begin{eqnarray}
&&E^{++++}_{\bk, \bk_3+\bk_4-\bk,\bk_3,\bk_4} = k_3+k_4-k\cr
&& \hspace{20pt} -\sqrt{(k_3+k_4-k)^2-2k_3k_4(1-\cos(\theta_3+\theta_4))+
2k_3k(1-\cos\theta_3)+2k_4k(1-\cos\theta_4)} \cr
&&\hspace{80pt} \simeq \btheta^\top H\btheta
\end{eqnarray} 
to quadratic order in $\btheta=(\theta_3,\theta_4),$ with the $2\times 2$ matrix    
\be
H =  \frac{1}{2(k_3+k_4-k)}
\begin{pmatrix}
 k_3(k_4-k) &  k_3k_4 \\
k_3k_4        & k_4(k_3-k)
\end{pmatrix}. 
\ee
Using $\det H = \frac{kk_3k_4}{4(k-k_3-k_4)}$, the integral over angles in eq.(\ref{q-pert3}) is then evaluated asymptotically for $t\rightarrow\infty$ by the 
method of stationary phase \cite{Fedoryuk71}, giving  
\begin{eqnarray}
&& \frac{d^2}{dt^2}[n_{s}(\bk, t) - n_{s}(\bk, 0)] \cr
&& \hspace{10pt} =
-\frac{\alpha^2}{2\pi^2}{\rm Re}\bigg[ \frac{2\pi}{t} \iint_{k_3>0,k_4>0,k_3+k_4>k} k_3k_4 \ \dbar k_3\dbar k_4\ |\det H|^{-1/2}\  M^{++++}(\theta_3=\theta_4=0)
\cr
&& \hspace{60pt} + \frac{2\pi}{t} e^{\frac{\pi}{2}i} \iint_{k_3>0,k_4>0,k_3+k_4<k}k_3k_4\ \dbar k_3\dbar k_4\ |\det H|^{-1/2}\  M^{++++}(\theta_3=\theta_4=0) \bigg] + O(t^{-2})
\cr
&& \hspace{10pt} 
= -\frac{2\alpha^2}{\pi} \frac{1}{t} \iint_{k_3>0,k_4>0,k_3+k_4>k} \dbar k_3\dbar k_4\ \sqrt{\frac{(k_3+k_4-k)k_3k_4}{k}} M^{++++}(\theta_3=\theta_4=0)
+ O(t^{-2}). 
\end{eqnarray}
This integrates to 
\begin{eqnarray}
&& n_{s}(\bk, t) - n_{s}(\bk, 0) \cr
&& \hspace{10pt} 
\sim -\frac{2\alpha^2}{\pi} t\ln t \iint_{k_3>0,k_4>0,k_3+k_4>k} \dbar k_3 \dbar k_4\ \sqrt{\frac{(k_3+k_4-k)k_3k_4}{k}} M^{++++}(\theta_3=\theta_4=0)
\end{eqnarray}
as $t\rightarrow\infty.$ The crucial point is that the leading secular growth is now faster than $t$ by a logarithmic factor
$\ln t$, as a consequence of the degeneracy of resonance. Note that the entire contribution arises from the critical subset 
of the resonant manifold, with all quartet wavevectors parallel to $\bk.$

Taking into account the electron-hole scattering changes this result only by appearance of an additional term, which is 
obtained by a very similar calculation. The complete asymptotics is given by  
\begin{eqnarray}
&& n_{s}(\bk, t) - n_{s}(\bk, 0) \cr
&& \hspace{10pt} 
\sim -\frac{2\alpha^2}{\pi} t\ln t \iint_{k_3>0,k_4>0,k_3+k_4>k} \dbar k_3\dbar k_4\ \sqrt{\frac{(k_3+k_4-k)k_3k_4}{k}} M(\theta_3=\theta_4=0)
\lb{q-sec} \end{eqnarray}
with 
$$ M(\bk_3,\bk_4) = M^{++++}(\bk_3,\bk_4) + M^{+-+-}(\bk_3,\bk_4) $$
and the electron-hole scattering contribution 
\begin{eqnarray}
&& M^{+-+-}(\bk_3,\bk_4) = R^{+-+-}_{\bk,-\bk_4,\bk_3,-\bk_2} \cr
&& \times \left[ n_{s}(\bk)  n_{-s}(-\bk_4)\wt{n}_{s}(\bk_3)\wt{n}_{-s}(-\bk_2) 
- \wt{n}_{s}(\bk)\wt{n}_{-s}(-\bk_4) n_{s}(\bk_3) n_{-s}(-\bk_2)\right], 
\end{eqnarray} 
\be
R^{+-+-}_{\bk\bk_2\bk_3\bk_4} =
|T^{+-+-}_{\bk\bk_2\bk_3\bk_4} - T^{+--+}_{\bk\bk_2\bk_4\bk_3} |^2 
+ (N-1)|T^{+-+-}_{\bk\bk_2\bk_3\bk_4} |^2+ (N-1)|T^{+--+}_{\bk\bk_2\bk_4\bk_3} |^2. 
\ee
We have made a change of variables $\bk_2\rightarrow -\bk_4$, $\bk_4\rightarrow -\bk_2$ in the integral for 
the electron-hole contribution so that the range of integration is the same as for the electron-electron 
contribution. The physics of the electron-hole scattering term is easy to understand, if one recalls that a hole excitation 
with wavenumber $\bk$ has a group velocity $\grad\omega_-(\bk)=-v_F\hat{\bk}$ which is opposite to the 
group velocity $\grad\omega_+(\bk)=+v_F\hat{\bk}$ for an electron excitation with the same wavenumber $\bk.$
Hence, non-dispersive interactions with identical group velocities for a quartet of modes requires 
that the holes have wavenumbers anti-parallel to the wavenumbers for the electrons. The fact that 
electron-hole scattering couples occupation numbers for anti-parallel wavenumbers will be seen below to 
have interesting consequences.  

The leading secular growth in (\ref{q-sec}) can be removed, following the ideas in \cite{NewellAucoin71}, by 
allowing the occupation numbers to evolve on the time-scale $\tau_1=\alpha^2 t\ln t$ according to the 
{\it singular kinetic equation}:
\begin{eqnarray}
&& dn_s(\bk)/d\tau_1 = -\frac{2}{\pi} \iint_{k_3>0,k_4>0,k_3+k_4>k} \dbar k_3\dbar k_4\ \sqrt{\frac{(k_3+k_4-k)k_3k_4}{k}} \cr
&& \times \bigg\{ R^{++++}_{\bk\bk_2\bk_3\bk_4}  \left[ n_{s}(\bk)  n_{s}(\bk_2)\wt{n}_{s}(\bk_3)\wt{n}_{s}(\bk_4) 
- \wt{n}_{s}(\bk)\wt{n}_{s}(\bk_2) n_{s}(\bk_3) n_{s}(\bk_4)\right] \cr
&& + R^{+-+-}_{\bk,-\bk_4,\bk_3,-\bk_2} \left[ n_{s}(\bk)  n_{-s}(-\bk_4)\wt{n}_{s}(\bk_3)\wt{n}_{-s}(-\bk_2) 
- \wt{n}_{s}(\bk)\wt{n}_{-s}(-\bk_4) n_{s}(\bk_3) n_{-s}(-\bk_2)\right]\bigg\}, 
\lb{sing-eq} \end{eqnarray} 
where in the collision integral all wavenumbers $\bk_i = k_i\hat{\bk}$ are parallel to $\bk$ and $k_2=k_3+k_4-k.$
Similarly as in the work of Newell \& Aucoin \cite{NewellAucoin71} on acoustic turbulence, this new kinetic equation   
is actually a continuum of uncoupled equations, one for each line along $\pm\hat{\bk}.$ Whereas 
the usual quantum kinetic equation holds on a time-scale $\tau_2=\alpha^2 t,$ the singular kinetic 
equation is valid at logarithmically shorter times. Put another way, for $\tau_1=O(1),$ $\tau_2=O(1)$ one 
finds $t\approx \tau_1/\alpha^2\ln(1/\alpha)\ll \tau_2/\alpha^2,$ when $\ln(1/\alpha)\gg 1.$ It is important 
to note that the singular kinetic equation (\ref{sing-eq}) coincides (up to a constant of proportionality) 
with the result previously derived for electron-hole kinetics in graphene by \cite{Kashuba08,Fritzetal08}, who made  
a leading-logarithm approximation to the divergent collision integral in the standard quantum Boltzmann equation. 
However, such a ``derivation'' of (\ref{sing-eq}) is inconsistent, taken literally, because it employs the quantum Boltzmann 
equation in a regime outside its validity. We discuss further below the derivation of \cite{Kashuba08,Fritzetal08}, 
which must be consistently understood within a proper field-theoretic framework. Both our derivation and that 
of  \cite{Kashuba08,Fritzetal08} have also made an {\it ad hoc} assumption that the Coulomb interaction is 
dynamically screened at very low waveumbers, in order to eliminate an infrared divergence of the collision 
integral \footnote{This divergence due to unscreened Coulomb interaction is simply exhibited in the coordinates 
used in Fig.~10, with $\bk_1=\bk,$ $\bk_2=\bp,$ $\bk_3=\bk+\bq,$ $\bk_4=\bp-\bq,$ for which the singular 
collision integral is with respect to the measure $\sqrt{\frac{p(k+q)(p-q)}{k}}dpdq$ over the range $p>0$ and $p>q>-k.$
Since the $d=2$ Fourier transform of the $d=3$ Coulomb potential is $\propto 1/q,$ the interaction coefficients
$R^{++++},R^{+-+-}\propto 1/q^2,$ leading to an integral over $dq/q^2$ divergent at $q=0$}, but a proper derivation 
of this effect requires a more sophisticated many-body theory.    
  
At a time $t=\tau_1/\alpha^2\ln(1/\alpha)$ with $\tau_1\gg 1$  (but with $t\ll \tau_2/\alpha^2$), the solutions of the singular 
kinetic equation should be expected to approach a {\it local equilibrium} separately along each line in directions $\pm \hat{\bk}.$
This is what occurs in the singular kinetics for acoustic turbulence \cite{NewellAucoin71} and was also 
argued to occur for electron-hole kinetics in graphene in \cite{Kashuba08,Fritzetal08}. The local equilibria 
of (\ref{sing-eq}) can be easily checked to be of a generalized Fermi-Dirac form   
\be n_s(\bk) = \frac{1}{\exp[ (\beta_s(\hat{\bk})sk+\mu_s(\hat{\bk}))]+1}, 
%n_s(\bk) = \frac{1}{\exp[(\beta_o(\hat{\bk})+\beta_e(\hat{\bk})s)k+\mu(\hat{\bk})+\chi(\hat{\bk})s]+1},
\,\,\,\, s=\pm 1, \lb{LFD} \ee 
where 
$$ \beta_s(\hat{\bk})=\beta_e(\hat{\bk})+\beta_o(\hat{\bk})s, \,\,\,\, \mu_s(\hat{\bk})=\mu(\hat{\bk})+\chi(\hat{\bk})s$$
are distinct inverse temperatures and chemical potentials, independently specified for each direction $\hat{\bk}$
and for electrons ($s=1$) and holes ($s=-1$), subject to the conditions that $\beta_e(\hat{\bk})$ must be even 
and $\beta_o(\hat{\bk})$ odd:
\be \beta_e(-\hat{\bk})=\beta_e(\hat{\bk}),\,\,\,\, \beta_o(-\hat{\bk})=-\beta_o(\hat{\bk}). \lb{restrict} \ee
This last restriction arises from the vanishing of the electron-hole contribution to the collision integral in (\ref{sing-eq}),
whereas the same local Fermi-Dirac distribution (\ref{LFD}) causes the electron-electron/hole-hole term to vanish
without any restriction on parameters. Note that $\beta_o(\hat{\bk})$ is the temperature asymmetry between 
electrons and holes, and $\chi(\hat{\bk})$ is the chemical potential asymmetry.  If one confines attention to solutions 
satisfying particle-hole symmetry appropriate to zero doping, $\tilde{n}_{-s}(\bk)=n_s(\bk)$, then the possible equilibria 
are reduced to 
\be n_s(\bk) = \frac{1}{\exp[ (\beta_e(\hat{\bk})k+\chi(\hat{\bk}))s]+1},
\,\,\,\, s=\pm 1. \lb{LFD-sym} \ee 
Nonzero values of symmetric chemical potential $\mu(\hat{\bk})$ or of temperature asymmetry $\beta_o(\hat{\bk})$ 
explicitly break particle-hole symmetry.
The results (\ref{LFD}),(\ref{restrict}),(\ref{LFD-sym}) do not seem to have been given earlier in the literature, although
they are implicit in the papers \cite{Kashuba08,Fritzetal08,Muelleretal09} on dissipative transport by electrons in 
graphene. These local equilibrium solutions 
are not only stationary solutions of (\ref{sing-eq}) but in fact should be global attractors. It is straightforward to show 
by standard arguments that the entropy 
$$ S(\hat{\bk}) =  -k_B \sum_s \int_0^\infty k\dbar k\ [n_s(\bk)\ln n_s(\bk) + \tilde{n}_s(\bk)\ln \tilde{n}_s(\bk) ] $$ 
satisfies an $H$-theorem of the form $d S(\hat{\bk})/d\tau_1\geq 0$ for solutions of the singular kinetic 
equation (\ref{sing-eq}), separately for each direction $\hat{\bk},$ and the only solutions with vanishing 
entropy production, $d S(\hat{\bk})/d\tau_1= 0,$ are the generalized Fermi-Dirac distributions (\ref{LFD}),(\ref{restrict}).   
 
Until now our discussion has been quite parallel to the case of acoustic turbulence \cite{NewellAucoin71}, but  
we now encounter a difference, because the critical set which dominates the singular kinetic equation 
(\ref{sing-eq}) is not the entire resonant manifold for electrons in graphene, unlike the situation for acoustic waves. 
Thus, there are additional secular terms $\propto \alpha^2 t$ in eq.(\ref{q-pert}) which arise from integration 
over the remainder of the resonant manifold. To remove those secularities, one should impose an additional 
dependence upon the time variable $\tau_2,$ equivalent to the condition that $n_s$ satisfies the ordinary quantum
Boltzmann equation on time-scales $\tau_2=O(1),$ or that $dn_s(\bk)/d\tau_2=C_{\bk s}(n)$ with the collision 
integral in (\ref{coll4}). This collision integral diverges for general distributions $n_s(\bk),$ but we can show 
that it is finite for local Fermi-Dirac distributions of the form (\ref{LFD}). (Details will be presented elsewhere \cite{ShiEyink15b}.) 
A simple picture thus arises for electron kinetics in graphene as a {\it three time-scale problem}. At the shortest 
times $t=\tau_0$ of order the linear wave period, the dynamics is dominated by the free part of the Hamiltonian,
whose dispersive character drives the system into a locally quasi-free state completely 
characterized by the occupation numbers $n_s(\bk).$ Cf. the discussion in \cite{Spohn06}, section 9. 
At times $t=\tau_1/\alpha^2\ln(1/\alpha)\gg \tau_0$ the occupation numbers evolve further into the local Fermi-Dirac form 
(\ref{LFD}), completely characterized by the local thermodynamic parameters $\beta_o(\hat{\bk}),\beta_e(\hat{\bk}),\mu(\hat{\bk}),\chi(\hat{\bk})$
along each wavevector direction. Finally, at times $t=\tau_2/\alpha^2\gg \tau_1/\alpha^2\ln(1/\alpha)$ the distribution further relaxes according 
to the standard quantum Boltzmann equation. Because the ratio of times $t$ with $\tau_2=O(1)$ and $\tau_1=O(1)$ is only logarithmically 
large, it is possible that the occupation numbers will not fully relax to a local Fermi-Dirac form (\ref{LFD})  
for times $\tau_2=O(1)$ and there may be corrections of order $1/\ln(1/\alpha).$ If there is no external driving 
or boundary conditions to keep the system in a dissipative non-equilibrium state, the subsequent evolution 
by the standard quantum Boltzmann equation will relax the system at times $\tau_2\gg 1$ to a global Fermi-Dirac
equilibrium 
$$  n_s(\bk) = \frac{1}{\exp( \beta s k+\mu)+1},
\,\,\,\, s=\pm 1, $$    
with uniform values of inverse temperature $\beta$ and chemical potential $\mu.$

So far we have discussed electron kinetics in graphene from the multi-time perturbation theory viewpoint 
developed by Newell-Aucoin \cite{NewellAucoin71} to describe semi-dispersive acoustic turbulence. 
There is however another point of view on singular kinetics for acoustic waves which was developed 
by L'vov et al. \cite{Lvovetal97}, based on a Martin-Siggia-Rose field-theoretic formulation. In this approach,  
the starting point is a set of Schwinger-Dyson integrodifferential equations, which are exact and non-perturbative 
but non-closed. By a set of rational approximations based on weak non-linearity and self-consistency, 
the authors of \cite{Lvovetal97} showed that the Schwinger-Dyson equations for acoustic wave turbulence 
can be simplified to a {\it generalized kinetic equation}. This has a form similar to the standard 3-wave 
kinetic equation 
\begin{eqnarray}
 \partial_t n(\bk,t)
%+\grad_\bk\omega(\bk,\bx)\cdot\grad_\bx W(\bk,\bx,t)-\grad_\bx\omega(\bk,\bx)\cdot\grad_\bk W(\bk,\bx,t)\cr
&=&36\pi \sum_{\ul{s}=(-1,s_2,s_3)} \int d^dk_2 \int d^dk_3\ |H^{\ul{s}}_{\ul{\bk}}|^2
\delta_{\Gamma(\ul{\bk})}(\ul{s}\cdot\omega(\ul{\bk}))\ddelta^d(\ul{s}\cdot\ul{\bk})\cr
&& \,\,\,\,\,\,\,\,\,\,\,\,\,\,\,\,\,\,\,\,\,\,\,\,\,\,\,
\times \Big\{n(\bk_2)n(\bk_3)-s_2 n(\bk)n(\bk_3)-s_3 n(\bk)n(\bk_2)\Big\}.  
\lb{wavkineq}\end{eqnarray}
but with $\delta_{\Gamma(\ul{\bk})}(\omega)$ a resonance-broadened delta-function or Lorentzian 
of the form (\ref{Lorentzian}), where
$$ \Gamma(\ul{\bk}) =\gamma(\bk)+\gamma(\bk_2)+\gamma(\bk_3) $$
is a triad-interaction decay rate. The individual rate $\gamma(\bk)$ is given by the imaginary part of the 
self-energy function $\Sigma(\bk)$, so that it must be determined self-consistently in terms of the solution 
$n(\bk)$ of the generalized kinetic equation and is $O(\epsilon^2)$ in the nonlinear interaction strength 
$\epsilon$. See eq.(B13) of \cite{Lvovetal97}. The collision integral of this generalized kinetic equation 
remains finite and free of any divergences due to the Van Hove-type singularities in the ``manifold" of exact 
resonances for acoustic waves.  

This field-theoretic point of view is very closely related to the previous derivations of the quantum 
Boltzmann equation for electron kinetics in graphene \cite{Kashuba08,Fritzetal08}, which were 
based on a nonequilibrium Schwinger-Keldysh field-theory approach (the quantum analogue 
of the Martin-Siggia-Rose field-theory for classical dynamics). In fact, the authors of \cite{Kashuba08,Fritzetal08}
assumed that self-energy corrections will cut off the divergence of the standard collision integral  due to the 
resonance Van Hove singularity, but without assuming an explicit form for these corrections. 
It is likely that the self-consistent approach of \cite{Lvovetal97} can be carried over to electron 
kinetics in graphene, e.g. with 
% As previously noted, a constant resonance broadening does not suffice to 
% remove the logarithmic divergence in phase measure for graphene, but we have shown that
a wavenumber-dependent broadening corresponding to a quartet-interaction time
$$ \Gamma(\ul{\bk}) =\gamma(\bk)+\gamma(\bk_2)+\gamma(\bk_3) +\gamma(\bk_4). $$
As discussed in section \ref{graph}, such a broadening should cure the logarithmic divergence, if 
the resonance width 
is non-zero in the vicinity of the critical set. There is considerable interest in investigating explicit 
self-energy regularizations, since it has been estimated that corrections to the leading-logarithm
approximation may make a 30\% change to the electrical conductivity of defectless graphene 
at experimentally realizable temperatures \cite{Kashuba08}. In addition to the interest for 
potential electronic applications of graphene, such a study would also help to assess
the validity of theoretical approximations for acoustic turbulence which, to our knowledge, 
have never been subjected to experimental test. More generally, electron kinetics in graphene 
is a problem which should illuminate the subject of singular wave kinetics, with applications 
to a wide variety of systems. We are currently pursuing such investigations \cite{ShiEyink15b}.  \\

%To summarize, the singular kinetic equation is given by
%\be
%\partial_{\tau_1} f_{s_1}^{\hat{\bk}} (k) = C_0[f_{s_1}^{\hat{\bk}}].
%\ee
%The $H$-theorem suggests that the system will thermalize along each ray to the local equilibrium solution. On the other hand, the full kinetic equation is given by
%\be
%\partial_{\tau_2} f_{s_1}(\bk) = C_1[f_{s_1}] + C_2[f_{s_1}].
%\ee
%Before $C_1[f_{s_1}]$ goes to to zero, the strong interaction along each ray may act as forcing on
%the full system. For large $\tau_1$, $C_1[f_{s_1}]$ vanishes (this may happen before the system
%reaches local equilibrium) and the full system is governed by
%\be
%\partial_{\tau_2} f_{s_1}(\bk) = C_2[f_{s_1}].
%\ee

%%%%%%%%%%%%%%%%%%%%%%%%%%%%%%%%%%%%%%%%%%%%%%
%%%%%%%%%%%%%%%%%%%%%%%%%%%%%%%%%%%%%%%%%%%%%%
%%%%%%%%%%%%%%%%%%%%%%%%%%%%%%%%%%%%%%%%%%%%%%
\section{Conclusion}
%%%%%%%%%%%%%%%%%%%%%%%%%%%%%%%%%%%%%%%%%%%%%%
%%%%%%%%%%%%%%%%%%%%%%%%%%%%%%%%%%%%%%%%%%%%%%
%%%%%%%%%%%%%%%%%%%%%%%%%%%%%%%%%%%%%%%%%%%%%%

While we have developed no comprehensive theory for existence of resonance Van Hove singularities, the examples presented 
in this work indicate that they occur rather commonly.  Their presence may be due to disparate causes, including periodicity of 
Fourier space, anisotropy of the wave dispersion relation, or intersection of the trivial and non-trivial parts of the resonant 
manifold for 4-wave resonances. The basic requirement for such critical points is that there be distinct wavevectors 
with the same group velocity, which is facilitated by dispersion laws with segments strictly linear in wavenumber or 
with inflection points. Our several examples have presumably not exhausted the possible mechanisms to produce
such resonance singularities.   

The effects of the singularities on the kinetic theory can range from none at all, to moderate, to quite destructive. 
As a general rule of thumb, the singularities are more threatening in low dimensions $(d=1,2)$.
%For example, the critical points in the 4-wave resonant manifold that result from intersection with the trivial part have
%generic dimension $d-1$ and this is insufficient to produce a divergence in the phase measure on the resonant manifold 
%in dimension $D=2d$ for $d>1.$ 
For example, the non-degenerate critical points for the gravity-capillary wave system in $d=1$ illustrated in Fig.~8 
produce a logarithmic divergence in the phase measure, whereas the same system for the physical dimension 
$d=2$ has a locally finite phase measure near the critical points. 
Likewise, the singularities will generally be less important for $N$-wave resonances 
with $N$ large, since what matters is the size of the phase-space dimension $D=(N-2)d.$ The cautionary remark to these 
general rules of thumb is that degeneracy degree $\delta>0$ can lead to a stronger singularity at the critical set,
and this may result in a divergence even when $D$ is larger than 2. This is what occurs in the cases of three-dimensional inertial-waves 
and electron-hole excitations in two-dimensional graphene, for example. 

We collect our case studies in the table below. The table shows space dimension $d,$
order of resonance $N,$  geometry of the critical set,  degree of degeneracy $\delta,$ effective dimensionality 
$D'=d(N-1)-\delta$ of the phase space, local finiteness of 
the phase measure at the singularity (if any), and divergence or not of the standard collision integral. Note 
that we consider only genuine critical points, not pseudo-critical points (which are usually harmless). When the critical 
set is empty, we take $\delta=0$ in the definition of $D'.$ If the phase measure is locally infinite near the singularity, 
we indicate the nature of the divergence: 

\vspace{20pt} 
\begin{center}
   \begin{tabular}{| l | c | c | l | c | c | l | l |}
   \hline
    {\bf Wave system}   & {\bf d} & {\bf N} &  {\bf critical set} & {\boldmath $\delta$} &  ${\bf D'}$ & {\bf phase measure} & {\bf collision integral}
    \\
    \hline
    Isotropic power-law,  $ \alpha>1$  & $\geq 2$  & 3  & $\emptyset$ & $-$  & $d$ & finite & finite
    \\
    \hline

    Acoustic waves       & \,3\, & 3 & line & 1 & 2  & ill-defined &  divergent
        \\
    \hline
    Rossby/drift waves &  2 & 3 & point & 0 & 2  & log-divergent & finite
    \\
    \hline
    Inertial waves   & 3 & 3 &  line & 1  & 2  & log-divergent & divergent
    \\
    \hline
    Internal gravity waves & 3 & 3 & $\emptyset$ & $-$ & 3 & finite & finite
    \\
    \hline
    Surface gravity waves & 2 & 4 & $\emptyset$ & $-$ & 4 & finite & finite
    \\
    \hline
    Surface gravity-capillary waves & 2 & 4 & point  & 0 & 4 & finite & finite 
     \\
    \hline
    Light waves in optical fiber  & 1 & 4 & point & 1 & 1  & linear divergent & finite (but large)
    \\
    \hline
    Electrons \& holes in graphene & 2 & 4 & surface & 2 & 2 & log-divergent & divergent
    \\
    \hline
   \end{tabular}
\end{center}

\vspace{20pt}
\noindent The diligent reader will recognize that the table is a simplification of the discussion in the text and glosses
over some of the finer points. (For example, a degenerate critical point with $\delta=1$ is possible for gravity-capillary
waves at the inflection point of the dispersion relation when $k=k_*$.) Nevertheless, the results presented in the table 
support the general lessons educed above. In particular, there tend to be serious consequences for the standard kinetic 
description when the critical set is non-empty and $D'\leq 2.$ In such cases, closer examination of the problem is 
warranted, to see whether there are any ameliorating circumstances (vanishing interaction coefficients, cancellations 
in the collision integral, etc.) or whether the standard kinetic equation indeed breaks down. Resonance Van Hove singularities 
and their potential effects should be generally recognized as a possibility in wave kinetics. 

\vspace{20pt}
\noindent {\bf Acknowledgements} Both authors wish to thank Alan Newell for useful conversations on the subject of this paper. 
GE thanks participants of the 2013 Eilat Workshop ``Turbulence \& Amorphous Materials'' for discussions, in particular 
G. Falkovich and S. Lukaschuk. We also acknowledge the Institute for Pure \& Applied Mathematics at UCLA for support 
during the program ``Mathematics of Turbulence", where we carried out some of the work presented here.    

\newpage

%%%%%%%%%%%%%%%%%%%%%%%%%%%%%%%%%%%%%%%%%%%%%%
%%%%%%%%%%%%%%%%%%%%%%%%%%%%%%%%%%%%%%%%%%%%%%
%%%%%%%%%%%%%%%%%%%%%%%%%%%%%%%%%%%%%%%%%%%%%%
\appendix
%%%%%%%%%%%%%%%%%%%%%%%%%%%%%%%%%%%%%%%%%%%%%%
%%%%%%%%%%%%%%%%%%%%%%%%%%%%%%%%%%%%%%%%%%%%%%
%%%%%%%%%%%%%%%%%%%%%%%%%%%%%%%%%%%%%%%%%%%%%%

%%%%%%%%%%%%%%%%%%%%%%%%%%%%%%%%%%%%%%%%%%%%%%
\section{Perturbative Derivation of the Quantum Boltzmann Equation}\lb{perturbation}
%%%%%%%%%%%%%%%%%%%%%%%%%%%%%%%%%%%%%%%%%%%%%%

The Heisenberg equation of motion for the Hamiltonian (\ref{H0}),(\ref{H1}) are
\begin{eqnarray}
&&i\dot{\gamma}_{sa}(\bk) =\omega_{s}(k) \gamma_{sa}(\bk) +  \frac{\alpha}{2}\sum_{b}\sum_{s_2s_3s_4}
\int  \dbar^dk_2 \dbar^dk_3 \dbar^dk_4 \ddelta^d(\bk_1+\bk_2-\bk_3-\bk_4)
\cr
&&\,\,\,\,\,\,\,\,\,\,\,\,\,\,\,\,\,\,\,\,\,\,\,\,\,\times
[ T^{s_4s_3s_2s}_{\bk_4\bk_3\bk_2\bk}\gamma_{s_2b}^\dagger(\bk_2)\gamma_{s_3b}(\bk_3)
\gamma_{s_4a}(\bk_4)
- T^{s_4s_3ss_2}_{\bk_4\bk_3\bk\bk_2}\gamma_{s_2b}^\dagger(\bk_2)\gamma_{s_3a}(\bk_3)
\gamma_{s_4b}(\bk_4)
].
\end{eqnarray}
We shall write this using a self-explanatory shorthand notation as 
\be
i\dot{\gamma}_{1a} = \omega_1\gamma_{1a} + \frac{\alpha}{2}\sum_{b}\int d_{234}\ \delta^{12}_{34}
\left[
T_{4321} \gamma_{2b}^\dagger \gamma_{3b}\gamma_{4a} - T_{4312}\gamma_{2b}^\dagger\gamma_{3a}\gamma_{4b}
\right].
\ee
Hermiticity of the interaction Hamiltonian requires 
$$   T^*_{1234}=T_{4321}  $$
and, using the canonical anti-commutation relations, one can also impose the symmetry
$$ T_{1234}=T_{2143}. $$  
One can check that these relations are satisfied for the coefficient $T_{1234}$ 
arising from the Coulomb interaction between electrons \& holes in graphene by means of the explicit 
expression given in \cite{Fritzetal08}, Eq.(3.9). Using these symmetries, we have
\be
i\dot{\gamma}_{1a} = \omega_1\gamma_{1a} +
\frac{\alpha}{2} \sum_{b}\int d_{234}\ \delta^{12}_{34}
\left[
T_{1234}^* \gamma_{2b}^\dagger \gamma_{3b}\gamma_{4a}  + (3\leftrightarrow 4)
\right].
\lb{Heq-two} \ee
Introducing
$
\wt{\gamma}_{1a} = e^{i\omega_1 t}\gamma_1
$,
we have
\be
i\frac{d}{dt}\wt{\gamma}_{1a} = \frac{\alpha}{2}  \sum_{b}
\int  d_{234}\ \delta^{12}_{34} e^{i\omega^{12}_{34}t} \left[T_{1234}^* \wt{\gamma}_{2b}^\dagger \wt{\gamma}_{3b}
\wt{\gamma}_{4a}+ (3\leftrightarrow 4)\right].
\ee
Here $\omega^{12}_{34}=\omega_{s_1}(k_1)+\omega_{s_2}(k_2)-\omega_{s_3}(k_3)-\omega_{s_4}(k_4)$.
For simplicity, we shall omit the tilde ``$\,\,\wt{\!\,}\,\,$'' from now on.
We shall calculate the mean occupation number $n_1$ perturbatively by expanding the creation/annihilation operators
into a power series 
\be\label{expansion}
\gamma_{1a} = \gamma_{1a}^{(0)} + \alpha  \gamma_{1a}^{(1)} + \alpha^2  \gamma_{1a}^{(2)} + o(\alpha^2).
\ee
A straightforward calculation gives
\begin{eqnarray}\lb{op-exp1}
\gamma_{1a}^{(0)} &=& \gamma_{1a}(0)
\\ \lb{op-exp2}
\gamma_{1a}^{(1)} &=& -\frac{i}{2} \sum_{b}\int d_{234}\ \delta^{12}_{34} \Delta_t(\omega^{12}_{34})
 \left[T_{1234}^* \gamma_{2b}^\dagger  \gamma_{3b}
\gamma_{4a}+ (3\leftrightarrow 4)\right].
\\ \lb{op-exp3}
\gamma_{1a}^{(2)} &=& \frac{1}{4}\sum_{b,c}\int d_{234567}\ \delta^{12}_{34} T_{1234}^*
\cr
&\times\bigg[ &
  E_t(\omega^{167}_{345};\omega^{12}_{34}) \delta^{25}_{67}
T_{2567} \gamma_{7b}^\dagger \gamma_{6c}^\dagger \gamma_{5c} \gamma_{3b}\gamma_{4a}  
- E_t(\omega^{125}_{467};\omega^{12}_{34}) \delta^{35}_{67}
 T_{3567}^*\gamma_{2b}^\dagger \gamma_{5c}^\dagger \gamma_{6c} \gamma_{7b} \gamma_{4a}  
\cr
&-&  E_t(\omega^{125}_{367};\omega^{12}_{34}) \delta^{45}_{67}
 T_{4567}^*\gamma_{2b}^\dagger \gamma_{3b} \gamma_{5c}^\dagger \gamma_{6c} \gamma_{7a}  
+(6\leftrightarrow 7) + (3\leftrightarrow 4) + (3\leftrightarrow 4, 6\leftrightarrow 7)
 \bigg ]
\end{eqnarray}
where we employ the standard definitions \cite{BenneyNewell69}:
\be
\Delta_t(x) = \int_0^t \exp(ixs) ds, \quad E_t(x,y) = \int_0^t \Delta_s(x-y)\exp(isy) ds.
\ee
We shall omit $(0)$-superscript below, when there is no possibility of confusion.
The mean occupation number is obtained perturbatively by substituting (\ref{expansion}) into $\langle\gamma_{1'a}^\dagger \gamma_{1a}\rangle$.
\be\lb{sp-exp}
\llangle \gamma_{1'a}^\dagger \gamma_{1a}\rrangle= \llangle \gamma_{1'a}^{(0)\dagger} \gamma_{1a}^{(0)}\rrangle
+\alpha \left( \llangle\gamma_{1'a}^{(0)\dagger}\gamma_{1a}^{(1)}\rrangle + c.c \right)
+\alpha^2 \left( \llangle\gamma_{1'a}^{(0)\dagger}\gamma_{1a}^{(2)}\rrangle + c.c +
 \llangle \gamma_{1'a}^{(1)\dagger} \gamma_{1a}^{(1)} \rrangle\right).
\ee
We assume as initial condition a fermionic quasi-free state with the 2nd-order correlations 
$\langle\gamma_{1'a}^\dagger \gamma_{1b}\rangle = n_1\delta^1_{1'}\delta^{s_1}_{s_1'}\delta^a_b$,
$\langle\gamma_{1'a} \gamma_{1b}^\dagger\rangle = \wt{n}_1\delta^1_{1'}\delta^{s_1}_{s_1'}\delta^a_b$ 
and $\wt{n}_1=1-n_1$. Substituting (\ref{op-exp1})-(\ref{op-exp3}) into (\ref{sp-exp}), we have
\begin{eqnarray}\lb{sp-exp1}
\llangle\gamma_{1'a}^{(0)\dagger}\gamma_{1a}^{(1)}\rrangle
&=& - \frac{i}{2} \sum_{b} \int d_{234}\ \Delta_t(\omega^{12}_{34}) 
\left[ T_{1234}^*\llangle\gamma_{1'a}^\dagger\gamma_{2b}^\dagger\gamma_{3b}\gamma_{4a}\rrangle
+(3\leftrightarrow 4)\right]
\\ \lb{sp-exp2}
\llangle\gamma_{1'a}^{(0)\dagger}\gamma_{1a}^{(2)}\rrangle
&=&  \frac{1}{4}\sum_{b,c}\int d_{234567}\ \delta^{12}_{34} T_{1234}^*
\cr
&\times\bigg[ &
  E_t(\omega^{167}_{345};\omega^{12}_{34}) \delta^{25}_{67}
T_{2567} 
\llangle\gamma_{1'a}^\dagger\gamma_{7b}^\dagger \gamma_{6c}^\dagger \gamma_{5c} \gamma_{3b}\gamma_{4a}\rrangle \hspace{24pt}  \bigg\} I
\cr
&-& E_t(\omega^{125}_{467};\omega^{12}_{34}) \delta^{35}_{67}
 T_{3567}^*
 \llangle\gamma_{1'a^\dagger}\gamma_{2b}^\dagger \gamma_{5c}^\dagger \gamma_{6c} \gamma_{7b} \gamma_{4a}\rrangle \hspace{20pt} \bigg\} II
\cr
&-&  E_t(\omega^{125}_{367};\omega^{12}_{34}) \delta^{45}_{67}
 T_{4567}^*
 \llangle\gamma_{1'a}^\dagger \gamma_{2b}^\dagger \gamma_{3b} \gamma_{5c}^\dagger \gamma_{6c} \gamma_{7a} \rrangle \hspace{23pt} \bigg\} III
 \cr
&+&(6\leftrightarrow 7) + (3\leftrightarrow 4) + (3\leftrightarrow 4, 6\leftrightarrow 7)
 \bigg ]
 \\ \lb{sp-exp3}
 \llangle\gamma_{1'a}^{(1)\dagger}\gamma_{1a}^{(1)}\rrangle
&=&  \frac{1}{4}\sum_{b,c}\int d_{234567}\ \delta^{1'2}_{34} \delta^{15}_{67} T_{1'234}
\Delta_t(-\omega^{1'2}_{34})\Delta_t(\omega^{12}_{34})
\cr
&& \hspace{-60pt} \times\bigg[
T_{1567}^* 
\llangle\gamma_{4a}^\dagger\gamma_{3b}^\dagger \gamma_{2b} \gamma_{5c}^\dagger \gamma_{6c}\gamma_{7a}\rrangle
+
(6\leftrightarrow 7) + (3\leftrightarrow 4) + (3\leftrightarrow 4, 6\leftrightarrow 7)
 \bigg ]
\end{eqnarray}
Using Wick's rule, we have 
\begin{eqnarray}\lb{gamma01}
\llangle\gamma_{1'a}^{(0)\dagger}\gamma_{1a}^{(1)}\rrangle
&=& it \delta^1_{1'} \Omega_1n_1
\\ \lb{gamma02}
\llangle\gamma_{1'a}^{(0)\dagger}\gamma_{1a}^{(2)}\rrangle
&=& \delta^1_{1'}  (\mbox{I}+\mbox{II}+\mbox{III})
\\ \lb{gamma11}
\llangle\gamma_{1'a}^{(1)\dagger}\gamma_{1a}^{(1)}\rrangle
&=&\delta^1_{1'} (\mbox{IV})
\end{eqnarray}
where 
\begin{eqnarray}
\mbox{I}
&=-&\int d_{234}\ \delta^{12}_{34} \  E_t(0;\omega^{12}_{34})
R_{1234} \ n_{1}n_3n_4
+ E_t(0;0)\int d_2\ \Omega_2(NT_{1221}-T_{1212}) n_2
\cr
\mbox{II}
&=&
\int d_{234}\ \delta^{12}_{34}  \ E_t(0;\omega^{12}_{34})
R_{1234} \ (n_1n_2n_4+n_1n_2n_3)
- E_t(0;0) \int d_2\ \frac{\Omega_1+\Omega_2}{2}(NT_{1221}-T_{1212})n_2
\cr
\mbox{III}
&=-&
\int d_{234}\ \delta^{12}_{34} \  E_t(0;\omega^{12}_{34})
R_{1234}\ (n_1n_2\wt{n}_3+n_1n_2\wt{n}_4)
- E_t(0;0) \int d_2\ \frac{\Omega_1+\Omega_2}{2}(NT_{1221}-T_{1212})n_2
\cr
\mbox{IV}
&=&
\int d_{234} \ \delta^{12}_{34} \ |\Delta_t(\omega^{12}_{34})|^2 R_{1234}\ \wt{n}_{2}n_3n_4 
+
\frac{1}{2}|\Delta_t(0)|^2 \Omega_1^2 n_1
\end{eqnarray}
with
\begin{eqnarray}
R_{1234} &=& \frac{1}{2}T_{1234}^* (NT_{1234}-T_{1243}) + (3\leftrightarrow 4)
\\
\Omega_1 &=& \int d_2\ (NT_{1221}-T_{1212})\ n_2.
\end{eqnarray}

Now, recalling the standard relations \cite{BenneyNewell69}
\begin{eqnarray}
E_t(0;0)=\frac{t^2}{2},
\quad
\Delta_t(0) = t,
\end{eqnarray}
we see that the terms which involve these factors in the previous expressions are undesirable. Their growth is $O(t^2)$
at long times, by far the most secular behavior, but they do not correspond to terms in the expected kinetic equation. Fortunately,
it is straightforward to check that the sum of all these undesirable terms arising from I-IV exactly cancel.  
Also, the $O(\alpha)$ term from (\ref{gamma01}) has vanishing real part and thus gives a zero contribution
to the evolution of the occupation numbers. Then, using 
\begin{eqnarray}
2 \ {\rm Re}\, E_t(0;x) = |\Delta_t(x)|^2 = \frac{2(1-\cos(xt))}{x^2},
\end{eqnarray}
and
\be
n_1n_3n_4-n_1n_2n_4-n_1n_2n_3+n_1n_2\wt{n}_3+n_1n_2\wt{n}_4-\wt{n}_2n_3n_4 = 
n_1n_2\wt{n}_3\wt{n}_4-\wt{n}_1\wt{n}_2n_3n_4, 
\ee
together with $\langle\gamma_{1'a}^\dagger\gamma_{1a}\rangle=n_1\delta^1_{1'},$ we 
can combine (\ref{gamma02})-(\ref{gamma11}) to get 
\be\lb{closure}
n_1(t) = n_1(0) - \alpha^2 \int d_{234}\ \delta^{12}_{34} \frac{2(1-\cos(\omega^{12}_{34}t))}{(\omega^{12}_{34})^2}R_{1234} \left(n_1n_2\wt{n}_3\wt{n}_4-\wt{n}_1\wt{n}_2n_3n_4 \right).
\ee
This is equivalent to the expression (\ref{q-pert}) in the text. 

There is another approach in the literature for dealing with the undesirable $O(t^2)$ terms which we should briefly mention.
It is possible to exactly remove those terms by a {\it frequency renormalization} \cite{Zakharovetal92,Nazarenko11,NewellRumpf11}, introducing  
$$\wt{\omega}_1 = \omega_1 + \alpha \Omega_1 $$
into the Heisenberg equations (\ref{Heq-two}), which then becomes  
\be
i\dot{\gamma}_{1a} = \tilde{\omega}_1\gamma_{1a} +
\frac{\alpha}{2} \sum_{b}\int d_{234}\ \delta^{12}_{34}
\left[
T_{1234}^* \left(\gamma_{2b}^\dagger \gamma_{3b}\gamma_{4a}
-\langle\gamma_{2b}^\dagger \gamma_{3b}\rangle \gamma_{4a}
+ \langle \gamma_{2b}^\dagger\gamma_{4a}\rangle \gamma_{3b}
\right)  + (3\leftrightarrow 4)
\right].
\lb{Heq-three} \ee
The rest of the derivation is as before, except that now one defines 
$
\wt{\gamma}_{1a} = e^{i\tilde{\omega}_1 t}\gamma_1
$
using the renormalized frequency. The counterterms which appear in the renormalized Heisenberg equations
of motion can be readily checked to cancel all of the $O(t^2)$ terms in each of the individual expressions 
I-IV, without the necessity of adding them together. The final result is the same as (\ref{closure})
except that $\omega^{12}_{34}$ is replaced with $\tilde{\omega}^{12}_{34}=\tilde{\omega}_1+\tilde{\omega}_2
-\tilde{\omega}_3-\tilde{\omega}_4.$ 

This alternate procedure yields the same kinetic equation as discussed in the text, but with bare frequencies 
replaced by renormalized frequencies. There might thus naively appear to be an inconsistency between the 
two approaches. In particular, the quantum Boltzmann equation for electrons in graphene obtained by 
the alternate procedure would disagree with that derived previously \cite{Kashuba08,Fritzetal08}, with the 
bare frequency $\omega_s(\bk)=sv_Fk$ undergoing an additional renormalization or, equivalently, 
with an additional renormalization of the Fermi velocity $v_F$. However, this inconsistency is only apparent.
As is well-known in the wave turbulence literature (e.g. \cite{Zakharovetal92} p.71), the two approaches lead to equivalent 
kinetic equations to order $O(\alpha^2),$ since the frequency renormalization is $O(\alpha)$ and thus corrects 
the kinetic equation only to order $O(\alpha^3).$ The frequency renormalization is therefore entirely optional 
in the derivation of the kinetic equation at order $O(\alpha^2).$ The consistency of the two approaches is further 
evidenced by the fact that the undesirable $O(t^2)$ terms cancel completely at order $O(\alpha^2)$ without any 
use of a frequency renormalization.

\bibliographystyle{model1-num-names}
\bibliography{kin-hier}

\begin{thebibliography}{40}
\expandafter\ifx\csname natexlab\endcsname\relax\def\natexlab#1{#1}\fi
\providecommand{\bibinfo}[2]{#2}
\ifx\xfnm\relax \def\xfnm[#1]{\unskip,\space#1}\fi
%Type = Article
\bibitem[{{Peierls}(1929)}]{Peierls29}
\bibinfo{author}{R.~{Peierls}},
\newblock \bibinfo{title}{{Zur kinetischen {T}heorie der {W\"a}rmeleitung in
  Kristallen}},
\newblock \bibinfo{journal}{Annalen der Physik} \bibinfo{volume}{395}
  (\bibinfo{year}{1929}) \bibinfo{pages}{1055--1101}.
%Type = Book
\bibitem[{Zakharov et~al.(1992)Zakharov, L'vov, and Falkovich}]{Zakharovetal92}
\bibinfo{author}{V.~E. Zakharov}, \bibinfo{author}{V.~S. L'vov},
  \bibinfo{author}{G.~Falkovich}, \bibinfo{title}{Kolmogorov spectra of
  turbulence I. {W}ave turbulence}, Springer series in nonlinear dynamics,
  \bibinfo{publisher}{Springer Berlin}, \bibinfo{year}{1992}.
%Type = Book
\bibitem[{{Nazarenko}(2011)}]{Nazarenko11}
\bibinfo{author}{S.~{Nazarenko}}, \bibinfo{title}{{Wave turbulence}}, volume
  \bibinfo{volume}{825} of \textit{\bibinfo{series}{Lecture Notes in Physics,
  Berlin Springer Verlag}}, \bibinfo{year}{2011}.
%Type = Article
\bibitem[{{Newell} and {Rumpf}(2011)}]{NewellRumpf11}
\bibinfo{author}{A.~C. {Newell}}, \bibinfo{author}{B.~{Rumpf}},
\newblock \bibinfo{title}{Wave turbulence},
\newblock \bibinfo{journal}{Annu. Rev. Fluid Mech.} \bibinfo{volume}{43}
  (\bibinfo{year}{2011}) \bibinfo{pages}{59--78}.
%Type = Article
\bibitem[{Eyink and Shi(2012)}]{EyinkShi12}
\bibinfo{author}{G.~L. Eyink}, \bibinfo{author}{Y.-K. Shi},
\newblock \bibinfo{title}{Kinetic wave turbulence},
\newblock \bibinfo{journal}{Physica D} \bibinfo{volume}{241, 18}
  (\bibinfo{year}{2012}) \bibinfo{pages}{1487Ð1511}.
%Type = Article
\bibitem[{Spohn(2006)}]{Spohn06}
\bibinfo{author}{H.~Spohn},
\newblock \bibinfo{title}{The phonon {B}oltzmann equation, properties and link
  to weakly anharmonic lattice dynamics},
\newblock \bibinfo{journal}{J. Stat. Phys.} \bibinfo{volume}{124}
  (\bibinfo{year}{2006}) \bibinfo{pages}{1041--1104}.
%Type = Book
\bibitem[{H\"{o}rmander(1983)}]{Hormander89}
\bibinfo{author}{L.~H\"{o}rmander}, \bibinfo{title}{The analysis of linear
  partial differential operators {I}: {D}istribution theory and {F}ourier
  analysis}, volume \bibinfo{volume}{256} of
  \textit{\bibinfo{series}{Grundlehren der mathematischen Wissenschaften}},
  \bibinfo{publisher}{Springer-Verlag Berlin Heidelberg New York},
  \bibinfo{year}{1983}.
%Type = Article
\bibitem[{{Van Hove}(1953)}]{VanHove53}
\bibinfo{author}{L.~{Van Hove}},
\newblock \bibinfo{title}{{The occurrence of singularities in the elastic
  frequency distribution of a crystal}},
\newblock \bibinfo{journal}{Phys. Rev.} \bibinfo{volume}{89}
  (\bibinfo{year}{1953}) \bibinfo{pages}{1189--1193}.
%Type = Book
\bibitem[{Milnor(1963)}]{Milnor63}
\bibinfo{author}{J.~M. Milnor}, \bibinfo{title}{Morse theory},
  volume~\bibinfo{volume}{51} of \textit{\bibinfo{series}{Annals of mathematics
  studies}}, \bibinfo{publisher}{Princeton University Press},
  \bibinfo{year}{1963}.
%Type = Book
\bibitem[{Lang(2012)}]{Lang12}
\bibinfo{author}{S.~Lang}, \bibinfo{title}{Differential manifolds},
  \bibinfo{publisher}{Springer-Verlag New York}, \bibinfo{edition}{2nd}
  edition, \bibinfo{year}{2012}.
%Type = Book
\bibitem[{Guillemin and Pollack(1974)}]{GuilleminPollack74}
\bibinfo{author}{V.~Guillemin}, \bibinfo{author}{A.~Pollack},
  \bibinfo{title}{Differential topology}, \bibinfo{publisher}{Prentice-Hall,
  Inc., Englewood Cliffs, New Jersey}, \bibinfo{year}{1974}.
%Type = Article
\bibitem[{Newell and Aucoin(1971)}]{NewellAucoin71}
\bibinfo{author}{A.~C. Newell}, \bibinfo{author}{P.~J. Aucoin},
\newblock \bibinfo{title}{Semidispersive wave systems},
\newblock \bibinfo{journal}{J. Fluid Mech.} \bibinfo{volume}{49}
  (\bibinfo{year}{1971}) \bibinfo{pages}{593--609}.
%Type = Article
\bibitem[{L'vov et~al.(1997)L'vov, L'vov, Newell, and Zakharov}]{Lvovetal97}
\bibinfo{author}{V.~S. L'vov}, \bibinfo{author}{Y.~L'vov},
  \bibinfo{author}{A.~C. Newell}, \bibinfo{author}{V.~Zakharov},
\newblock \bibinfo{title}{Statistical description of acoustic turbulence},
\newblock \bibinfo{journal}{Phys. Rev. E} \bibinfo{volume}{56}
  (\bibinfo{year}{1997}) \bibinfo{pages}{390--405}.
%Type = Article
\bibitem[{{Kashuba}(2008)}]{Kashuba08}
\bibinfo{author}{A.~B. {Kashuba}},
\newblock \bibinfo{title}{{Conductivity of defectless graphene}},
\newblock \bibinfo{journal}{Phys. Rev. B} \bibinfo{volume}{78}
  (\bibinfo{year}{2008}) \bibinfo{pages}{085415}.
%Type = Article
\bibitem[{{Fritz} et~al.(2008){Fritz}, {Schmalian}, {M{\"u}ller}, and
  {Sachdev}}]{Fritzetal08}
\bibinfo{author}{L.~{Fritz}}, \bibinfo{author}{J.~{Schmalian}},
  \bibinfo{author}{M.~{M{\"u}ller}}, \bibinfo{author}{S.~{Sachdev}},
\newblock \bibinfo{title}{{Quantum critical transport in clean graphene}},
\newblock \bibinfo{journal}{Phys. Rev. B} \bibinfo{volume}{78}
  (\bibinfo{year}{2008}) \bibinfo{pages}{085416}.
%Type = Article
\bibitem[{{Zakharov} and {Sagdeev}(1970)}]{ZakharovSagdeev70}
\bibinfo{author}{V.~E. {Zakharov}}, \bibinfo{author}{R.~Z. {Sagdeev}},
\newblock \bibinfo{title}{{Spectrum of acoustic turbulence}},
\newblock \bibinfo{journal}{Sov. Phys. Dok.} \bibinfo{volume}{15}
  (\bibinfo{year}{1970}) \bibinfo{pages}{439}.
%Type = Article
\bibitem[{Balk et~al.(1990)Balk, Nazarenko, and Zakharov}]{Balketal90}
\bibinfo{author}{A.~M. Balk}, \bibinfo{author}{S.~V. Nazarenko},
  \bibinfo{author}{V.~Zakharov},
\newblock \bibinfo{title}{{Nonlocal turbulence of drift waves}},
\newblock \bibinfo{journal}{Sov. Phys. JETP} \bibinfo{volume}{71}
  (\bibinfo{year}{1990}) \bibinfo{pages}{249--260}.
%Type = Article
\bibitem[{Newell et~al.(2001)Newell, Nazarenko, and Biven}]{Newelletal01}
\bibinfo{author}{A.~C. Newell}, \bibinfo{author}{S.~Nazarenko},
  \bibinfo{author}{L.~Biven},
\newblock \bibinfo{title}{Wave turbulence and intermittency},
\newblock \bibinfo{journal}{Physica D} \bibinfo{volume}{152-153}
  (\bibinfo{year}{2001}) \bibinfo{pages}{520--550}.
%Type = Article
\bibitem[{Biven et~al.(2001)Biven, Nazarenko, and Newell}]{Bivenetal01}
\bibinfo{author}{L.~Biven}, \bibinfo{author}{S.~Nazarenko},
  \bibinfo{author}{A.~Newell},
\newblock \bibinfo{title}{Breakdown of wave turbulence and the onset of
  intermittency},
\newblock \bibinfo{journal}{Phys. Lett. A} \bibinfo{volume}{280}
  (\bibinfo{year}{2001}) \bibinfo{pages}{28--32}.
%Type = Article
\bibitem[{Galtier(2003)}]{Galtier03}
\bibinfo{author}{S.~Galtier},
\newblock \bibinfo{title}{{Weak inertial-wave turbulence theory}},
\newblock \bibinfo{journal}{Phys. Rev. E} \bibinfo{volume}{68}
  (\bibinfo{year}{2003}) \bibinfo{pages}{015301}.
%Type = Article
\bibitem[{Caillol and Zeitlin(2000)}]{CaillolZeitlin00}
\bibinfo{author}{P.~Caillol}, \bibinfo{author}{V.~Zeitlin},
\newblock \bibinfo{title}{{Kinetic equations and stationary energy spectra of
  weakly nonlinear internal gravity waves}},
\newblock \bibinfo{journal}{Dynam. Atmos. Oceans} \bibinfo{volume}{32}
  (\bibinfo{year}{2000}) \bibinfo{pages}{81--112}.
%Type = Article
\bibitem[{{Dyachenko} and {Zakharov}(1994)}]{DyachenkoZakharov94}
\bibinfo{author}{A.~I. {Dyachenko}}, \bibinfo{author}{V.~E. {Zakharov}},
\newblock \bibinfo{title}{{Is free-surface hydrodynamics an integrable
  system?}},
\newblock \bibinfo{journal}{Phys. Lett. A} \bibinfo{volume}{190}
  (\bibinfo{year}{1994}) \bibinfo{pages}{144--148}.
%Type = Article
\bibitem[{Newell and Zakharov(2008)}]{NewellZakharov08}
\bibinfo{author}{A.~C. Newell}, \bibinfo{author}{V.~E. Zakharov},
\newblock \bibinfo{title}{The role of the generalized {P}hillips' spectrum in
  wave turbulence},
\newblock \bibinfo{journal}{Phys. Lett. A} \bibinfo{volume}{372}
  (\bibinfo{year}{2008}) \bibinfo{pages}{4230--4233}.
%Type = Article
\bibitem[{Michel et~al.(2011)Michel, Garnier, Surret, Randoux, and
  Picozzi}]{Micheletal10}
\bibinfo{author}{C.~Michel}, \bibinfo{author}{J.~Garnier},
  \bibinfo{author}{P.~Surret}, \bibinfo{author}{S.~Randoux},
  \bibinfo{author}{A.~Picozzi},
\newblock \bibinfo{title}{Kinetic description of random optical waves and
  anomalous thermalization of a nearly integrable wave system},
\newblock \bibinfo{journal}{Lett. Math. Phys.} \bibinfo{volume}{96}
  (\bibinfo{year}{2011}) \bibinfo{pages}{415--447}.
%Type = Article
\bibitem[{Suret et~al.(2010)Suret, Randoux, Jauslin, and Picozzi}]{Suretetal10}
\bibinfo{author}{P.~Suret}, \bibinfo{author}{S.~Randoux},
  \bibinfo{author}{H.~Jauslin}, \bibinfo{author}{A.~Picozzi},
\newblock \bibinfo{title}{Anomalous thermalization of nonlinear wave systems},
\newblock \bibinfo{journal}{Phys. Rev. Lett.} \bibinfo{volume}{104}
  (\bibinfo{year}{2010}).
%Type = Article
\bibitem[{{M\"uller} et~al.(2009){M\"uller}, {Schmalian}, and
  {Fritz}}]{Muelleretal09}
\bibinfo{author}{M.~{M\"uller}}, \bibinfo{author}{J.~{Schmalian}},
  \bibinfo{author}{L.~{Fritz}},
\newblock \bibinfo{title}{{ Graphene: a nearly perfect fluid}},
\newblock \bibinfo{journal}{Phys. Rev. Lett.} \bibinfo{volume}{301}
  (\bibinfo{year}{2009}) \bibinfo{pages}{025301}.
%Type = Article
\bibitem[{Sachdev(1998)}]{Sachdev98}
\bibinfo{author}{S.~Sachdev},
\newblock \bibinfo{title}{{Nonzero-temperature transport near fractional
  quantum Hall critical points}},
\newblock \bibinfo{journal}{Phys. Rev. B} \bibinfo{volume}{57}
  (\bibinfo{year}{1998}) \bibinfo{pages}{7157}.
%Type = Book
\bibitem[{Lewin(1981)}]{Lewin81}
\bibinfo{author}{L.~Lewin}, \bibinfo{title}{Polylogarithms and associated
  functions}, \bibinfo{publisher}{North Holland}, \bibinfo{year}{1981}.
%Type = Article
\bibitem[{Benney and Newell(1969)}]{BenneyNewell69}
\bibinfo{author}{D.~J. Benney}, \bibinfo{author}{A.~C. Newell},
\newblock \bibinfo{title}{Random wave closures},
\newblock \bibinfo{journal}{Stud. Appl. Math.} \bibinfo{volume}{48}
  (\bibinfo{year}{1969}) \bibinfo{pages}{29--53}.
%Type = Article
\bibitem[{Shi and Eyink(2015)}]{ShiEyink15a}
\bibinfo{author}{Y.-K. Shi}, \bibinfo{author}{G.~L. Eyink},
\newblock \bibinfo{title}{Local well-posedness of the wave kinetic hierarchy},
\newblock \bibinfo{journal}{in preparation}  (\bibinfo{year}{2015}).
%Type = Article
\bibitem[{Lukkarinen and Spohn(2007)}]{LukkarinenSpohn07}
\bibinfo{author}{J.~Lukkarinen}, \bibinfo{author}{H.~Spohn},
\newblock \bibinfo{title}{Kinetic limit for wave propagation in a random
  medium},
\newblock \bibinfo{journal}{Arch. Rational Mech. Anal.} \bibinfo{volume}{183}
  (\bibinfo{year}{2007}) \bibinfo{pages}{93--162}.
%Type = Article
\bibitem[{de~Verdiere and Vey(1979)}]{ColinVey79}
\bibinfo{author}{Y.~C. de~Verdiere}, \bibinfo{author}{J.~Vey},
\newblock \bibinfo{title}{Le lemme de {M}orse isochore},
\newblock \bibinfo{journal}{Topology} \bibinfo{volume}{18}
  (\bibinfo{year}{1979}) \bibinfo{pages}{283 -- 293}.
%Type = Book
\bibitem[{Federer(1969)}]{Federer69}
\bibinfo{author}{H.~Federer}, \bibinfo{title}{Geometric measure theory},
  Grundlehren der mathematischen Wissenschaften, \bibinfo{publisher}{Springer},
  \bibinfo{year}{1969}.
%Type = Book
\bibitem[{Arnold et~al.(2012)Arnold, Varchenko, and Gusein-Zade}]{Arnoldetal12}
\bibinfo{author}{V.~I. Arnold}, \bibinfo{author}{A.~Varchenko},
  \bibinfo{author}{S.~M. Gusein-Zade}, \bibinfo{title}{Singularities of
  differentiable maps: {V}olume {I}: {T}he classification of critical points,
  caustics, and wave fronts}, Monographs in Mathematics,
  \bibinfo{publisher}{Birkh{\"a}user Boston}, \bibinfo{year}{2012}.
%Type = Article
\bibitem[{Gonz{\'a}lez et~al.(1999)Gonz{\'a}lez, Guinea, and
  Vozmediano}]{Gonzalezetal99}
\bibinfo{author}{J.~Gonz{\'a}lez}, \bibinfo{author}{F.~Guinea},
  \bibinfo{author}{M.~Vozmediano},
\newblock \bibinfo{title}{Marginal-fermi-liquid behavior from two-dimensional
  {C}oulomb interaction},
\newblock \bibinfo{journal}{Phys. Rev. B} \bibinfo{volume}{59}
  (\bibinfo{year}{1999}) \bibinfo{pages}{R2474}.
%Type = Article
\bibitem[{Son(2007)}]{Son07}
\bibinfo{author}{D.~T. Son},
\newblock \bibinfo{title}{Quantum critical point in graphene approached in the
  limit of infinitely strong {C}oulomb interaction},
\newblock \bibinfo{journal}{Phys. Rev. B} \bibinfo{volume}{75}
  (\bibinfo{year}{2007}) \bibinfo{pages}{235423}.
%Type = Article
\bibitem[{Hofmann et~al.(2014)Hofmann, Barnes, and Sarma}]{Hofmannetal14}
\bibinfo{author}{J.~Hofmann}, \bibinfo{author}{E.~Barnes},
  \bibinfo{author}{S.~D. Sarma},
\newblock \bibinfo{title}{Why does graphene behave as a weakly interacting
  system?},
\newblock \bibinfo{journal}{Phys. Rev, Lett.} \bibinfo{volume}{113}
  (\bibinfo{year}{2014}) \bibinfo{pages}{105502}.
%Type = Article
\bibitem[{Kotov et~al.(2012)Kotov, Uchoa, Pereira, Guinea, and
  Neto}]{Kotovetal12}
\bibinfo{author}{V.~N. Kotov}, \bibinfo{author}{B.~Uchoa},
  \bibinfo{author}{V.~M. Pereira}, \bibinfo{author}{F.~Guinea},
  \bibinfo{author}{A.~C. Neto},
\newblock \bibinfo{title}{Electron-electron interactions in graphene: Current
  status and perspectives},
\newblock \bibinfo{journal}{Rev. Mod. Phys.} \bibinfo{volume}{84}
  (\bibinfo{year}{2012}) \bibinfo{pages}{1067}.
%Type = Article
\bibitem[{Fedoryuk(1971)}]{Fedoryuk71}
\bibinfo{author}{M.~V. Fedoryuk},
\newblock \bibinfo{title}{The stationary phase method and pseudodifferential
  operators},
\newblock \bibinfo{journal}{Russ. Math. Surv.} \bibinfo{volume}{26}
  (\bibinfo{year}{1971}) \bibinfo{pages}{65--115}.
%Type = Article
\bibitem[{Shi and Eyink(2015)}]{ShiEyink15b}
\bibinfo{author}{Y.-K. Shi}, \bibinfo{author}{G.~L. Eyink},
\newblock \bibinfo{title}{Singular wave kinetics and electron transport in
  defectless graphene},
\newblock \bibinfo{journal}{in preparation}  (\bibinfo{year}{2015}).

\end{thebibliography}

\end{document}